\documentclass[aps,pra,twocolumn,superscriptaddress,floatfix,10pt,showpacs,final]{revtex4}
\usepackage[usenames]{color}
\usepackage{amsmath}
\usepackage{euscript}
\usepackage{graphicx}

\begin{document}
\title{Dissociative electron attachment to the H$_2$O molecule. I. Complex-valued potential-energy surfaces for the $^2B_1$, $^2A_1$, and $^2B_2$ metastable states of the water anion}
\author{Daniel J. Haxton}
\altaffiliation{Present address: Department of Physics and JILA, University of Colorado, Boulder, CO 80309, USA.}
\affiliation{Department of Chemistry, University of California, Berkeley, California 94720}
\affiliation{Lawrence Berkeley National Laboratory, Chemical Sciences, Berkeley, California 94720}

\author{C. W. McCurdy}
\affiliation{Lawrence Berkeley National Laboratory, Chemical Sciences, Berkeley, California 94720}
\affiliation{Departments of Applied Science and Chemistry, University of California, Davis, California 95616}

\author{T. N. Rescigno}
\affiliation{Lawrence Berkeley National Laboratory, Chemical Sciences, Berkeley, California 94720}

\pacs{34.80.Ht}

\begin{abstract}

We present the results of calculations defining global, three-dimensional representations of the complex-valued potential-energy surfaces of the $^2B_1$, $^2A_1$, and $^2B_2$  metastable  states of the water anion that underlie the physical process of
dissociative electron attachment to water.  The real part of the resonance energies is obtained from configuration-interaction calculations performed in a restricted Hilbert space, while the imaginary part of the energies (the widths) is derived from complex Kohn scattering calculations. A diabatization is performed on the $^2A_1$
and $^2B_2$ surfaces, due to the presence of a conical intersection between
them.  We discuss the implications that the shapes of the constructed
potential-energy surfaces will have upon the nuclear dynamics of 
dissociative electron attachment to H$_2$O.  

\end{abstract}

\maketitle

\section{Introduction }

Dissociative electron attachment (DEA) to the water molecule  proceeds through 
a number of channels, each with a different energetic threshold,
\begin{equation}
\mathrm{H}_2\mathrm{O} + e^- \to 
\begin{cases} 
\ \mathrm{H} + \mathrm{O}\mathrm{H}^- & 3.27\ \mathrm{eV}\\
\ \mathrm{H}_2 + \mathrm{O}^- & 3.56 \ \mathrm{eV}\\
\ \mathrm{H}^- + \mathrm{O}\mathrm{H} \quad (X \ ^2\Pi) & 4.35 \ \mathrm{eV}\\
\ \mathrm{H} + \mathrm{H} + \mathrm{O}^- & 8.04 \ \mathrm{eV}\\
\ \mathrm{H}^- + \mathrm{O}\mathrm{H}^* \quad (^2\Sigma) & 8.38 \ \mathrm{eV} \\
\ \mathrm{H}^- + \mathrm{H} + \mathrm{O}  & 8.75 \ \mathrm{eV}
\end{cases}
\label{asymptotes}
\end{equation}
The production of these species occurs 
via three metastable Born-Oppenheimer electronic
states of the H$_2$O$^-$ system, whose vertical transition energies therefore determine the 
incident energies at which DEA occurs.  Those electronic states of the anion are the
$^2B_1$, $^2A_1$, and $^2B_2$ Feshbach resonances, and they are responsible
for the three distinct peaks in the DEA cross section.  
Their potential-energy surfaces contain asymptotes corresponding
to the product channels listed in Eq.(\ref{asymptotes}), with
the exception of the H+OH$^-$ channel; this product is a result
of nonadiabatic effects.

Here we report the construction of the complex-valued adiabatic potential-energy surfaces associated with these resonance states, which will be used
within the local
complex potential (LCP) model \cite{BirtwistleHerzenberg,DubeHerzenberg,BardsleyWadehra,OmalleyTaylor,Omalley}
to calculate the nuclear dynamics leading to dissociation.  
The present study is followed
by a second paper\cite{paper2}, to which we will refer as paper II, in which we present the results
of nuclear dynamics calculations under the LCP model using the
calculated surfaces.

Dissociative electron attachment to water was studied as
early as 1930, in the experiment of Lozier\cite{Lozier},
and as recently as 2006, in the study
by Fedor \textit{et al.}\cite{fedor}.
These two experiments, along with the rest of the prior 
theoretical and experimental
work on this subject\cite{Buchel,Schultz,compton,Melton,sancheschultz,tjhall,belic,curtiswalker,
Claydon,Jungen,Gil,Morgan,Gorfinkiel}, have succeeded in characterizing
each of the product channels of Eq.(\ref{asymptotes}) and the three
Feshbach resonances involved in their production.  However, prior to our recent theoretical study of DEA to water via the lowest-energy
$^2B_1$ resonance\cite{haxton1,haxton2}, there had been no complete
theoretical treatment of this process, nor, in fact, any \textit{ab initio}
treatment of dissociative attachment to any molecule,
 involving more than one nuclear
degree of freedom.  We will give a more complete summary of the prior
theoretical and experimental results concerning the dynamics of this process in paper II.

The present treatment supersedes our previous study of DEA via the
lowest-energy $^2B_1$ state.
We have also studied the angular dependence of DEA
to H$_2$O and H$_2$S via the $^2B_1$ state of either anion~\cite{haxton4}.
Additionally,
we previously presented a qualitative study of the topology of the potential-energy surfaces of these three electronic states~\cite{haxton3}, including the many intersections that these surfaces exhibit.  That qualitative study
informs the present study, in which we construct quantitative surfaces.

\begin{figure}
\begin{center}
\includegraphics*[width=0.98\columnwidth]{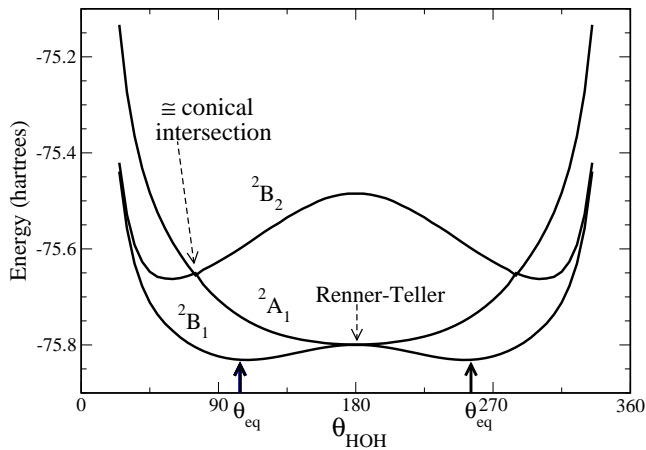}
\end{center}
\caption{Real parts of resonance energies, in units of hartrees, for OH bond
distance = 1.81$a_0$  in $C_{2v}$ geometry, plotted with respect
to bending angle, in degrees.}
\label{c2vfig} 
\end{figure}

Two of the most significant features of these surfaces can be seen in Fig. \ref{c2vfig},
reproduced here from Ref.\cite{haxton3}.
This figure depicts the behavior of the resonance energies with
respect to the H-O-H bending angle $\theta_{HOH}$, fixing the
OH bond lengths at $r_1$=$r_2$=1.81$a_0$.  The $^2B_1$ and
$^2A_1$ resonances are degenerate at linear geometry; this degeneracy will 
lead to Renner-Teller coupling between the two states.  In addition, there
is a conical intersection between the $^2A_1$ and $^2B_2$ surface
located at approximately $\theta_{HOH}$=73$^\circ$.  This conical
intersection plays a crucial role for the nuclear dynamics of
DEA via the uppermost $^2B_2$ resonance state.

The presence of a conical intersection between the adiabatic 
$^2A_1$ and $^2B_2$ states means that the nuclear dynamics
calculations that employ these potential-energy surfaces must either explicitly 
include the coupling between them, or be performed
in a diabatic basis with a smooth coupling term.  We perform an 
approximate diabatization of the 
calculated adiabatic $^2A_1$ and $^2B_2$ (1 and 2 $^2A'$) 
states to produce diabatic $^2A_1$ and
$^2B_2$ surfaces.  This diabatization is based upon the diagonalization 
of a particular symmetry operation described in Sec. \ref{sec:diab},
not the explicit minimization of derivative matrix elements.

A resonance state may be characterized by a
width, $\Gamma$, and an energy, $E_R$,
which are functions of the internuclear geometry $\vec{q}$.
These quantities define a complex potential surface $V(\vec{q})$, 
\begin{equation}
\label{lcppot}
V(\vec{q}) = E_R(\vec{q}) - i\frac{\Gamma(\vec{q})}{2}.
\end{equation}
The width $\Gamma$ is related to the lifetime as $\mathbf{\tau} = 1/\Gamma$ 
(we use atomic units throughout).
For a triatomic, the internal coordinates $\vec{q}$ may be
the set of bond-angle coordinates ($r_1$, $r_2$, $\theta$).  

We use separate techniques to define the two components of the potential-energy surfaces, $E_R(\vec{q})$ and $\Gamma(\vec{q})$.
We employ the complex Kohn variational 
method~\cite{kohn1, kohn2, kohn3, Kurilla, kohn5, kohn6, kohn7, kohn8, rmo95,rlm95} 
to perform scattering calculations which include the effect of the electronic
continuum upon the anion state.
These calculations yield both $E_R$ and $\Gamma$,
but we discard the value of $E_ R$ obtained from them.  In its place, we use bound-state
configuration-interaction (CI) calculations to obtain $E_R$.  

In the asymptotic regions,
the resonances become bound electronic states, and therefore the use of bound-state methods 
is entirely appropriate in those regions of nuclear geometry.  Near the Franck-Condon region where these states are resonances,
our CI treatment  restricts the included
configuration space to  eliminate the ground state electronic continuum
from the calculation.  Thus our CI calculations neglect the shift in $E_R$ due to coupling with that continuum, a well-known effect explained by the Feshbach resonance formalism \cite{Feshbach}.   Since the shift in $E_R$ by this coupling is generally of the same order as the width,   $\Gamma$,  this is an excellent approximation in regions where  the resonance is narrow.  The three resonance states we treat here have as their dominant configuration an electron attached to a singly excited configuration of the neutral target (Feshbach resonances), and thus tend to have small widths.  However, there are 
a some geometries where the width of one or the other of the two upper resonances is of the order of a few tenths of an eV, and in those regions the error in our calculated values of $E_R$ for that resonance are larger.

We also define a ground-state neutral H$_2$O potential surface. The resulting energetics of the anion surfaces relative to the ground state 
appear to reproduce the vertical transition energies and all but one of the two-body asymptotes
of these resonances very well, without a relative vertical adjustment
between the neutral and anion calculations.  However, as we will describe in 
detail, the CI calculations
fail to produce one of the two-body breakup asymptotes on one resonance surface, 
and we are forced to employ an \textit{ad hoc} patching procedure with another CI to
correct the unphysical behavior.  A similar patching procedure was necessary to correctly describe the 
three body breakup asymptotes, but dynamics leading to them is not part of the study 
we will present in paper II.

The data points calculated at a large set of nuclear geometries are assembled
into global representations of the potential-energy surfaces.  We construct
the global representations of the real part, $E_R$, and the width, $\Gamma$,
separately.
The global representations of the real parts of the 
adiabatic $^2B_1$ and the diabatic $^2A_1$ and $^2B_2$ surfaces
are defined by a sum of an
analytic fit and a spline of the residual error of this fit.  
We also define a global representation
of the off-diagonal coupling between the diabatic states.
The global representations of $E_R$ for the adiabatic 
$^2A_1$ and $^2B_2$  states
are obtained as the eigenvalues of the 2$\times$2 Hamiltonian matrix
defined by the fitted diabatic surfaces.  The adiabatic-to-diabatic transformation angle
is applied to the constructed width surfaces to obtain diabatic
widths.  The final result is a complete set of complex potential surfaces for the full dynamics calculations we report in paper II.

The outline of this paper is as follows.  We begin in  Sec. \ref{elecsect} with
a description of the electronic structure of the resonance states and the
basic features of their potential-energy surfaces, reviewing the
results on the topology of those surfaces of Ref.~\cite{haxton3}.  
In Sec. \ref{scatsect} we discuss the complex Kohn scattering
calculations from which we obtain $\Gamma(\vec{q})$, present the results of
these calculations, and describe
the construction
of the global potential-energy surfaces from the individual data points.
In Sec. \ref{CIsect} we do the same for the
configuration-interaction calculations defining $E_R(\vec{q})$.
In Sec. \ref{surfsect} we present the constructed potential-energy
surfaces and discuss their features.

\section{Electronic structure of the resonance states} \label{elecsect}

The electronic states that are primarily involved in
dissociative electron attachment to H$_2$O are
the $^2B_1$, $^2A_1$, and $^2B_2$ Feshbach resonances  \cite{Lozier,fedor,Buchel,Schultz,compton,Melton,sancheschultz,tjhall,belic,curtiswalker,
Claydon,Jungen,Gil,Morgan,Gorfinkiel,haxton1,haxton2,haxton3,haxton4}.  At the equilibrium geometry of neutral
H$_2$O, $r_1=r_2=1.81a_0$, $\theta=104.5^\circ$, where $a_0$ is the Bohr radius
0.529189379 $\times$ 10$^{-10}$ m, these states are characterized by their dominant electronic configurations
\begin{description}

\item \qquad $^2B_1$ : \quad 1$a_1^2$ 2$a_1^2$ 1$b_2^2$ 3$a_1^2$ 1$b_1^1$ 4$a_1^2$ 

\item \qquad $^2A_1$ : \quad 1$a_1^2$ 2$a_1^2$ 1$b_2^2$ 3$a_1^1$ 1$b_1^2$ 4$a_1^2$ 

\item \qquad $^2B_2$ : \quad 1$a_1^2$ 2$a_1^2$ 1$b_2^1$ 3$a_1^2$ 1$b_1^2$ 4$a_1^2$ 

\end{description}
such that each is described as [H$_2$O]$^{-1}$4$a_1^2$.  The parent state
of each corresponds to the analagous configuration [H$_2$O]$^{-1}$4$a_1^1$,
which may be a singlet or a triplet.
Since the resonance states are doublets, both the singlet and
triplet states of the neutral may be considered the parent state to which the incident electron is attached.

The peaks
in the DEA cross section occur at
approximately 6.5, 8.4, and 11.8eV,
respectively, at the equilibrium geometry of the neutral; these values
approximate the vertical transition energies of the resonances.
The ionization energy of H$_2$O is 12.621eV.  The vertical transition
energy to excite the $^2B_2$ resonance is close this value and we cannot 
eliminate the possibility that it is physically above the ionization energy.
It is possible, therefore, that the parent state of the  $^2B_2$ resonance is 
an autoionizing excited state of neutral H$_2$O\textemdash at least at some geometries. 

From the equilibrium geometry of the neutral, the resonance states may
be followed adiabatically toward the two-body breakup arrangements in which 
the electron is attached to one of the fragments  or H-OH or O-H$_2$.  
We identified the proper asymptotes
in these two arrangements previously~\cite{haxton3}.  In the first of these arrangements, 
the $^2B_1$ and $^2A_1$
states correlate with H$^-$+OH ($^2\Pi$), while the $^2B_2$ state correlates
with H$^-$+OH ($^2\Sigma$), leaving the OH fragment excited.  In the second, the $^2B_1$ state correlates with
ground-state H$_2$+O$^-$.  The $^2A_1$ state is found at a much higher energy, as
H$_2$(triplet 1$\sigma_g$1$\sigma_u$)+O$^-$, and does not have a bound two-body
asymptote in this arrangement.  

An additional complication is that the $^2B_2$ surface is inherently double-valued.
As described in Ref.~\cite{haxton3}, 
its lower asymptote is ground-state H$_2$+O$^-$, while the upper asymptote of this
surface in this arrangement is not electronically bound, and corresponds to
O ($^1D$) plus a resonant state of H$_2^-$.   In our previous study~\cite{haxton3},
we did not determine which resonant state of H$_2^-$ is involved.  
The double-valuedness of the $^2B_2$ surface presents a problem for our
treatment because we are unable to fully characterize both sheets of this
surface for all geometries.   We instead define
a single surface which interpolates between the two sheets within the three-body
breakup region. We will attempt to identify any discrepancies between the results here and
the experimental ones which we might ascribe to this omission.  However, it
is likely that this feature of the adiabatic manifold only plays a significant
role for three-body breakup, a channel which we will not consider in paper II.

We do, however, include the $^2A_1$-$^2B_2$ conical intersection in the
current treatment, because it is critical for a description of the dynamics
of dissociative electron attachment via the $^2B_2$ state.  This conical
intersection is a consequence of the crossing of the 1$b_2$ and 3$a_1$ orbital
energies as the bond angle is varied, and has analogs in both the singlet
and triplet $A_1$ and $B_2$ states of the neutral, which exhibit conical intersections
near the present one.

The symmetry labels $^2B_1$, $^2A_1$, and $^2B_2$ are appropriate 
when the H$_2$O molecule has $C_{2v}$
symmetry, i.e., when the OH bond lengths are equal.  When the OH bond lengths
are unequal, the molecule belongs to the C$_s$ point group.
The appropriate symmetry labels in those geometries  are $^2A''$, 1~$^2A'$, and 2 $^2A'$.  Due to the
conical intersection, the lower (1) and upper (2) $^2A'$ states may each 
correspond to either $^2A_1$ or $^2B_2$, depending on the bond angle.  A complete 
discussion of the topology of the anion surfaces is given in Ref. \cite{haxton3}.

\section{Fixed-nuclei electron scattering calculations}  \label{scatsect}

The resonance positions and widths are extracted from the results of fixed-nuclei scattering calculations carried out at physical (real) energies. The scattering calculations yield an \textit{S} matrix whose energy dependence is analyzed to determine the location of the resonance pole in the complex energy plane as a function of nuclear geometry. These calculations are fully \textit{ab initio}.  We use the complex Kohn variational
method~\cite{kohn1, kohn2, kohn3, Kurilla, kohn5, kohn6, kohn7, kohn8,
  rmo95,rlm95}, which provides a stationary principle for the \textit{S} matrix.  
Since detailed descriptions of the method have been given
elsewhere \cite{rmo95,rlm95},  we
will limit ourselves here to a brief summary to establish the terminology
we will use to describe our numerical calculations.

The ($N+1$)-electron scattering
wavefunction is represented explicitly in this calculation, using the standard methods of
quantum chemistry: one-electron molecular orbitals are assembled as sums of multicenter contracted
Cartesian Gaussian functions, products of which define $N$-electron configuration state
functions used to construct the target states. This basis, augmented with additional Gaussian functions and numerical continuum functions to describe the scattered waves, is also used to expand the ($N+1$)-electron wave function.

Thus, key components of this calculation include the appropriate choice of one-electron orbital 
and multielectron configuration bases defining the target states and the
resonance state.  Not only must the resonance state be accurately represented, but also the
target states into which it decays.  This requirement becomes more difficult with increasing
resonance energy, as more target states become energetically accessible as decay channels.

\subsection{Representation of the wavefunction and matrix elements  }

In our implementation of the complex Kohn variational method,
the ($N+1$)-electron scattering wavefunction is expanded as
\begin{equation}
\begin{split}
\Psi^{(+)}_{\Gamma_0} = \EuScript{A} \bigg[ & 
\sum_\Gamma 
\chi_\Gamma(
\mathbf{r}_1 ... \mathbf{r}_N) 
F^{(+)}_{\Gamma\Gamma_0}(\mathbf{r}_{N+1}) \\
 & +
\sum_\mu
d_\mu^{\Gamma_0}\Theta_\mu
(
\mathbf{r}_1 ... \mathbf{r}_{N+1} 
)\bigg] .
\end{split}
\label{eq:trialwf}
\end{equation}
The first sum in Eq. (\ref{eq:trialwf}) is over target states $\chi_\Gamma(\mathbf{r}_1 ... \mathbf{r}_N)$
explicitly included
in a close-coupling expansion, which may be energetically open or closed.
The antisymmetrizer is denoted
by  $\EuScript{A}$, and the scattered wave associated with channel $\Gamma$ is further expanded as
\begin{equation}
\begin{split}
F^{(+)}_{\Gamma\Gamma_0}(\mathbf{r}) = &  \sum_i c_i^{\Gamma\Gamma_0}\varphi_i(\mathbf{r}) + \delta_{\Gamma\Gamma_0}\sum_{lm} 
\Big[
J_{lm}(k_\Gamma {\mathbf r})\delta_{ll_0}\delta_{mm_0}  \\
&+
T_{ll_0mm_0}^{\Gamma\Gamma_0} G_{l m}^{(+)}(k_\Gamma, {\mathbf r})
\Big]
/{k_\Gamma^{\frac{1}{2}}r}
\end{split}
\label{eq:channelfcn}
\end{equation}
for incoming boundary conditions in the target channel $\Gamma_0$ with initial
$l_0m_0$ quantum numbers for the incident electron.
The functions $\varphi_i(\mathbf{r})$ in Eq.(\ref{eq:channelfcn}),
which we denote as ``scattering orbitals'', are Gaussian molecular
orbitals that are orthogonal to the ``target orbitals'' used to
construct the target states $\chi_\Gamma$.  Thus, one of the first steps in the complex
Kohn calculation is the partitioning of the one-electron Hilbert space into the sets of 
target and scattering orbitals.  The functions $J_{lm}$ and $G_{l m}^{(+)}$ in Eq.(\ref{eq:channelfcn})
are constructed from products of radial functions 
($j_{l}$ and $g_{l }^{(+)}$) times  angular functions that are real-valued combinations of spherical harmonics $ Y_{lm}$ consistent with the spatial symmetry of the anion. These functions are then Schmidt orthogonalized to the target and scattering molecular orbitals. 
The function $j_{l}$ is the regular
Ricatti-Bessel function, while  $g_{ l }^{(+)}$
is a numerically generated continuum function that is regular at
the origin and behaves asymptotically like the outgoing Riccatti-Hankel
function:
\begin{equation}
g^{(+)}_{\Gamma lm}(k_\Gamma , r) \underset {r\rightarrow \infty} {\longrightarrow}
h^{(+)}_l(k_\Gamma r).
\end{equation}
It is obtained by solving the driven radial equation,
\begin{equation}
\left( k_\Gamma^2 - \frac{\partial^2}{\partial r^2} +\frac{l(l+1)}{r^2}\right) g_{\Gamma l}^{(+)}(k_\Gamma ,r) = 
       r j_l(k_\Gamma r) \exp(-\alpha r^2),
\end{equation}
subject to the stated boundary conditions. For all calculations presented here, we use
$\alpha=0.04$.

For energetically closed channels, only the scattering orbitals $\varphi_i(\mathbf{r})$
are included in the sum in Eq.(\ref{eq:channelfcn}); the 
continuum functions $j_{lm}$
and $g_{\Gamma l m}^{(+)}$ are not included.  Thus, the calculation can give \textit{S} matrices that are discontinuous across channel
thresholds.  To minimize this problem, we include additional diffuse Gaussian functions in the scattering
orbital basis to represent the wavefunction in 
barely-bound channels.  In any case, the
resonance energies can be formally discontinuous across channel thresholds;
their discontinuities are not due to errors in the calculation,
but instead are properties of the true resonance states~\cite{Newton}.

The second sum in Eq.(\ref{eq:trialwf}) is over square-integrable
($N+1$)-electron configurations $\Theta_\mu$ constructed exclusively from
target orbitals.  For
convenience we refer to the ($N+1$)-electron configurations
as the ``\textit{Q} space'' and to the close-coupling part of the expansion
of the wavefunction as the ``\textit{P} space'' of the calculation.  The \textit{P} space 
may be further divided into the bound component\textemdash
built from ($N+1$)-electron configurations incorporating only Gaussian orbitals\textemdash
and the ``free"  component, corresponding to the target states times continuum functions.

Two approximations are made in calculating the matrix elements of the electronic
Hamiltonian with respect to the antisymmetrized basis functions of Eq.(\ref{eq:trialwf}). The 
exchange portion of all matrix elements within \textit{P} space that involve  
``free" components, as well as the matrix elements between \textit{Q} space and the free components of \textit{P} space, are assumed to be zero.  These approximations follow from the orthogonalization
of the free functions to the bound molecular orbitals and  the assumed completeness of the combined sets of target + scattering orbitals over the restricted region of space spanned by the target orbitals. 
Errors associated with these approximations are minimized by keeping the target orbitals  compact and 
by augmenting the set of scattering orbitals with functions that extend beyond the target orbitals so that
the orthogonalized continuum functions are separated from the target orbitals by
a large region of space. Further details concerning these approximations can be
found in Refs. \cite{rmo95,rlm95}.

\subsection{Target states and basis }

The description of a complex Kohn calculation requires the
specification of the target states $\chi_\Gamma$, the
\textit{Q} space configurations $\Theta_\mu$, and, for the expansion of the
channel eigenfunctions, the scattering orbitals $\varphi_i$
and the $lm$ pairs included in the asymptotic partial wave expansion.
We first turn our attention to the target orbitals and states.

\begin{table*}

\begin{ruledtabular}
\caption{\label{statestable}  Fifteen lowest ${\rm H_2O}$ target states
for Kohn calculation: energies at the equilibrium geometry
$r_1$=$r_2$=1.8$a_0$, $\theta=104.5^\circ$, and coefficient of dominant
configuration in CI expansion, compared to energies from van Harrevelt and van
Hemert~\cite{robsurf} and from our previous complex Kohn study\cite{haxton1}.}
\begin{tabular}{c|c|ccc|cc}
  State  &  Energy (hartree) & \multicolumn{3}{c}{\underline{{\qquad \qquad Excitation energy (eV) \qquad \qquad }}} & \multicolumn{2}{c}{Dominant config.} \\
  {}  &  Current & Current  &  Prev\cite{haxton1} & Ref.~\cite{robsurf} & Config. & Coef.  \\
\hline  
      ${ ^1A_1}$       &  -76.0417104    & 0.0   &    0.0       &    0.0    &      & 0.9883  \\
      ${ ^3B_1}$       &  -75.8042082    &         &    6.463   &           &   1$b_1$ $\rightarrow$ 4$a_1$   & 0.9944  \\
      ${ ^1B_1}$       &  -75.7907517    & 6.829   &    7.932   &    7.63   &   1$b_1$ $\rightarrow$ 4$a_1$   & 0.9960  \\
      ${ ^3A_2}$       &  -75.7294891    & 8.496   &    9.511   &           &   1$b_1$ $\rightarrow$ 2$b_2$   & 0.9966  \\
      ${ ^1A_2}$       &  -75.7268244    & 8.568   &    9.611   &    9.60  &   1$b_1$ $\rightarrow$ 2$b_2$   & 0.9963  \\
      ${ ^3A_1}$       &  -75.7196285    & 8.764   &    9.926   &           &   3$a_1$ $\rightarrow$ 4$a_1$   & 0.9588  \\
      ${ ^3A_1}$       &  -75.7114283    & 8.987   &            &           &   1$b_1$ $\rightarrow$ 2$b_1$   & 0.9593  \\
      ${ ^1A_1}$       &  -75.7100565    & 9.025   &   10.534   &    9.95   &   1$b_1$ $\rightarrow$ 2$b_1$   & 0.7224  \\

      ${ ^3B_1}$       &  -75.7072318    & 9.101   &            &      &    1$b_1$ $\rightarrow$ 5$a_1$   & 0.9959  \\
      ${ ^1B_1}$       &  -75.7060192    & 9.134   &            &      &    1$b_1$ $\rightarrow$ 5$a_1$   & 0.9946  \\
      ${ ^1A_1}$       &  -75.6876243    & 9.635   &            &      &    3$a_1$ $\rightarrow$ 4$a_1$   & 0.7012  \\
      ${ ^3B_1}$       &  -75.6863850    & 9.669   &            &      &    1$b_1$ $\rightarrow$ 6$a_1$   & 0.9936  \\
      ${ ^1B_1}$       &  -75.6826701    & 9.770   &            &      &    1$b_1$ $\rightarrow$ 6$a_1$   & 0.9951  \\
      ${ ^3B_2}$       &  -75.6691870    & 10.137  &            &      &    1$b_1$ $\rightarrow$ 1$a_2$   & 0.9952  \\
      ${ ^1B_2}$       &  -75.6688136    & 10.147  &            &      &   1$b_1$ $\rightarrow$ 1$a_2$   & 0.9940  \\
\end{tabular}
\end{ruledtabular}
\end{table*}  

The one-electron orbital basis was constructed entirely from
SCF orbitals.  We began with the following primitive Gaussian
basis.  On the oxygen, we used Dunning's triple-$\zeta$ basis~\cite{Dunning}
plus polarization ($d$) and Rydberg ($s$, $p$, and $d$) functions, augmented
with an $s$ function with exponent 0.0955 and a $p$ function with exponent
0.774.  On the hydrogen, we modified the double-$\zeta$ plus diffuse basis
of Chipman~\cite{Chipman}.  Chipman's basis consists
of four contracted $s$ functions and one $p$ function.  The most diffuse $s$ function
has exponent 0.0483.  We replaced the single $p$ function with two $p$ functions with
exponents 0.55 and 0.13, which were chosen to minimize the energy of the H$^-$ anion
given by full CI in this basis, which was -0.52190 hartrees, corresponding to a hydrogen
electron affinity of 0.59591eV, which recovers most of the experimental value of 0.75419eV.
The total size of the contracted target basis was 54.

Using this basis, we first performed a two-shell, generalized SCF calculation corresponding to the average of a ten-electron (neutral H$_2$O)
and nine-electron (cation H$_2$O$^+$) SCF calculation.
The purpose of such a calculation is to obtain a basis that can describe both the neutral and Feshbach resonance wavefunctions, recalling from the previous section that the Feshbach
resonances may be approximately described as two 4$a_1$ electrons bound to different
states of the cation core.  Further details about the generalized SCF calculation can be found in the EPAPS archive\cite{epaps}.

The set of target orbitals used to construct the neutral states $\chi_\Gamma$ included a total of eleven orbitals\textemdash the five SCF orbitals plus six of 
the virtual orbitals: three $a_1$, labelled 4$a_1$ - 6$a_1$; the 1$a_2$; the 2$b_1$; and the
2$b_2$.  The target states were obtained from a restricted configuration-interaction 
calculation within this 11-orbital space, in which the 1$a_1$ and 2$a_1$ orbitals were
constrained to be doubly occupied, and all single and double excitations from
the 1$b_2$, 3$a_1$, and 1$b_1$ orbitals into the set of six virtual orbitals were
included.  This ``all singles and doubles'' calculation is designed to account for the
single excitations describing the dominant configurations of the excited states,
plus relaxation of the remaining orbitals to first order; it also can describe correlation
in the target, particularly in the ground state.

Near the equilibrium geometry of the neutral, 
this description of the target states puts the excited states slightly lower
in energy, relative to the ground state, than appropriate, but otherwise represents key
features of their potential-energy surfaces well.  In Table \ref{statestable}, we list the
energies of the low-lying target states at the equilibrium geometry of the neutral, along
with the results of van Harrevelt and van Hemert~\cite{robsurf}.  

For the close-coupling expansion of the scattering wavefunctions, we selected sets of target states $\chi_\Gamma$
based on spin and spatial symmetry.  At higher energies, a larger number of
target states was required to converge the complex Kohn calculation, and therefore we used a different
number of target states for calculations on the $^2A''$ ($^2B_1$), the 1~$^2A'$ ($^2A_1$ or $^2B_2$), and the 2 $^2A'$ ($^2B_2$ or $^2A_1$) states.
For calculations of the $^2A''$ resonance, we included 15 target states, which near equilibrium geometry are the 15 lowest energy states listed
in Table \ref{statestable}.  For the 1~$^2A'$ state calculation, we added 14 additional states of $A'$ symmetry, for a total of 29 states. For the 2 $^2A'$ calculation, we added an additional eight $A''$ states, for a total of 37 states. The calculations are all converged with respect to the target state expansion at the equilibrium geometry
of the neutral, but there are some geometries at which the calculations are not fully converged.

The ($N+1$)-electron configurations $\Theta_\mu$ are constructed from the target orbitals and 
describe correlation, relaxation, and the penetration
of the incident electron into the target.  For the present calculation, they include the dominant configurations which contribute to the resonant states.  This set of configurations was again obtained by keeping the 1$a_1$ and 2$a_1$ orbitals  doubly occupied and distributing the other seven electrons over the remaining target orbitals, including at least one but no more than three electrons  in the (4$a_1$, 5$a_1$, 6$a_1$, 1$a_2$, 1$b_1$, 2$b_2$) virtual space.

\subsection{Scattering orbitals and insertion basis }

Of the original 54 molecular orbitals, 11 were used for the description of the target and
the remaining 43 orbitals were included in the set of scattering orbitals
$\varphi_i$.  To these 43 orbitals we added an additional set of contracted
Gaussian functions, all centered on the oxygen.  This "insertion basis" is included to improve the two approximations mentioned above associated with  the 
neglect of certain bound-free and free-free matrix elements.

The insertion basis was constructed from even-tempered  sets of eight $p$- and $d$-type 
primitive Gaussian functions, $\eta_i$. (To avoid linear dependence, we do not include $s$-type Gaussians separately, but rather include all Cartesian components of the $d$-type functions.)  The exponents started at 0.07 for the $p$-wave set and 0.075 for the $d$-wave set, and in both cases the ratio of consecutiveexponents was 0.8.  From these sets of primitive functions, we constructed a set of six $p$ and six $d$-type contracted, orthogonal functions that were designed
to have the minimum overlap with the most diffuse functions in the target basis.
To this end, we constructed a matrix representation, in the primitive insertion basis, of the operator $P$ which projects onto the space spanned by the most diffuse functions included in the target basis, i.e.,
\begin{equation}
P_{ij} = \sum_k \langle \eta_i \vert \phi_k \rangle \langle \phi_k \vert \eta_j \rangle,
\end{equation}
where the $\phi_k$ are the most diffuse functions of the target basis, Schmidt-orthogonalized.  These functions comprised the two $s$- , two $p$- and one $d$-type target functions with the smallest exponents.

Diagonalization of $P$ gave two $p$- and two $d$-type vectors with eigenvalues close to unity, 
which were discarded. The other six $p$- and six $d$-type functions $\Psi_n$  were retained  as the insertion basis, giving a total of 6 $\times$ (3+6) = 54 additional Cartesian Gaussian functions and a total (target plus augmented scattering) set of 108 functions. The exponents and contraction coefficients of the insertion basis are listed in the EPAPS archive\cite{epaps}.

\begin{figure}
\resizebox{1.0\columnwidth}{!}{\includegraphics{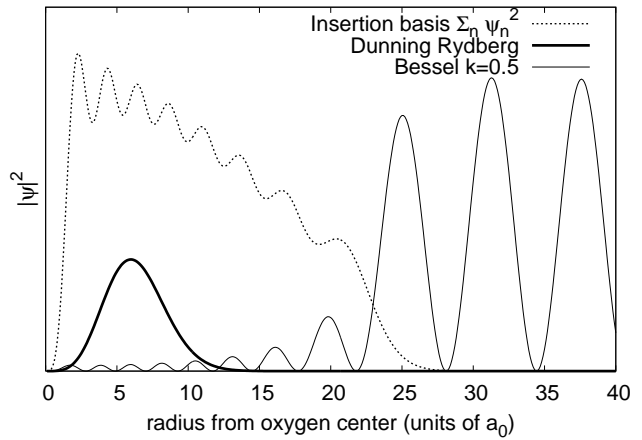}}
\caption{\label{basisfigure}
Modulus squared of diffuse basis functions for $l$=1 ($p$ waves) used in the Kohn calculations, centered on oxygen nucleus:
sum of insertion basis (dashed line), Rydberg p-function with exponent 0.028 (thick solid line), and orthogonalized Bessel function $j_1(kr)$
(thin solid line).  The Bessel function has $k$=0.5$a_0^{-1}$ and arbitrary normalization.
See text for further explanation.}
\end{figure}

We should remark that the insertion basis we use is very similar to
the exterior basis used by Nestmann and others~\cite{nestmann} for use
in \textit{R}-matrix calculations.  In the \textit{R}-matrix method,  the Gaussian basis must represent the outgoing
wave within the finite \textit{R}-matrix box.
Nestmann's basis is designed for a 20-bohr box and represents the continuum functions
well for 
energies up to about 12 eV.  In the current work, the exponents were chosen to provide
as complete a basis as possible within approximately 20 atomic units from
the oxygen center.

A pictorial representation of the basis set used in the Kohn calculations,
which demonstrates the extent of the diffuse target, insertion, and continuum 
basis functions, is shown in Fig. \ref{basisfigure}.  In this figure, the
squared modulus of various members of the diffuse and continuum basis for $l$=1
($p$ waves) is
plotted with respect to the radial distance from the oxygen nucleus.  The most diffuse
basis function used in the target calculations, with exponent 0.028, is plotted
along with the sum of the squared moduli of the insertion basis, $\sum_n \Psi_n^2$,
and that of the $l$=1 Bessel function, orthogonalized to the insertion basis.

\subsection{Extracting resonance parameters }

The complex Kohn calculation was performed at 10 to 20 energies around the resonance location,
and the \textit{S} matrices produced were fitted to a Breit-Wigner form,
\begin{equation}
S_{nlm,n'l'm'}(E) = S^{bg}_{nlm,n'l'm'}(E) +
\frac{\gamma_{nlm}\gamma_{n'l'm'}}{E-E_R+i\frac{\Gamma}{2}},
\label{bwform}
\end{equation}
where the background $S^{bg}$ is either linear or quadratic in E, 
to obtain the resonance energy $E_R$ and the width $\Gamma$ at each geometry, as well
as the partial amplitudes $\gamma_{nlm}$, labeled by the decay channel index $nlm$, where $n$
is electronic target state and $lm$ label the angular momentum of the emitted electron.
The modulus squared of the partial amplitudes corresponds to partial widths, and in particular,
the partial width with respect to decay into the electronic channel $n$ is
\begin{equation}
\Gamma_n = \sum_{lm} \left\vert \gamma_{nlm} \right\vert^2.
\end{equation}
Unitarity of the \textit{S} matrix implies $\sum_n \Gamma_n = \Gamma$.

Dissociative electron attachment to water occurs at incident electron
energies sufficient to excite multiple states of the neutral
H$_2$O target.  
The definition of the resonance location and width in terms of the
pole in the \textit{S} matrix, as per Eq.(\ref{bwform}), 
poses a problem near target state thresholds.
In our Born-Oppenheimer treatment of this process, 
we examine the behavior of the fixed-nuclei resonance width and location
as the nuclear geometry is varied.  At some nuclear geometries,
the resonance location may cross an excited state threshold.

According to analytic \textit{S} matrix theory as described by Newton \cite{Newton}, in such a situation it is 
generally not the same pole of the \textit{S} matrix that is responsible for the resonance
feature in the cross section both above and below a channel threshold.  
In such a case, in accordance with 
the formal theory, we observe that one pole 
quickly replaces another as the geometry is varied, so that the 
location of the pole identified with the resonance is effectively discontinuous
near such a threshold.  In contrast, the description of dissociative
electron attachment under the Local Complex Potential 
model\cite{BirtwistleHerzenberg,DubeHerzenberg,BardsleyWadehra,OmalleyTaylor,Omalley}
requires continuous potential-energy surfaces.  As a result,
we construct a global representation of the resonance width
that smoothly interpolates through such discontinuities.

\section{Results of scattering calculations }

\begin{figure*}
\begin{center} 
\begin{tabular}{cc}
\resizebox{0.95\columnwidth}{!}{\includegraphics{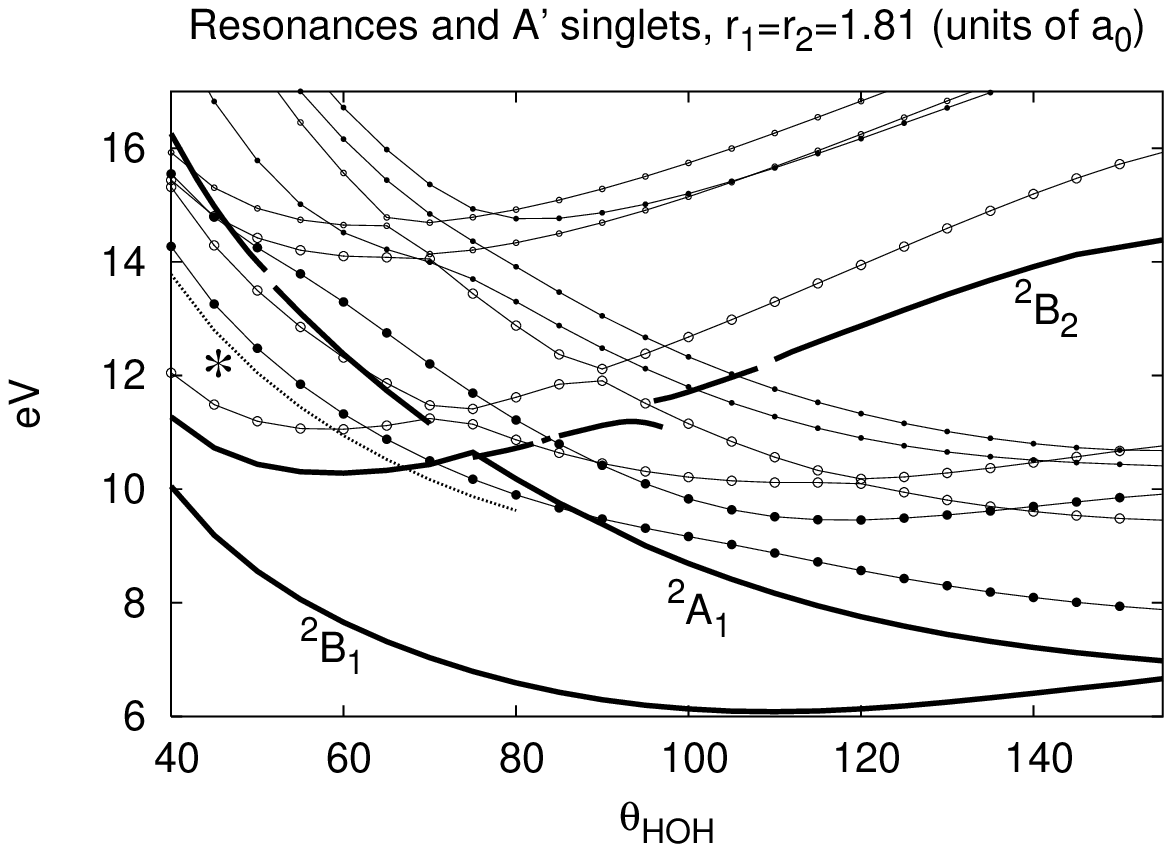}}  &
\resizebox{0.95\columnwidth}{!}{\includegraphics{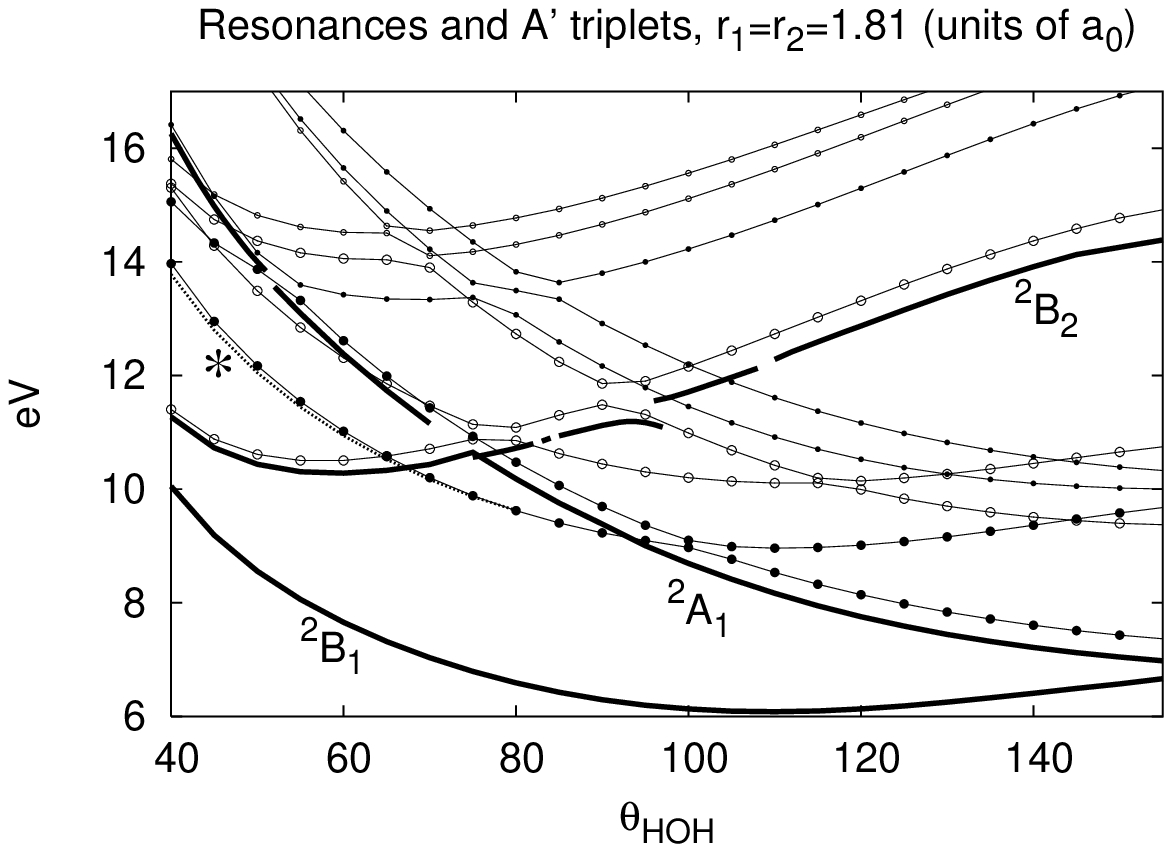}} \\ 
\resizebox{0.95\columnwidth}{!}{\includegraphics{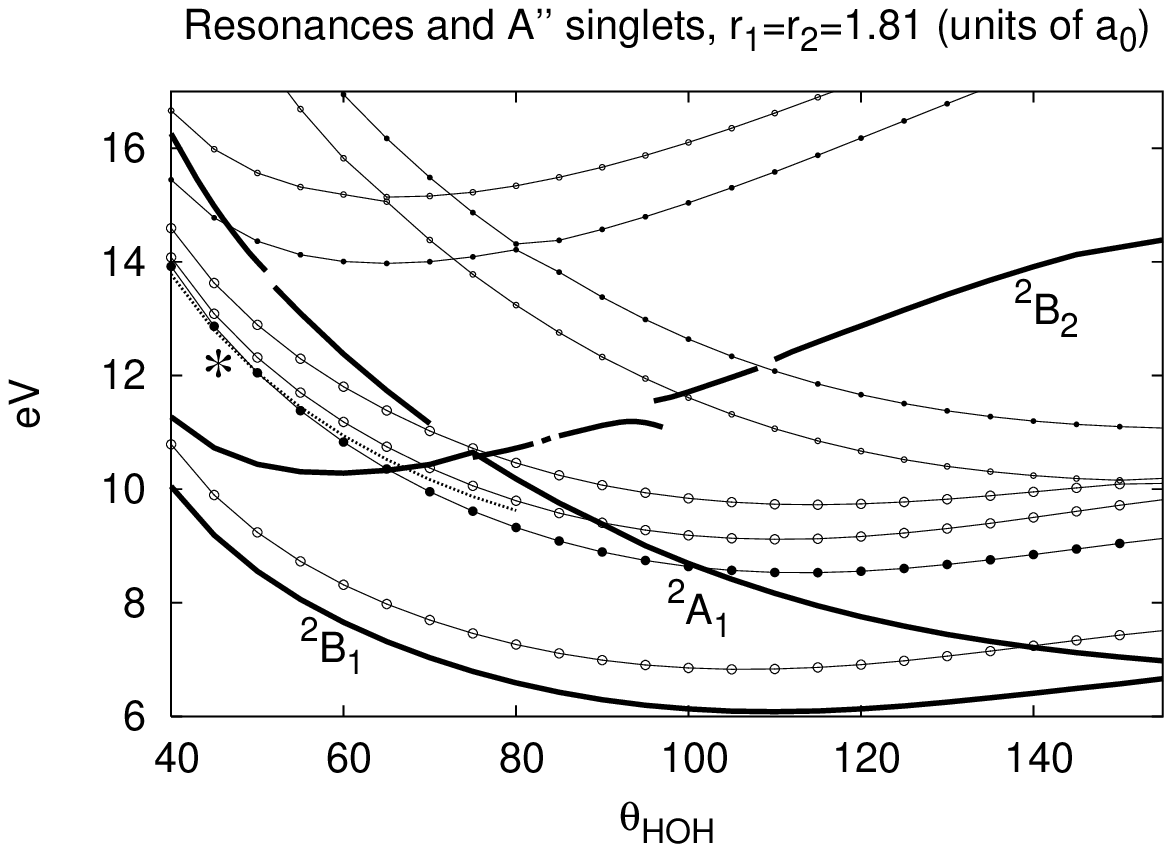}}  &
\resizebox{0.95\columnwidth}{!}{\includegraphics{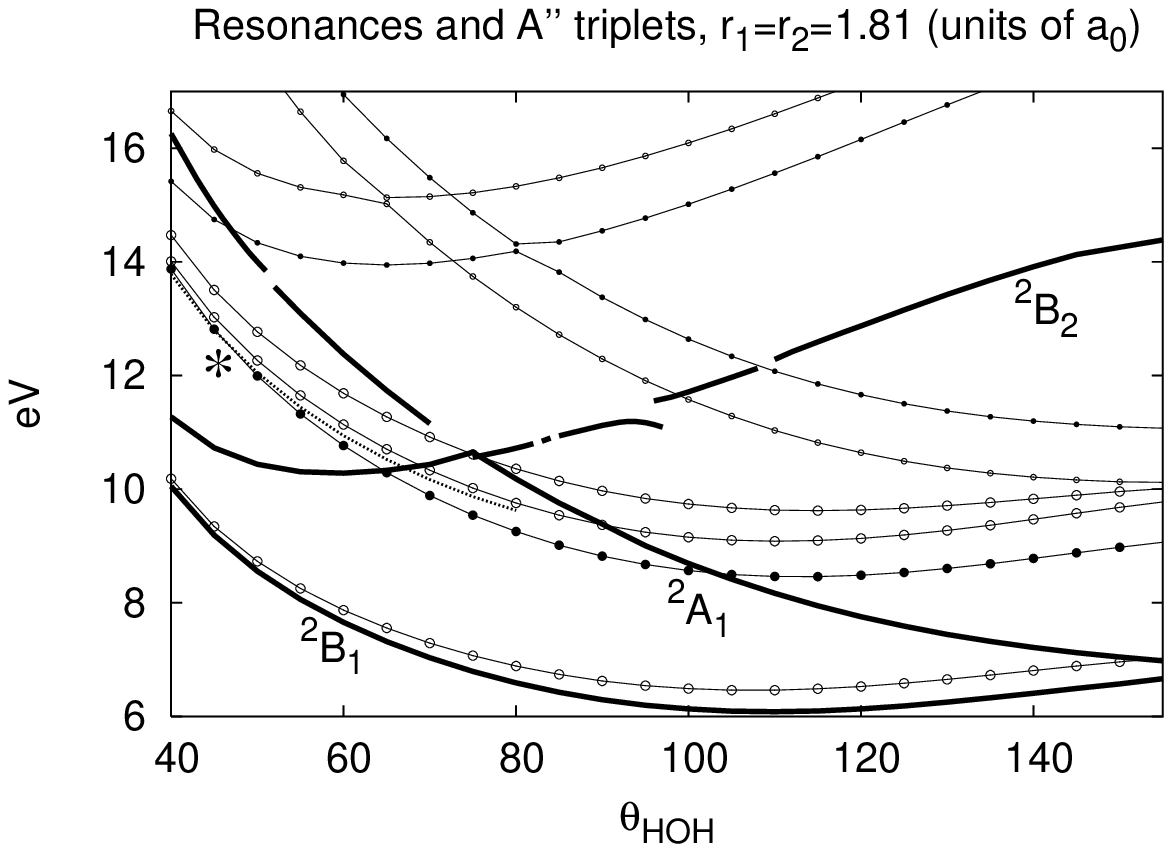}} 
\end{tabular}
\end{center}
\caption{\label{targbend181} 
Location of resonances (solid curves) and target states (dotted curves) at $r_1$=$r_2$=1.81 (units of $a_0$), as function of bending
angle $\theta_{HOH}$.  Filled dots, $A_1$ and $A_2$ target states; empty dots, $B_1$ and $B_2$ target states. The dotted curves, marked by an asterisk, refer to a second $^2A_1$ resonance discussed in the text.}
\end{figure*}

At the equilibrium geometry of the neutral, the resonance positions of the $^2B_1$, $^2A_1$, and $^2B_2$  states were calculated to be 
6.09, 8.41, and 11.97~eV, respectively, with widths of 10.31, 28.8, and 193~meV and
partial widths with respect to decay to the ground
state of 10.31, 10.30, and 9.135~meV.
At equilibrium target geometry,  each resonance lies below its parent state.
 The $^2B_1$ lies 370 meV below its $^3B_1$ parent.  The $^2A_1$ lies 351 meV below its triplet parent, which in turn is only 223 meV below the next $^3A_1$
state (1$b_1^{-1}$ 2$b_1^1$).  (These two $^3A_1$ states are in fact on the edge of an avoided
crossing, at the equilibrium geometry of the neutral, and thus the binding energy of the $A_1$
resonance versus its parent is slightly reduced by this avoided crossing.)  
The $^2B_2$ lies 1.831 eV above the first $^3B_2$ state, which
has a dominant configuration 3$a_1^{-1}$2$b_2^1$, and 471 meV below its triplet
1$b_2^{-1}$4$a_1^1$ parent.  The latter state is the 26th root of the target CI, and 
has a vertical energy of 12.438eV; this value is near the ionization potential of 
water, 12.621eV, which calls into question whether the $^3B_2$ parent is a true bound
electronic state of the target at all, at the equilibrium geometry of the neutral.  If not, it is likely that it exists instead 
as a low-lying Feshbach resonance in $e^-$+H$_2$O$^+$ Coulomb scattering.

We performed  scattering calculations along 13 distinct one-dimensional cuts within
nuclear configuration space. A dense set of points was chosen along each of these cuts  to give
an accurate and descriptive picture of the behavior of the resonances.  The
upper 2 $^2A'$ resonance was examined along nine of the 13 cuts.
Six of the 13 total cuts lie in $C_{2v}$ geometry: symmetric stretch ($r_1=r_2$) for $\theta_{HOH}$=75$^\circ$,
105$^\circ$, and 150$^\circ$; bend for $r_1$=$r_2$=1.81$a_0$ and 2.41$a_0$; and versus the Jacobi
coordinate $R$ for constant HH bond length $r_{HH}$=1.40$a_0$.  The remaining cuts are bend at $r_1$=1.81,
$r_1$=2.41$a_0$, and single-bond stretch at the six combinations of $r_1$=\{ 1.81, 2.41$a_0$ \} and 
$\theta_{HOH}$=\{ 75, 105, 150$^\circ$ \}.  For the 2 $^2A'$ surface we did not calculate the
three cuts at 150$^\circ$, nor the bending cut at ($r_1$=1.81, $r_2$=2.41).

We present
results for the three cuts which pass through the equilibrium geometry of the neutral:
as a function of bending angle $\theta$ for $r_1$=$r_2$=1.81$a_0$; as a function of
$r_2$ for $r_1$=1.81$a_0$ and $\theta$=105$^\circ$; and as a function of the equal 
bond lengths for symmetric stretch at $\theta$=105$^\circ$.   Also we present data for the
cut in Jacobi coordinates  at $r_{HH}$=1.4$a_0$, $C_{2v}$
symmetry ($\gamma$=$\pi/2$), which does not intersect the target equilibrium geometry.

These calculations test the limits of our implementation of the complex Kohn method,
and therefore it is useful to have a measure of the performance of the calculation.
One such measure is the modulus of the
\textit{S} matrices produced.  The physical \textit{S} matrix is unitary, with eigenphases having modulus
1.  At the equilibrium geometry of the neutral, at the center of  the $^2B_1$ resonance,
the modulus of the most nonunitary calculated \textit{S} matrix eigenphase was 1.005.

The $^2A_1$ calculation performed even
better, yielding a value of 1.004 for the modulus of most nonunitary eigenphase.  The corresponding value for the $^2B_2$ calculation, at higher
energy with 24 open electronic states of the target and a 46$\times$46 \textit{S} matrix,
was 1.2.  When in $C_s$ symmetry the on-resonance $^2B_2$ calculation
obtains a largest size of 120$\times$120 channels at $r_1$=1.61$a_0$, $r_2$=1.81$a_0$, 
$\theta_{HOH}$=105$^\circ$, at which point the \textit{S} matrix is also significantly nonunitary.
We note in passing that in all cases, the unitarity of the calculation is much improved for energies off-resonance.

\subsection{Bend, $\mathbf{r_1=r_2 = 1.81a_0}$}

Plots of the resonance positions and target states energies included in the Kohn
calculation for this cut are shown in Fig.
\ref{targbend181}.  The
resonance energies are lines, and the target energies are connected dots.  The large-dotted
lines are those target states included in all calculations; the small-dotted lines
are not included in the $^2B_1$ calculation (small dotted A' states) or not included in both
the $^2B_1$ and 1~$^2A'$ calculation (small dotted A'' states).  
Dots are filled for $A_1$ and $A_2$ symmetries, and open for $B_1$ and $B_2$ symmetries.

Within this cut, the target state curves compare well with the high-quality CI results of
Harrevelt and van Hemert\cite{robsurf}. 
As previously mentioned and shown in Table \ref{statestable}, 
at the equilibrium geometry of the neutral  
the Kohn target states are about 1eV below their proper location.
However, aside from this shift, the shapes of these bending potentials are
quite comperable to those of Ref. \cite{robsurf}. The 1~$^1B_1$ curves from both calculations are quite similar. Both calculations show an avoided crossing 
between the 2~$^1A_1$ and 3~$^1A_1$ states (near $\theta_{HOH}=105^\circ$ in our calculations), where they change character between 1$b_1^{-1}$2$b_1^1$ and 3$a_1^{-1}$4$a_1^1$, although the crossing is broader in our calculations. 
The 1~$^1B_2$ state undergoes several changes in character as it passes
several avoided crossings, but it seems that the Kohn target curve is again simply
shifted by about 1eV lower versus the results of Ref \cite{robsurf}.  The 1~$^1B_2$ state is
predominantly 3$a_1^{-1}$2$b_2^1$ at $\theta_{HOH}=180^\circ$ and changes character to 1$b_1^{-1}$2$a_2^1$ at the sharp avoided crossing near 120$^\circ$.  The 1~$^1A_2$ state nearly parallels the
1~$^1B_1$ state, slightly less than 2eV above it, for both calculations.

\begin{figure}[b]
\resizebox{0.95\columnwidth}{!}{\includegraphics{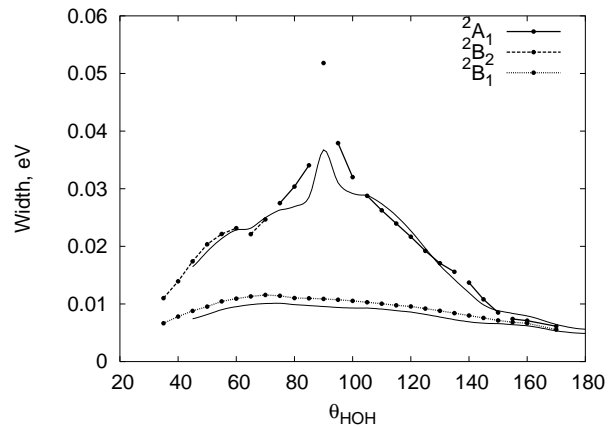}}  
\caption{\label{width1apbend}
Width of the $^2B_1$ and 1~$^2A'$ resonances (dots) with respect to bending angle $\theta$ for $r_1$=$r_2$=1.81
(units of $a_0$), with
interpolated global representation (plain lines).  }
\end{figure}

Along the cut shown in Fig. \ref{targbend181}, the $^2B_1$ Feshbach resonance stays below its $^3B_1$ parent, except at very small
bending angles $\theta_{HOH}$, and therefore is below all target states except
the ground electronic state.  
The $^2A_1$ state follows its triplet parent, $^3A_1$ [H$_2$O]3$a_1^{-1}$4$a_1^1$
configuration.  At most angles this configuration describes well one of the target
states, except where there are avoided crossings.  There are two avoided crossings
involving this state: the one with the [H$_2$O]1$b_1^{-1}$2$b_1^1$, around $\theta_{HOH}$=100$^\circ$ and 9.5eV, and also one with
the [H$_2$O]1$b_2^{-1}$2$b_2^1$ state near
$\theta_{HOH}$=55$^\circ$ and 13.5eV.  Near both of these crossings, there are discontinuities as the resonances cross the avoided state threshold.
For the former avoided crossing, the discontinuity in the resonance position is not visible on the scale plotted in Fig. \ref{targbend181}, but there is a  significant discontinuity in the 
$^2A_1$ width at $\theta_{HOH}$=95$^\circ$;
for the latter crossing, the discontinuity in position near $\theta_{HOH}$=50$^\circ$ 
is apparent as well. The $^2B_2$ resonance also follows its triplet parent configuration and parent state
through avoided crossings with the [H$_2$O]1$a_1^{-1}$2$b_2^1$ state (avoided 
crossing at $\theta_{HOH}$=92$^\circ$; discontinuity at $\theta_{HOH}$=96$^\circ$)
and [H$_2$O]1$b_1^{-1}$1$a_2^1$ state (avoided 
crossing at $\theta_{HOH}$=79$^\circ$; small discontinuity at $\theta_{HOH}$=83$^\circ$).
The former discontinuity is quite large ($\sim$ 0.25eV).

\begin{figure}
\begin{center} 
\begin{tabular}{c}
\resizebox{0.95\columnwidth}{!}{\includegraphics{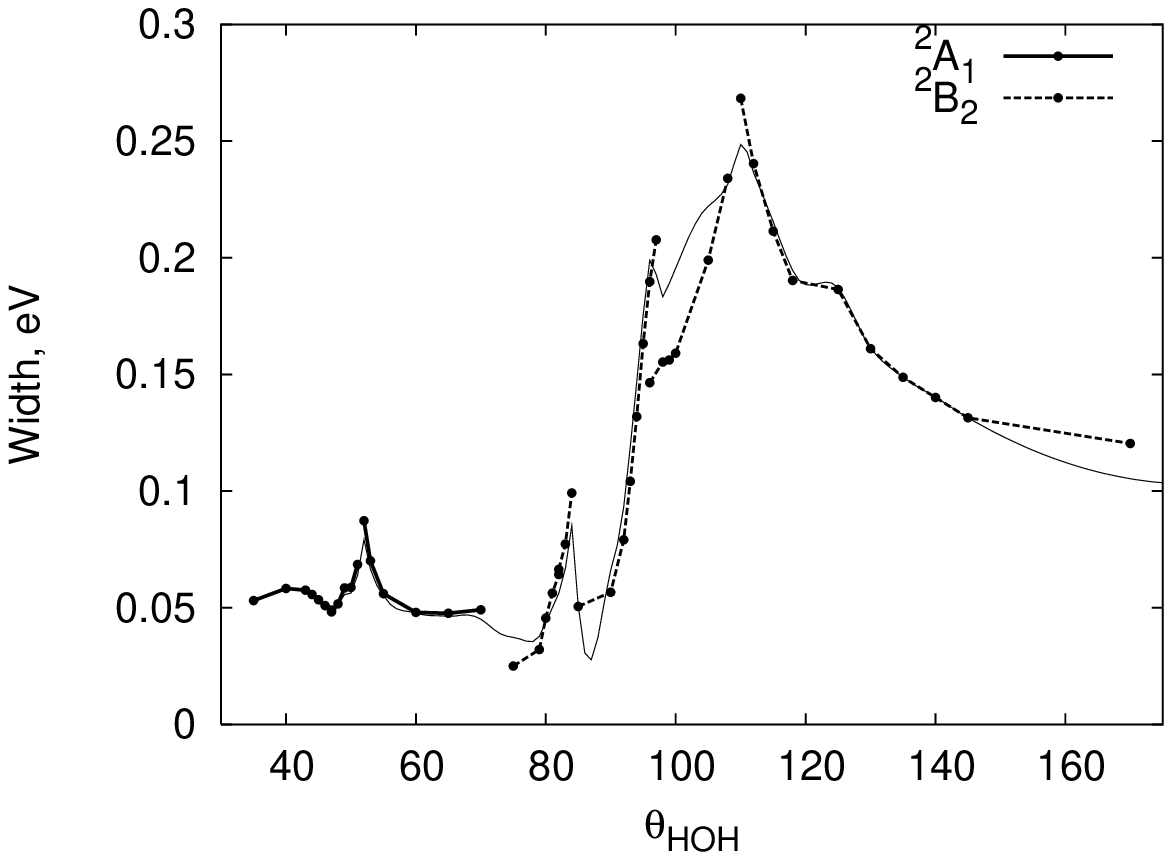}}  \\
\resizebox{0.95\columnwidth}{!}{\includegraphics{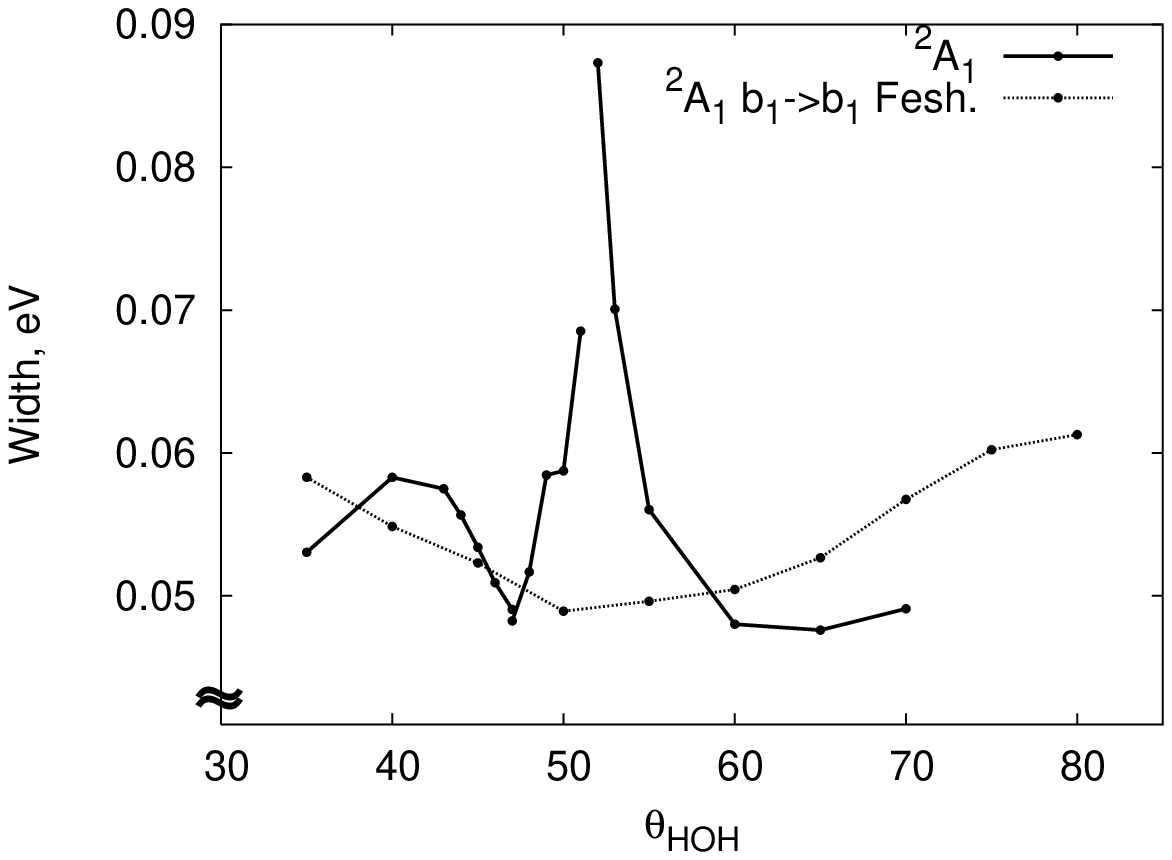}}  
\end{tabular}
\end{center}
\caption{\label{width2apbend}
Top, width of the 2 $^2A'$ resonance (dots) with respect to bending angle $\theta$ for $r_1$=$r_2$=1.81 
(units of $a_0$), with
interpolated global representation (plain line).  Bottom, zoom of width in $^2A_1$ symmetry,
with width of novel $^2A_1$ ([H$_2$O]1$b_1^{-1}$2$b_1^1$4$a_1^1$) Feshbach resonance.}
\end{figure}

The widths for the $^2B_1$ and 1~$^2A'$ ($^2B_2$ or $^2A_1$) states 
along this cut are shown in Fig. \ref{width1apbend}.
Both the raw Kohn results (dotted lines)
and an interpolated version (plain lines, defined later, in Sec. \ref{widthglobal}) are
plotted.
The width of the $^2B_1$ state is relatively constant. 
The $^2A_1$ state attains a relatively
large width around $\theta$=95$^\circ$ as it crosses multiple target states, but
then the width decreases before the $^2A_1$ state intersects the $^2B_2$ state between
$\theta_{HOH}$=70 and 75$^\circ$.   At this geometry the widths of the $^2A_1$ and $^2B_2$
Feshbach resonances happen to be nearly equal, and so there is only a small discontinuity
as the 1~$^2A'$ resonance changes symmetry from $^2A_1$ to $^2B_2$ as the bond angle is decreased.
In Fig. \ref{width1apbend} adjacent data points are connected by line segments only
if no target states are crossed as the geometry is varied between them.  Therefore one
can see that at most crossings on the 1~$^2A'$ surface the discontinuity in the resonance energy is
in fact small.  

The width of the 2 $^2A'$ ($^2A_1$ or $^2B_2$) resonance is plotted in Fig. \ref{width2apbend}.  
The $^2B_2$ state attains a large width (0.25eV, $\tau$=2.5fs) as it becomes the 2 $^2A'$ resonance,
and in particular as it rises above the $B_2$ target states near $\theta_{HOH}$=82$^\circ$
and 95$^\circ$, at which points there are significant discontinuities in the location
of the physical resonance pole.  In particular, at 95$^\circ$ there is a large discontinuity
in the real part of the resonance location:
the real parts of the energies of the two poles avoid each other by approximately 0.3eV,
whereas the widths avoid by only 0.025eV.
In Fig. \ref{width2apbend} adjacent data points have been connected if the resonance does
not cross a target state of the same spatial symmetry, or otherwise exhibit a large discontinuity.
For the $^2B_2$ Feshbach resonance, there are large discontinuities for non-$B_2$ crossings
at the 2 $^1A_1$ crossing near $\theta_{HOH}$=84$^\circ$, and at the crossing of both the 5 $^3B_1$
and 5 $^1B_1$ near $\theta_{HOH}$=108$^\circ$.

The calculation supports a different Feshbach resonance of $^2A_1$ symmetry for $\theta_{HOH} < 80^\circ$
along this cut, whose parent is the [H$_2$O]1$b_1^{-1}$2$b_1^1$ $^3A_1$ target state,
and which would have the dominant configuration [H$_2$O]1$b_1^{-1}$2$b_1^1$4$a_1^1$.
The location of this Feshbach resonance is also plotted  in Fig. \ref{targbend181}.
For these geometries it is 
impossible to say whether this is an artifact of the calculation or a physical
state, since along this cut, this Feshbach resonance is extremely narrow and is not bound by more than 25meV. A  small upward perturbation of the calculated resonance location would lead to its
disappearance above its parent. This state is only present when the
[H$_2$O]1$b_1^{-1}$2$b_1^1$ configuration is the lowest-energy $^3A_1$ target configuration, for
then it cannot decay to the triplet or singlet $B_1$ ([H$_2$O]1$b_1^{-1}$4$a_1^1$) target
states.  This is the case when the bond lengths are modest and the bond angle is small.
The Jacobi coordinate cut, which we present in Sec. \ref{jacobisubsub}, gives
a better view of the interaction between this Feshbach resonance and the other $^2A_1$
resonance and demonstrates that in fact this [H$_2$O]1$b_1^{-1}$2$b_1^1$4$a_1^1$
Feshbach resonance may exist as an asymptote of the [H$_2$O]3$a_1^{-1}$4$a_1^2$
Feshbach resonance, due to branching of the adiabatic PES.

\begin{figure*}
\begin{center} 
\begin{tabular}{cc}
\resizebox{0.95\columnwidth}{!}{\includegraphics{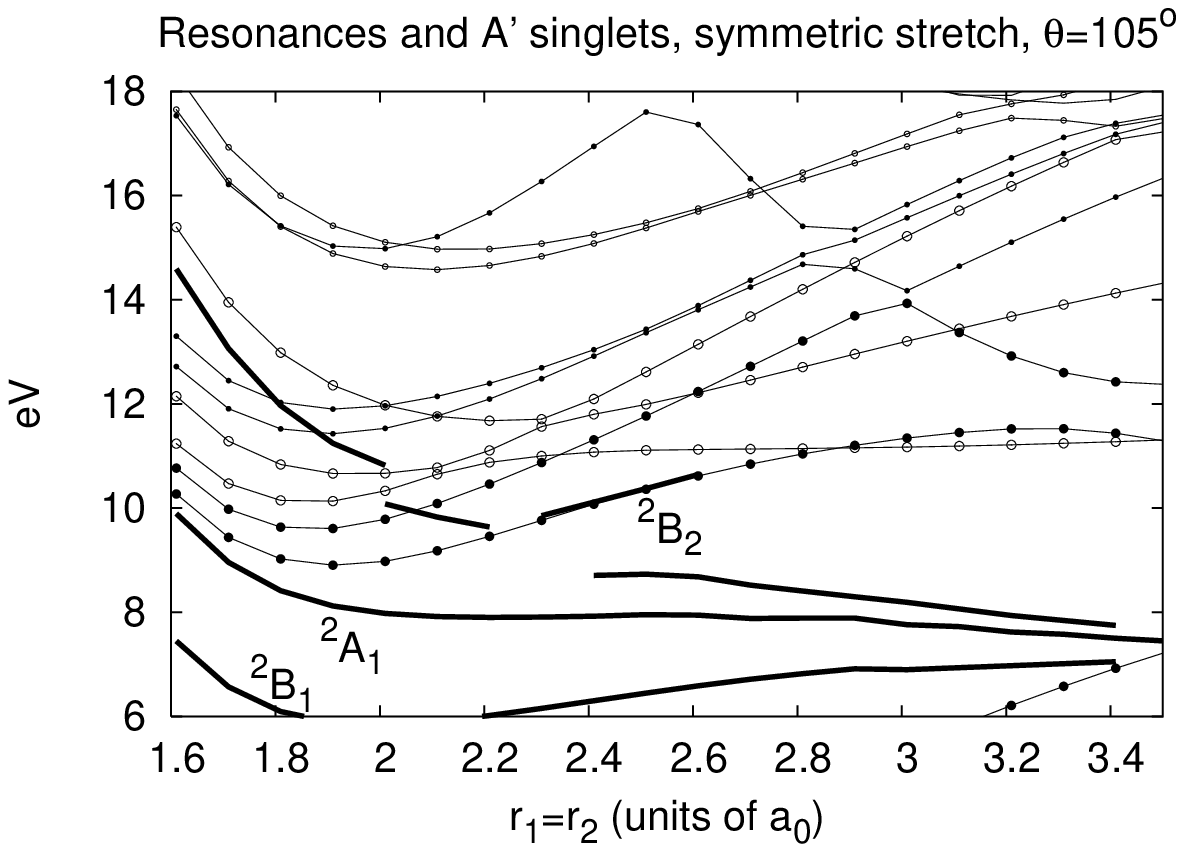}}  &
\resizebox{0.95\columnwidth}{!}{\includegraphics{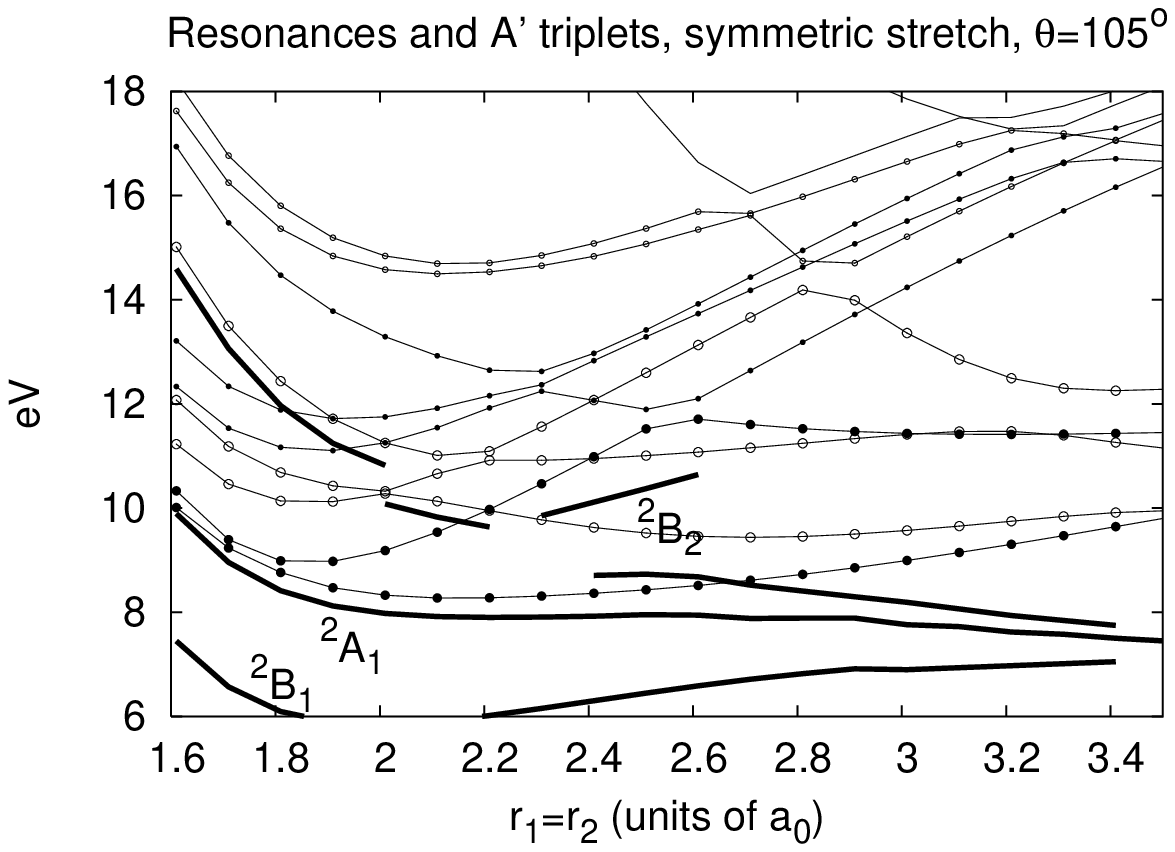}} \\ 
\resizebox{0.95\columnwidth}{!}{\includegraphics{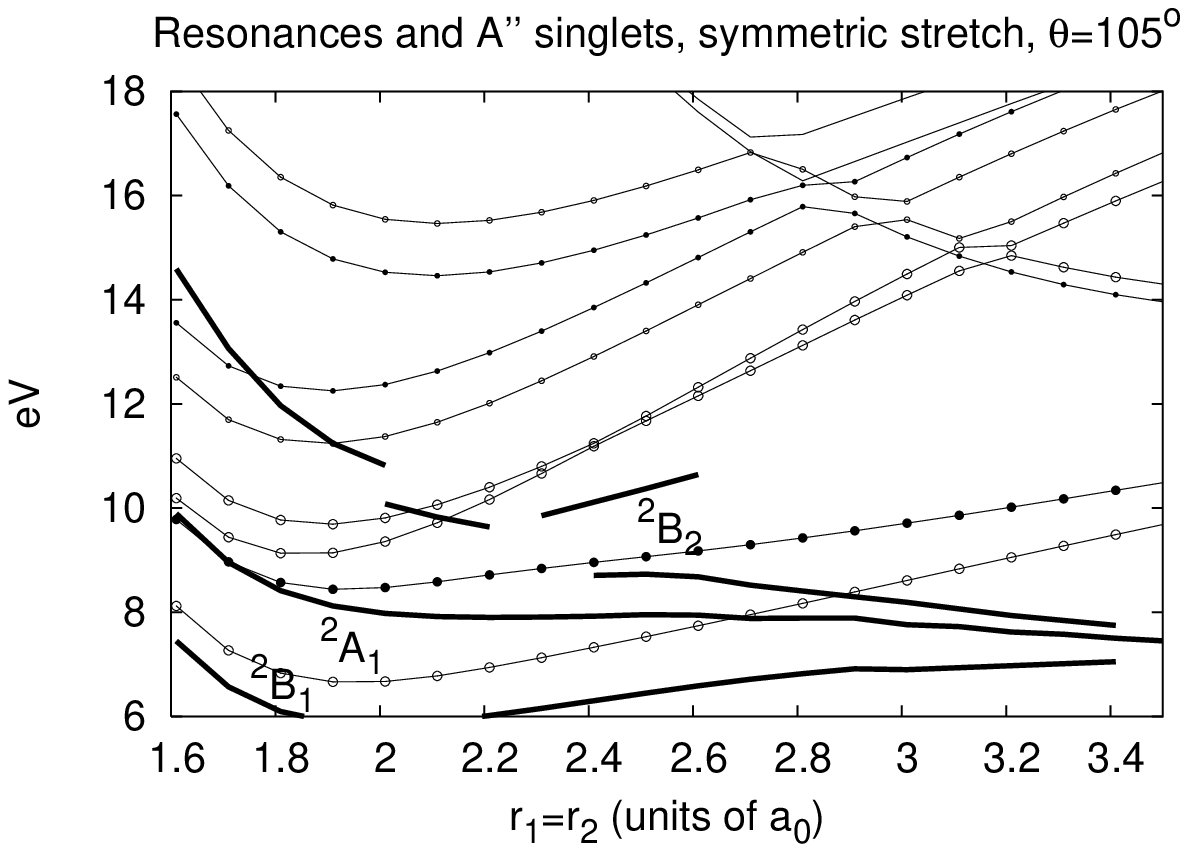}}  &
\resizebox{0.95\columnwidth}{!}{\includegraphics{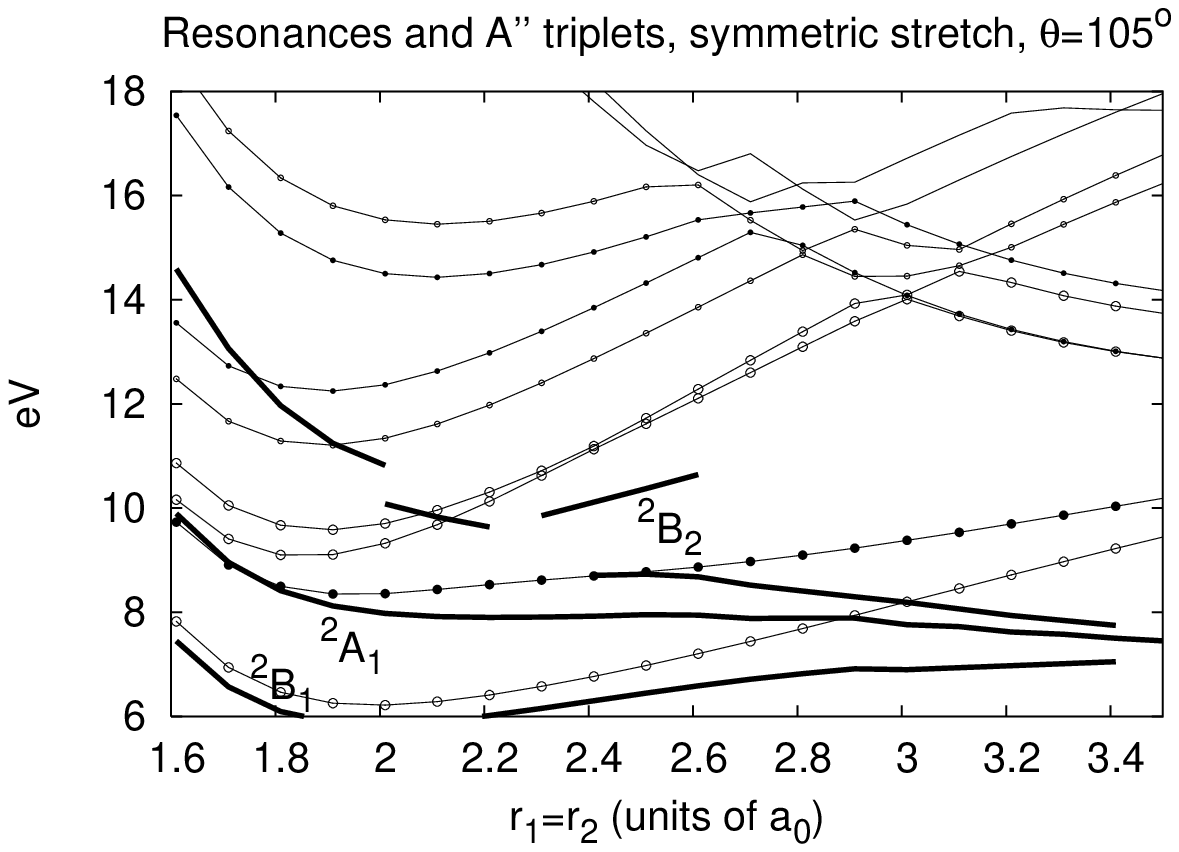}} \\ 
\end{tabular}
\end{center}
\caption{\label{targsym} 
Location of resonances (solid curves) and target states (dotted curves) at $\theta$=105$^\circ$, as function of 
symmetric stretch $r_1$=$r_2$.
Filled dots, $A_1$ and $A_2$ target states; empty dots, $B_1$ and $B_2$.}
\end{figure*}

\subsection{Symmetric stretch, $\mathbf{r_1=r_2, \theta=105^\circ}$}

The target energies and resonance positions along this cut are shown in Fig.
\ref{targsym}, with the same conventions as Fig. \ref{targbend181}.
The (undotted) curves near the top right part of each panel correspond to the first target state of each symmetry that is not included 
in any of the Kohn calculations.  The 
1~$^1B_1$ curve is very similar to that of Ref.~\cite{robsurf}, except that the minimum
is at 1.95$a_0$, not 2.1$a_0$.  The 1~$^1A_2$ asymptote of the Kohn target curve is too high
by $\sim$1~eV.  The 2 $^1A_1$ and 1~$^1B_2$ state curves are similar to those of Ref.~\cite{robsurf},
but shifted down $\sim$0.75~eV, and in fact duplicate the apparent change of character of 
the 1~$^1B_2$ state at about $r_1$=$r_2$=2.1$a_0$.

As the symmetric stretch coordinate $r_1$=$r_2$ is increased
from the equilibrium geometry at 1.81$a_0$, the $^2B_1$ and $^2A_1$ (1~$^2A'$) resonances approach the
bound O+H+H$^-$ asymptote of the three-body system and
both become bound at approximately $r_1$=$r_2$=3.5$a_0$.
The $^2B_2$ (2 $^2A'$) resonance exhibits large discontinuities, due to both target state
crossings and the interaction with the $^2B_2$ shape resonance as discussed in Ref.~\cite{haxton3}.
At $r_1$=$r_2$$>$2.4$a_0$, the lower-energy branch of the $^2B_2$ shape-Feshbach
system is uncovered beneath the 1~$^3A_2$ state of the target, which correlates to O+H+H.

\begin{figure}[b]
\resizebox{0.95\columnwidth}{!}{\includegraphics{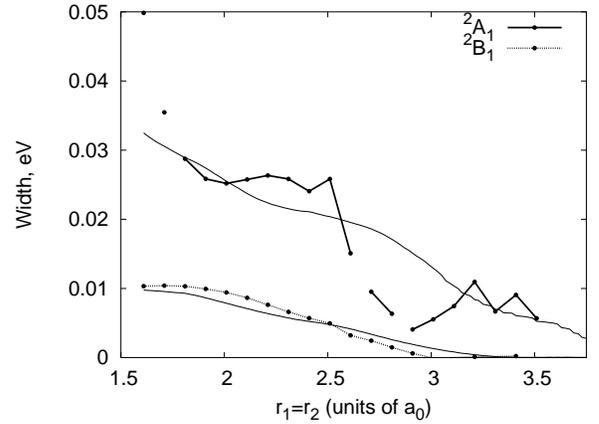}}  
\caption{\label{width1apsym}
Width of the $^2B_1$ and $^2A_1$ (1~$^2A'$) resonances (dots) with respect to symmetric stretch $r_1$=$r_2$, 
at $\theta$=105$^\circ$, 
with interpolated global representation (plain line).}
\end{figure}

\begin{figure}[t]
\resizebox{0.95\columnwidth}{!}{\includegraphics{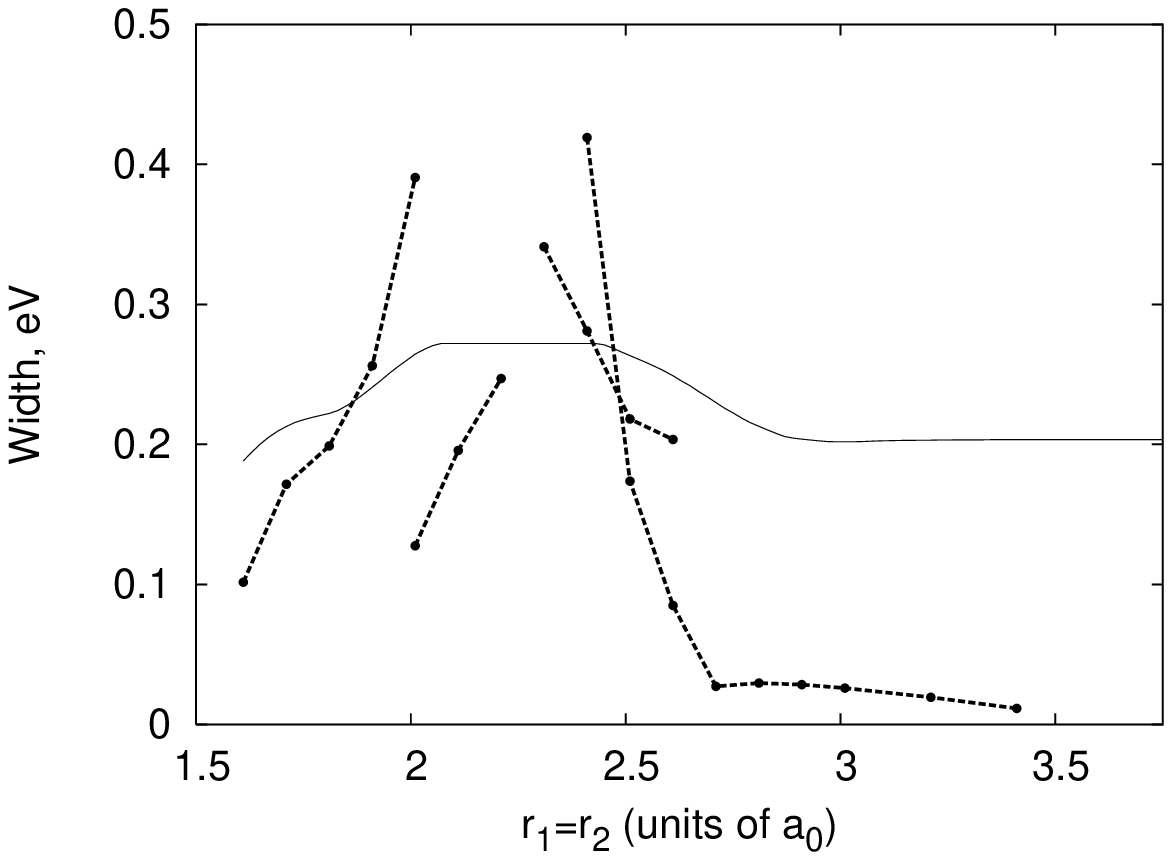}}  
\caption{\label{width2apsym}
Width of the $^2B_2$ (2 $^2A'$) resonance (dots) with respect to symmetric
 stretch, $r_1$=$r_2$,
at $\theta$=105$^\circ$,
with interpolated global representation (plain line).}
\end{figure}

\begin{figure*}
\begin{center} 
\begin{tabular}{cc}
\resizebox{0.95\columnwidth}{!}{\includegraphics{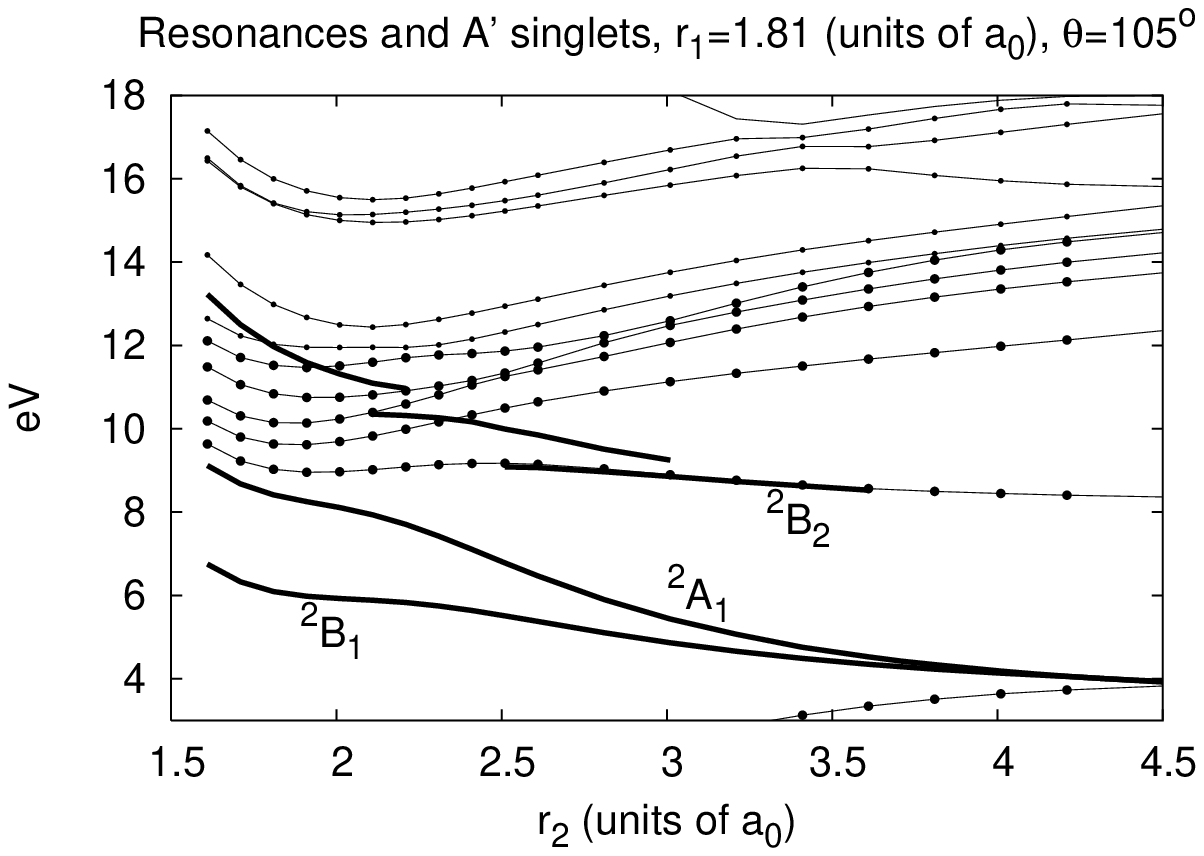}}  &
\resizebox{0.95\columnwidth}{!}{\includegraphics{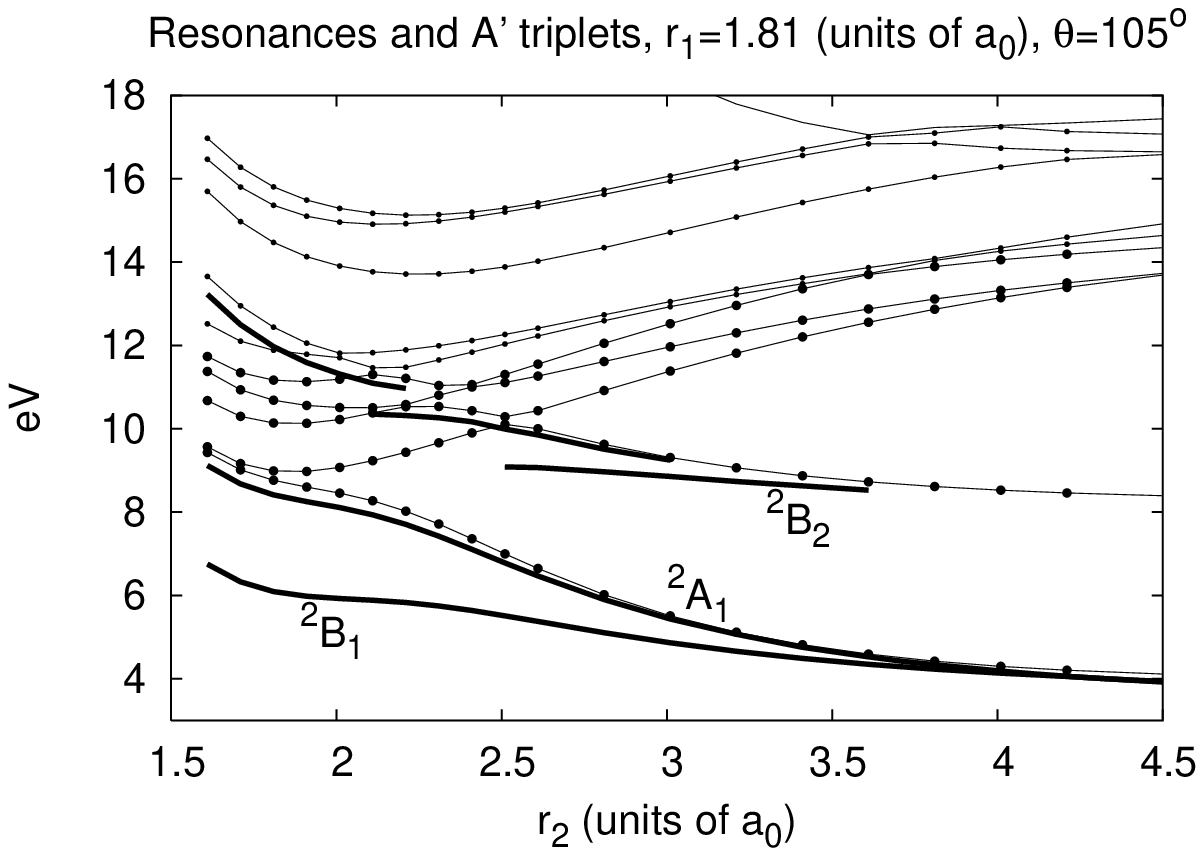}} \\ 
\resizebox{0.95\columnwidth}{!}{\includegraphics{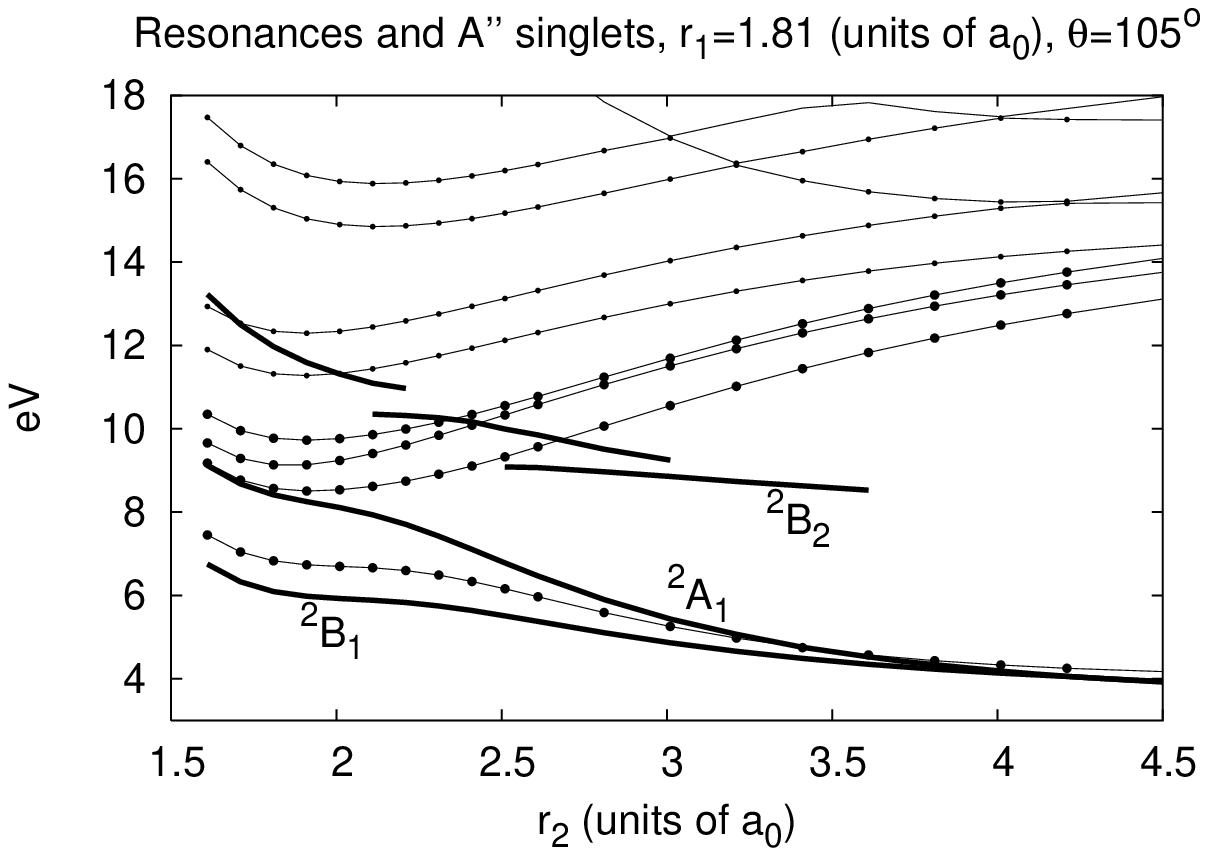}}  &
\resizebox{0.95\columnwidth}{!}{\includegraphics{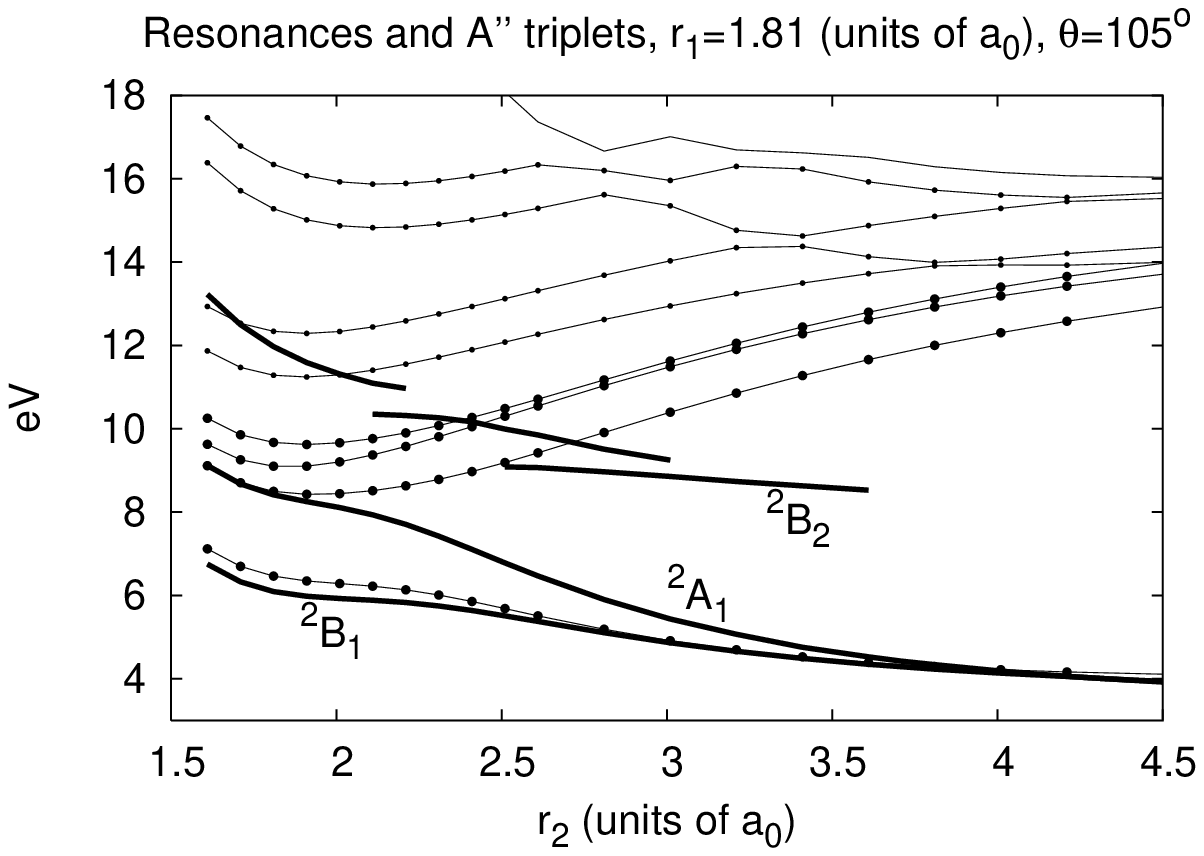}} \\ 
\end{tabular}
\end{center}
\caption{\label{targ181105} 
Location of resonances (solid curves) and target states (dotted curves) at $r_1$=1.81$a_0$, 
$\theta$=105$^\circ$, as function of $r_2$.}
\end{figure*}

The widths for these cuts are shown in Figs. \ref{width1apsym} and \ref{width2apsym},
with the same conventions as in Figs.  \ref{width1apbend} and \ref{width2apbend}.
For the 2 $^2A'$ state, the final branch of the $^2B_2$ state was not included in the global fit,
and thus the interpolated value levels off near 0.2eV, corresponding to the width of the
penultimate branch at $r_1$=$r_2$=2.6~$a_0$.

\subsection{Single bond stretch, $\mathbf{r_1=1.81a_0, \theta=105^\circ}$}

\begin{figure}[b]
\resizebox{0.95\columnwidth}{!}{\includegraphics{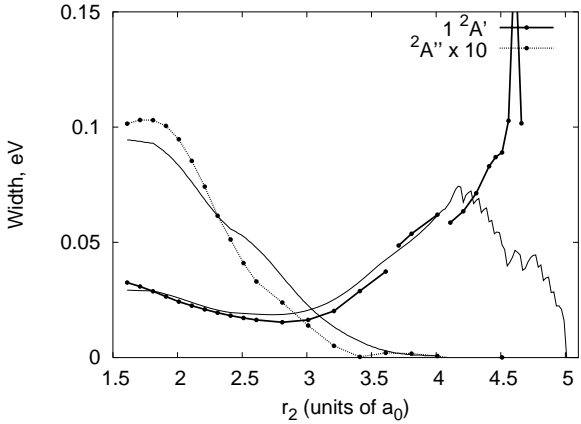}} 
\caption{\label{width1apr2}
Width of the $^2A''$ ($^2B_1$) ($\times$ 10) and 1~$^2A'$ resonances (dots) with respect to $r_2$ for $r_1$=1.81$a_0$, $\theta$=105$^\circ$,
with interpolated global representation (plain line).}
\end{figure}

\begin{figure}[b]
\resizebox{0.95\columnwidth}{!}{\includegraphics{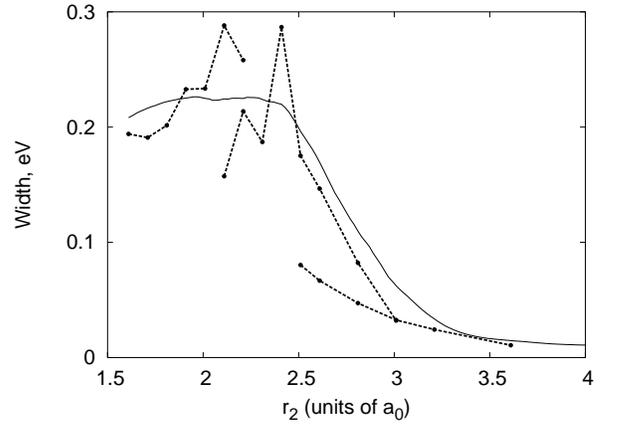}}  
\caption{\label{width2apr2}
Width of the 2 $^2A'$ resonance (dots) with respect to $r_2$ for $r_1$=1.81$a_0$, $\theta$=105$^\circ$,
with interpolated global representation (plain line).}
\end{figure}

The resonance locations and Kohn target state energies along this cut are shown in Fig.
~\ref{targ181105}.  Along this cut towards increasing $r_2$, the 1~$^3A''$, 1~$^1A''$, 
and 1~$^3A'$ Kohn target states, which correspond to the 1~$^3B_1$, 1~$^1B_1$, and 1~$^3A_1$ 
states at the equilibrium
geometry of the neutral, move downward in energy as they approach the ground-state
H+OH asymptote of the fragments.  
The 2 $^1A'$ (2 $^1A_1$) state surface is repelled by the
ground state and is essentially constant, starting at 9~eV, with a gentle maximum at 9.25~eV
at $r_2$=2.5$a_0$, and approaching 8~eV at the OH($^2\Sigma$)+H asymptote.  The difference
between this asymptote and the OH($^2\Pi$)+H ground state of the fragments is 4eV for
the Kohn states; this may be compared with an earlier calculated value~\cite{OH1974} of 5.27eV at $r_1$=1.80$a_0$.
The results for the excited state singlets are 
consistently $\sim$1eV below
the results of Ref.~\cite{robsurf}.  
Thus, the ground X $^1A_1$  Kohn target state curve dissociation energy is 
underestimated by about 1eV along this cut.

Moving towards increasing $r_2$ along this cut, the $^2A''$ and 1~$^2A'$ Feshbach resonances
fall in energy and become bound as H$^-$+OH(X $^2\Pi$) near $r_2$=4.5$a_0$.
The 2 $^2A'$ state correlates to the H$^-$+OH($\Sigma$) asymptote, which lies
above the ground state of the neutral, H+OH(X~$^2\Pi$); its width therefore goes to
zero only asymptotically.  The 2 $^2A'$ state has two large discontinuities along
this cut.  The first, near $r_2$=2.15$a_0$, is associated with an
avoided crossing between the $^3A'$ parent state and another state that becomes
the new parent, correlating to OH($^2\Sigma$)+H, triplet coupled.  
As $r_2$ is increased further, the binding energy of the Feshbach resonance with
respect to this state decreases to zero at $r_2$=3.0$a_0$, and it disappears for $r_2$$>$3.0$a_0$,
but by this time a distinct resonance pole which follows the corresponding 
singlet state [correlating to OH($^2\Sigma$)+H, singlet coupled] has already appeared
to take its place.

Along this cut, the binding energies of the Feshbach resonances 
(which are plotted in the EPAPS archive)
with respect to their
parents all reach a minimum around $r_2$=3.25$a_0$.
This is also the geometry at which the middle 2~$^2A'$ branch disappears.
It is interesting that this
is the case not only for the electronically similar $^2A''$ ($^2B_1$) and 1~$^2A'$ ($^2A_1$) states
but also for the final branch of the 2~$^2A'$ state.
In each case this minimum is approximately 50meV,
which for the 1 and  2~$^2A'$ states is only twice their width, $\Gamma \approx $ 25meV.
We certainly cannot be confident that our treatment of the $N$- and ($N+1$)-
electron systems is balanced at the 50meV level.
This value is an order of magnitude smaller than the binding energies
at the equilibrium geometry of the neutral, and calls into question
whether the physical states in fact may rise above their parents near
$r_2$=3.25$a_0$.

The widths for these cuts are shown in Figs. \ref{width1apr2} and \ref{width2apr2},
with the same conventions as in Figs.  \ref{width1apbend} and \ref{width2apbend}.
The width of the $^2A''$ state goes smoothly to zero as the resonance becomes bound.
In contrast, the 1~$^2A'$ state achieves a large width as $r_2$ increases, 
as high as 0.18eV at $r_2$=4.55$a_0$.  It may do so because it is connected
to the ground state by an $s$-wave matrix element; we suspect that it does do so
because at intermediate geometries, the electronic structure is highly correlated, and
the orbital description of the resonance is likely to be different from that
of the target states, muddling the distinction between Feshbach and shape resonances.
In other words, as the 1~$^2A'$ Feshbach resonance dissociates it takes on an increasing
degree of shape resonance character, leading to an increase in width.
We discuss this issue further in paper II.

\begin{figure}
\resizebox{0.95\columnwidth}{!}{\includegraphics{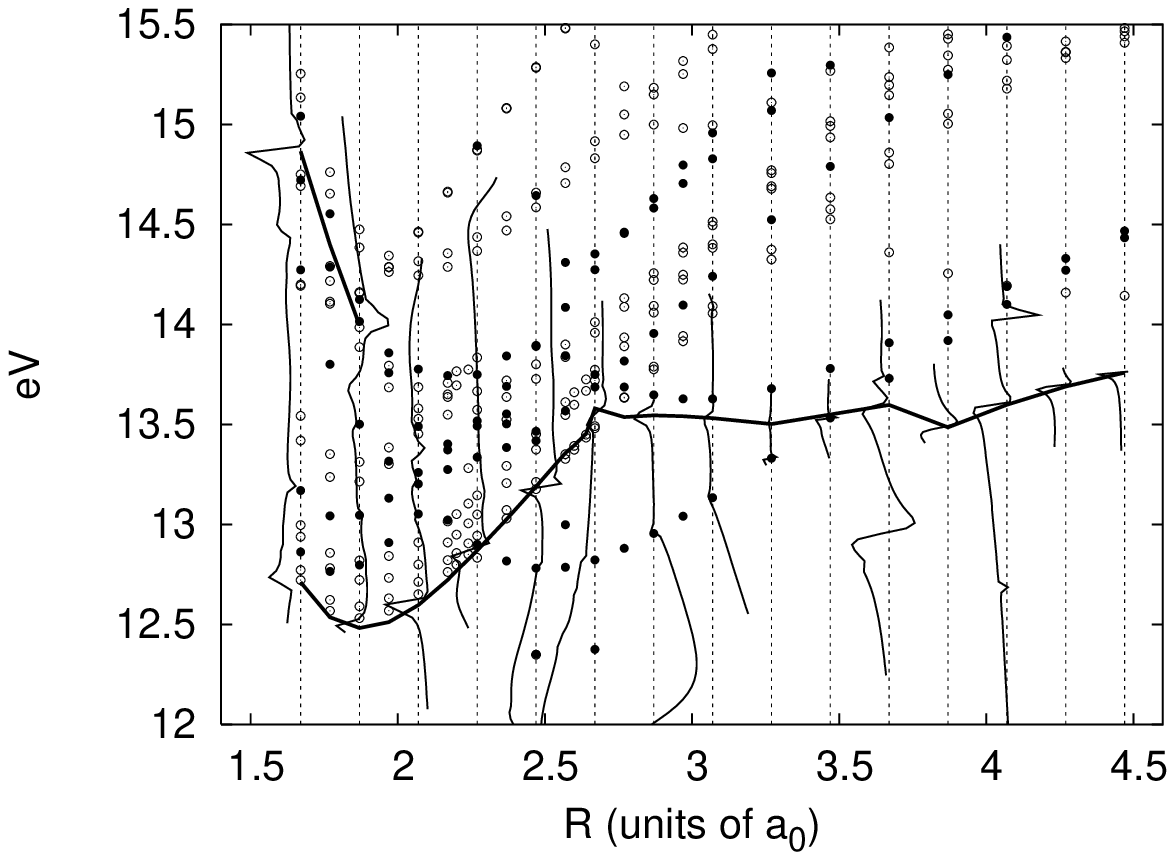}}
\caption{\label{smatrixJ2}
Location of $^2A_1$ resonances and raw complex Kohn data 
at $r_{HH}$=1.4$a_0$, $C_{2v}$ symmetry ($\gamma$=90$^\circ$)
as functions of Jacobi coordinate R.  Solid thick line: resonance location.  Filled dots: $A_1$ target
states.  Empty dots: other target states.  Thin lines: real part of $s\rightarrow s$ 
\textit{S} matrix element, on an arbitrary scale,
as a function of energy at various $R$, plotted with ordinate and abscissa
reversed.}
\end{figure}

\subsection{\label{jacobisubsub} Jacobi coordinates, $\mathbf{r_{HH}=1.4a_0, \gamma=90^\circ}$ }

One may define a Jacobi coordinate system for H$_2$O in which
$r_{HH}$ is the H$-$H bond
length, $R$ is the distance between the oxygen and the H$_2$ center of mass, 
and $\gamma$ is the angle between the two corresponding vectors.
Here we present data along the cut $r_{HH}$=1.40$a_0$, $\gamma$=90$^\circ$
($C_{2v}$ geometry), plotted as a function of $R$.
This cut traces
the resonances from a squeezed geometry (small $\theta_{HOH}$) towards the O+H$_2$ arrangement
in $C_{2v}$ symmetry.  It is uninteresting except for the
interaction between the  $^2A_1$ (2 $^2A'$) Feshbach resonance with configuration 
[H$_2$O]3$a_1^{-1}$4$a_1^2$ and the second $^2A_1$ Feshbach resonance mentioned above, with
configuration [H$_2$O]1$b_1^{-1}$2$b_1^1$4$a_1^1$.  

In Fig. \ref{smatrixJ2}, we show complex Kohn \textit{S} matrix elements and fitted resonance
locations along this cut.  This figure requires some explanation.  The target state energies
are plotted as dots, and the resonance location is plotted as a single, thick solid line.
At 15 of the plotted values of $R$ we have also plotted the real part of the
$s\rightarrow s$ 
\textit{S} matrix element calculated using the complex Kohn method
but with ordinate and abscissa
reversed.  The scale and origin in the horizontal (ordinate) direction is arbitrary for these
\textit{S} matrix elements.

The leftmost data in this figure are calculated at $R$=1.67$a_0$, a geometry
that in bond-angle coordinates 
is approximately ($r_1$=$r_2$=1.81$a_0$, $\theta_{HOH}$=45$^\circ$).
At this geometry, both the [H$_2$O]3$a_1^{-1}$4$a_1^2$ Feshbach resonance, at $\sim$14.9eV, and the [H$_2$O]1$b_1^{-1}$2$b_1^1$4$a_1^1$ Feshbach resonance, at $\sim$12.75eV,
are present and their Breit-Wigner profiles may be seen in the behavior of the \textit{S} matrix element.
As the coordinate $R$ is increased beyond 1.87$a_0$, however, the main [H$_2$O]3$a_1^{-1}$4$a_1^2$ resonance is lost within
a web of avoided and actual target state crossings around 14eV.  The lower [H$_2$O]1$b_1^{-1}$2$b_1^1$4$a_1^1$ 
Feshbach resonance is still present, however, and is bound by 0.5eV with respect to its [H$_2$O]1$b_1^{-1}$2$b_1^1$
$^3A_1$ parent state at $R$=1.97$a_0$.  This large binding energy brings up the possiblity that this
resonance is in fact physical, at least at this geometry, and not simply an artifact of the
calculation supported by recorrelation of the target.  This Feshbach resonance follows the
energy of its $^3A_1$ parent through an avoided crossing near 2.1$a_0$ and through actual crossings
with two $^1A_1$ states at approximately 2.45 and 2.55$a_0$.

\begin{figure}
\resizebox{0.95\columnwidth}{!}{\includegraphics{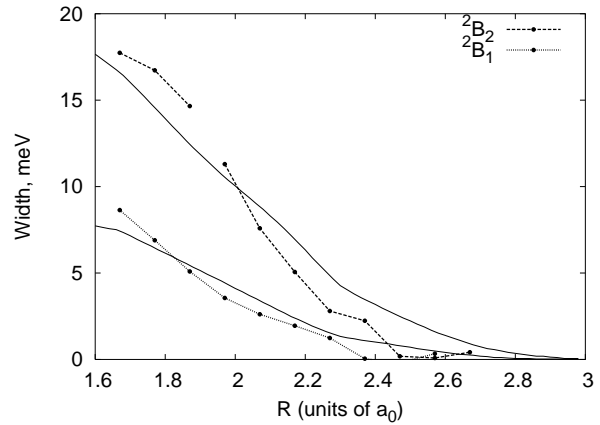}}  
\caption{\label{widthJ2}
Width of the $^2B_1$ and $^2B_2$ (1~$^2A'$) resonances (dots) in $C_{2v}$ symmetry with respect to the Jacobi coordinate $R$ for $r_{HH}$=1.4$a_0$,
with interpolated global representation (plain line).}
\end{figure}

At approximately $R$=2.67$a_0$ there is an actual crossing between the $^3A_1$ parent and
the $^1A_1$ state, which is a parent of the familiar [H$_2$O]3$a_1^{-1}$4$a_1^2$ Feshbach resonance,
and at this point, the [H$_2$O]1$b_1^{-1}$2$b_1^1$4$a_1^1$ resonance appears to change character
to that of the more familiar resonance, as there is a kink in the resonance trajectory and the
resonance pole then follows the $^1A_1$ parent as it dissociates toward H$_2$ (1$\sigma_g$1$\sigma_u$) +
O ($^3$P).

Thus, it appears that the $^2A_1$ Feshbach resonance may undergo an interaction with a different Feshbach
resonance of a sort similar to that which occurs~\cite{haxton3} within the $B_2$ manifold between 
the [H$_2$O]1$b_2^{-1}$4$a_1^2$$^2B_2$
Feshbach resonance and a $^2B_2$ shape resonance.  The difference would be that in the current case, the 
topology is supported by the disappearance and appearance of different branches of the adiabatic
manifold due to the crossing of $N$-electron target states, whereas in the $^2B_2$ case, 
it is supported by the underlying ($N+1$)-electron Hamiltonian.  Another difference is that the
topological complications that may occur on the $^2A_1$ potential-energy surface seem to do so at geometries
not sampled by the propagating DEA wavepacket, making them irrelevant to the physical problem,
although we did not attempt a comprehensive analysis of this issue.
Along this cut, the $^2B_1$ and $^2B_2$ (1~$^2A'$) states quickly
become bound as O$^-$+H$_2$, and we present the widths of these states in Fig. \ref{widthJ2}.

\subsection{Global representation of the widths\label{widthglobal}}

For the purpose of performing nuclear dynamics calculations, a global representation of the width $\Gamma$ is
required.  Global representations were constructed separately for the $^2A''$, 1~$^2A'$, and 2 $^2A'$ states.

The first step in constructing each of these global representations was to define a continuous representation
along each of the 13 one-dimensional cuts listed above, which was obtained via cubic splines 
in the coordinates $r_1$, $r_2$, and cos($\theta_{HOH})$.  Along these lines
but beyond the last data points of the cuts in the $r_1$, $r_2$, or symmetric stretch
directions, the width was either
set to the terminal value, or to zero in the case of the large-$r$ region for the asymptotically bound
1~$^2A'$ or $^2B_1$ resonances.

The global representation of the fit was obtained as follows. Each one-dimensional cut is represented by a curve in the three dimensional space of coordinates $\vec{q}=(r_1, r_2, \theta)$. For a desired geometry $\vec{q_0}$, we first calculate for each cut (by interpolation, as explained above) the width  at the point which is closest to $\vec{q_0}$. We then take an average of the values at these 13 points, each inversely weighted according to  its distance from $\vec{q_0}$.
Examples of the interpolated witdh surfaces are shown in the EPAPS archive\cite{epaps}.

\section{CI calculations for the real part of resonance surfaces and the neutral
  surface } \label{CIsect}

We construct the real part of the resonance energy $E_R$ as 
a function of the internal nuclear geometry of H$_2$O using
bound-state configuration-interaction calculations.  Our
task in doing so is to accurately reproduce real-valued component of
the physical
potential-energy surfaces, taking into account the numerous features
of these surfaces which were described in Ref.~\cite{haxton3},
and, in particular, the conical intersection between the
$^2B_2$ and $^2A_1$ surfaces, for which we require a diabatization.

The main configuration-interaction
calculations on the resonance states and the ground 
state of the neutral described below in Secs. \ref{sec:mainci} and \ref{sec:gndci} 
reproduce the vertical transition energies and all but one of
the two-body 
breakup asymptotes correctly.  However, the three-body asymptotes
of the main configuration-interaction calculation are all too
high in energy (by as much as more than 1eV), and the asymptote
of the diabatic $^2A_1$ state is far below its proper 
O$^-$+H$_2$($\sigma_g^1 \sigma_u^1$) $^3\Sigma_u$ asymptote, instead being degenerate
with the other resonances as O$^-$+H$_2$($\sigma_g^2$).
For this reason, we must patch
the \textit{ab initio} surfaces produced from the main CI calculation.

We construct global representations of the adiabatic
$^2B_1$ and the diabatic $^2A_1$ and $^2B_2$ states, as well as 
the coupling, by fitting the data points produced from the main CI
calculation at each nuclear geometry.  The global representation of the 
adiabatic 1~$^2A'$ surface
is then defined  as the lower eigenvalue obtained
by diagonalizing the 2$\times$2 Hamiltonian matrix of the
global representations of the diabatic surfaces and coupling.
The 1~$^2A'$ surface will be used for calculations on DEA
via the $^2A_1$ (1~$^2A'$) Feshbach resonance, because in that
case the conical intersection is not expected to play a large
role in the dyamics, and therefore the adiabatic basis is sufficient. 
(The corresponding vector potential is not included.) 
The diabatic surfaces and coupling are used
to calculate cross sections for DEA via the higher-energy $^2B_2$
(2 $^2A'$) state.

The constructed global representations of the diabatic $^2A_1$ and
$^2B_2$ and the adiabatic $^2B_1$ and 1~$^2A'$ surfaces include
the errors that we have already mentioned: the three-body asymptotes are
too high, and the two-body asymptote of the diabatic $^2A_1$ surface in the
H$_2$+O$^-$ arrangement is too low.  In order to fix these errors,
we combined these global representations with other global surfaces
designed to reproduce the correct behavior in the region in question.
In each case a single patching surface is combined with a single CI
surface such that the final surface reflects the correct behavior.
The surfaces are combined by taking either the higher-energy of
the two surfaces (for the $^2A_1$ patching) or the lower-energy
of the two surfaces (for the three-body patching), and smoothing
the resulting cusps with a simple algebraic formula.

The patching surface for the $^2A_1$ diabatic state must correlate
to the proper H$_2$(1$\sigma_g$1$\sigma_u$)+O$^-$ asymptote of this
state. We perform another CI calculation that correlates to this
asymptote, and construct a global patching surface from the 
results, using the combined analytic fit + spline technique we 
employed for the resonance surfaces.  The patching
surfaces for the three-body asymptotes of the adiabatic $^2B_1$ and
1~$^2A'$ surfaces, as well as that of the diabatic $^2B_2$ surface,
take the form of a simple analytic potential.

The diabatization of the 1 and 2 $^2A'$ adiabatic CI roots is
a requirement dictated by the nuclear dynamics calculations.
In the adiabatic basis, there are singular derivative
couplings near the 
conical intersection between the 1 and 2 $^2A'$ surfaces.
We have not calculated these couplings from our CI wave functions, and in preparation 
for dynamics calculations on the coupled surfaces, we therefore 
perform a diabatization upon the 1 and 2 $^2A'$ CI roots 
to produce the set of diabatic $^2A_1$ and $^2B_2$ curves and
the accompanying coupling potential.  

In Ref.\cite{haxton3}, we described how
a full characterization of the
manifold of Feshbach resonances must also include a $^2B_2$ shape resonance whose
potential-energy curve intersects that of the $^2B_2$ Feshbach resonance in
branch-point fashion.  
We do not include this $^2B_2$ shape-Feshbach intersection, and instead define
a single surface which interpolates between the two sheets within the three-body
breakup region.  Therefore, the nuclear dynamics on this surface is unlikely
to accurately represent the dynamics leading to three-body breakup.
The results that we will present in paper II indicate that
three-body breakup probably comprises a large component of the
cross section for dissociative attachment via the $^2B_2$ state.

\subsection{Orbital basis}

We constructed a single orbital basis for all the CI calculations on the resonances.  
We began by augmenting the contracted Gaussian basis of Gil \textit{et al.}~\cite{Gil} 
with the following additional Gaussian functions: on the hydrogens, two $s$ functions 
with exponents 0.08 and 0.0333, and two $p$ functions with exponents 0.2 and 0.05; on the
oxygen, one $s$ function with exponent 0.0316, and one $p$ function with exponent 0.254.  The
basis comprised 77 contracted Gaussians total.

We first obtained an orbital basis by performing a symmetry-restricted SCF calculation on the
$^2B_1$ resonance, which yielded the 1$a_1$, 2$a_1$, 1$b_2$, 3$a_1$, 1$b_1$, and 4$a_1$ orbitals, 
labeled 1$a'$, 2$a'$, 3$a'$, 4$a'$, 1$a''$, and 5$a'$ in $C_s$ symmetries.  This SCF calculation is bound, i.e., it cannot decay
to H$_2$O+$e^-$ by a symmetry-conserving rotation among the orbitals, because the 1$b_1$ orbital is
restricted to be singly occupied.  The same statement is not true of the other resonances,
which are described by a hole in an $a'$ orbital, and therefore we were forced to use this
SCF orbital basis for CI calculations on all three resonances, not just $^2B_1$.

These SCF orbitals have avoided crossings (see the graph in the EPAPS archive\cite{epaps}).
When the resonance is fully
dissociated, i.e., in the arrangement O$^-$+H+H, the 4$a_1$ (5$a'$) SCF orbital is best described as
$\sqrt{1/2}(\Psi_{H_a 1s} + \Psi_{H_b 1s})$, i.e., the bonding combination of the two hydrogen 1$s$ orbitals. These 1$s$ orbitals are more similar to the hydride 1$s$ orbital than to the hydrogen 1$s$ orbital.
As the hydrogens are brought closer, the energy of this orbital decreases and eventually crosses
the 3$a_1$ (4$a'$) and 2$b_2$ (3$a'$) orbitals.  
At finite O$^-$+H$_2$ separation, these crossings are avoided.  
There are avoided crossings between the 3$a'$ and 4$a'$ (3$a_1$ and 4$a_1$) orbitals near
$\theta_{HOH}$=48$^\circ$, and between the 4$a'$ and 5$a'$ (4$a_1$ and 2$b_2$)
orbitals near $\theta_{HOH}$=61$^\circ$.  In $C_{2v}$ geometries (e.g., $r_1$=$r_2$=7.0$a_0$)
the crossing between the 4$a_1$ and 2$b_2$ orbital is an actual crossing.

In the definition of the multielectron configuration space that we use for the CI calculation,
we treat the 4$a_1$ (5$a'$) orbital differently from the 3$a_1$ or 2$b_2$ orbitals.  Thus, it was useful
to define a ``diabatic'' 4$a_1$ orbital whose energy smoothly crosses  that of the
others
and is continuous in character.
To this end, we performed a rotation among these three orbitals wherein we replaced the oxygen
nucleus with an uncharged center, by obtaining the lowest eigenvalue of the 
H$_2^+$ Hamiltonian within the space of the 3$a_1$, 4$a_1$, and 2$b_2$ orbitals.  
The ground state eigenvector of this calculation 
was defined as the diabatic 4$a_1$ orbital, and its complement as the diabatic 3$a_1$ and 2$b_2$ orbitals.
The 4$a_1$ diabatic orbital is thus the ``H$_2$-like'' orbital.  Fortuitously, this rotation leaves the
4$a_1$ orbital virtually unchanged at the equilibrium geometry of the neutral.
In the OH+H$^-$ 
asymptote, the 4$a_1$ orbital is also left unchanged as the hydride 1$s$.  The expectation of the
Fock operator upon this diabatic 4$a_1$ orbital (plotted in the EPAPS archive\cite{epaps}) passes smoothly through the avoided crossings at $\theta_{HOH}$
= 48$^\circ$ and 61$^\circ$.
We used this rotated SCF basis for CI calculations on the three resonances,
and in the description of these calculations below, the notation ``4$a_1$'' refers to the diabatic 4$a_1$ orbital.

\subsection{Main CI calculation}
\label{sec:mainci}

\begin{figure}
\resizebox{0.9\columnwidth}{!}{\includegraphics*[0.8in,0.6in][5.6in,3.8in]{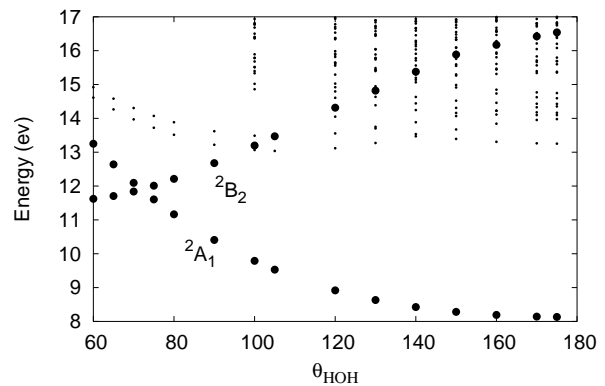}}
\caption{$^2A'$ roots of the main CI calculation at $r_1$=$r_2$=1.81$a_0$, as a function
of bending angle $\theta$. \label{roots}}
\end{figure}

The main configuration-interaction (CI) calculation on the three Feshbach resonances
employs a 
configuration space that is restricted to the configurations which contribute most to the
resonances under study.  
This space was defined by the three resonance configurations
[H$_2$O]1$b_1^{-1}$4$a_1^2$ , [H$_2$O]3$a_1^{-1}$4$a_1^2$ , and [H$_2$O]1$b_2^{-1}$4$a_1^2$,
plus all singles and doubles into the virtual space, with the 1$a_1$ orbital restricted to
be doubly occupied.  Thus, we excluded from the configuration space all references wherein
the combined occupancy of (1$b_2$ 3$a_1$ 1$b_1$) was six.  This restriction excludes configurations  of the form [H$_2$O]$n^1$, which correspond to discretized continuum states of H$_2$O +
$e^-$.  The total size of this configuration space is 111792 in $A'$ symmetry, and 106110
in $A''$ symmetry.

The restriction placed on the configuration space was critical in that it produced the resonance
energies as low-lying roots of the CI.  The resonances were identified by their dominant configuration.
The $^2B_1$ (1~$^2A''$) energy was found to be always the lowest
root of its CI, and the $^2A_1$ (1~$^2A'$) state was almost always the lowest root of the CI of that symmetry.  Therefore,
the energies thereby obtained for these resonances were smoothly varying functions of the
nuclear geometry, not suffering from avoided crossings with discretized continuum states.
The energy of the $^2B_2$ state rises above discretized continuum states of the type [H$_2$O]2$b_1^{-1}n^1m^1$
and [H$_2$O]3$a_1^{-1}n^1m^1$ at certain geometries, namely small OH bond lengths and near-linear
H-O-H geometry.  However, the resulting avoided crossings were generally observed to be very narrow.
The discretized continuum states of excited [H$_2$O$^*$]$n^1$ are
undercorrelated, relative to the resonance roots, by this multireference,
all-doubles treatment, and therefore occur at a higher energy than they would
otherwise.

In Fig. \ref{roots}, we plot
the $^2A'$ roots of this configuration-interaction calculation as a function of bending angle
$\theta_{HOH}$, at the equilibrium bond lengths $r_1$=$r_2$=1.81$a_0$, showing the behavior 
of the CI roots near the conical intersection.  Also apparent in this figure are the discretized 
continuum states, many of which lie below the $^2B_2$ resonance root when the bond angle is large.

The orbital and configuration basis for this CI calculation is designed to describe well the
resonances at the equilibrium geometry of the neutral and in the two-body dissociation channels.   
This calculation is not designed to reproduce the
three-body asymptotes, and makes a large error in these regions.  
The 4$a_1$ and 2$b_2$ orbitals have significant hydride 1$s$ character on both hydrogen
centers, as opposed to hydrogen 1$s$; since there is no relaxation of the
hydride 1$s$ orbital included in the reference space, the double excitations
into the virtual space must play that role, and the three-body asymptotes are 
therefore undercorrelated
and unphysically high in energy.  We have not attempted a precise characterization of the CI
roots in the three-body breakup region.

We performed these CI calculations at various geometries on a grid based on H-H-O bond angle coordinates
(not H-O-H) $r_{HH}$, $r_{OH}$, and $\theta_{HHO}$.  We used these coordinates so that the H$_2$+O$^-$
exit well (along with one of the OH+H$^-$ exit wells) would be well-represented by the spline proceedure
we use.  In our previous treatment\cite{haxton1} of the $^2B_1$ resonance, we used H-O-H bond-angle coordinates,
which led to an unphysically corrugated spline representation of the H$_2$+O$^-$ well.

For this calculation, we defined a full grid of 23 $r$ points between 1.0$a_0$ and 12.0$a_0$ and 29 $\theta$ points between 1$^\circ$ and 175$^\circ$.  This 23$\times$23$\times$29 grid
includes 15341 points.  The CI calculations were each performed at roughly 4200 appropriately
chosen points on this grid.

The energies of the resonance CI roots at the equilibrium geometry
($r_1$=$r_2$=1.81$a_0$, $\theta$=105$^\circ$) were -76.030888 hartree ($^2B_1$),  -75.943508 hartree
($^2A_1$ or 1~$^2A'$), and  -75.802877 hartree ($^2B_2$ or 2 $^2A'$).

\subsection{Diabatization}
\label{sec:diab}

Because the $^2A_1$ and $^2B_2$ states have a conical intersection, and since we 
have not calculated the derivative couplings between them, a diabatization~\cite{diareview} 
is required for the
nuclear dynamics calculations.  Our method for performing this diabatization is an approximate method,
based not on the explicit minimization of first-derivative matrix elements\cite{yarkonyconsequence,
yarkonyCW, atdtpoisson,meadrev,baerrev}
but upon the diagonalization of a property~\cite{propertydia} to obtain smoothly behaved diabatic states.
Our technique is thus analagous to a diabatization via the diagonalization of the dipole operator
between states which undergo a charge-transfer avoided crossing~\cite{dipoledia}, 
or the diagonalization of the $l_z$ angular momentum operator
between adiabatic states which have a $\Sigma$-$\Pi$ conical intersection at
linear nuclear geometry\cite{robsurf, aband_photo}.

The property we use for our diabatization is a symmetry operation\textemdash a
reflection perpendicular to the molecular plane\textemdash which is already diagonal in the adiabatic basis both in $C_{2v}$
geometries and in the asymptotic OH+H arrangement channel.  In $C_{2v}$
geometries, this reflection is that defined by the plane that contains the
$C_{2v}$ axis, and that is perpendicular to the molecular plane.  In such
geometries, the $^2A_1$ state is an eigenfunction of 
this symmetry operation with a  eigenvalue of +1 with respect to this reflection,
and the $^2B_2$ state has a eigenvalue -1.  In the asymptotic OH+H arrangement,
this reflection is defined by the plane perpendicular to the OH axis (which
again is perpendicular to the molecular plane) that crosses through the
oxygen nucleus.  
In these geometries, the matrix representation of the reflection operator in the basis of these
states is diagonal, with 
the 1~$^2A'$ state, which correlates to H$^-$+OH($^2\Pi$),
having a positive diagonal matrix element, and the 2 $^2A'$ state, which
correlates to
H$^-$+OH($^2\Sigma$), having a slightly negative diagonal matrix element.

\begin{figure}
\resizebox{0.6\columnwidth}{!}{\includegraphics{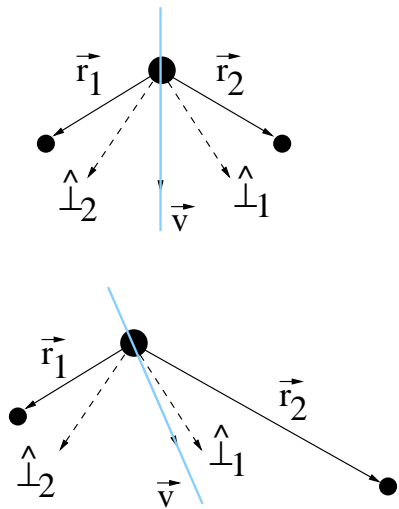}}
\caption{\label{diabat}
Vectors involved in defining the reflection operator whose diagonalization provides the
$^2A_1$ and $^2B_2$ diabatic states from the 1 and 2 $^2A'$ adiabatic states.  
The reflection plane is marked by a wide blue (grey) line and contains the vector $\vec{v}$.}
\end{figure}

In Fig. \ref{diabat}, we show the reflection plane for one $C_{2v}$ geometry and for 
one geometry approaching the H$^-$+OH arrangement.  The reflection plane is
that plane that contains the vector $\vec{v}$ and which is perpendicular to
the molecular plane.  The vector $\vec{v}$ is a weighted sum of the unit
vectors $\widehat{\perp_1}$ and $\widehat{\perp_2}$.  These vectors are
defined as the unit vectors that are perpendicular to the corresponding OH
bond vectors $\vec{r_1}$ and $\vec{r_2}$, and that are contained in the HOH
bond angle.  The expression for $\vec{v}$ is 
\begin{equation}
\vec{v}=\exp(-r_1/r_0)\widehat{\perp_2} + \exp(-r_2/r_0)\widehat{\perp_1},
\end{equation}
where the length parameter $r_0$ is 1 bohr.  This value was chosen on the
basis of the dimensions of the current molecular system, and by inspecting the
continuity of the diabatic potential-energy surfaces produced.

\subsection{CI calculation for the ground state potential surface}
\label{sec:gndci}

The Born-Oppenheimer potential-energy surface for the neutral
molecule is required for the nuclear dynamics calculations, and to 
set the zero of energy for the 
dissociative attachment cross sections thereby produced.

For the calculation of the neutral potential surface, we 
followed a prescription similar to that used in generating the resonance
surfaces.  An SCF calculation on the neutral was performed, followed
by a configuration-interaction calculation with all singles and
doubles from the SCF configuration, keeping the 1$a_1$ orbital doubly
occupied.  The size of this CI calculation is 22215 configurations
in $C_{2v}$ symmetry.

For the neutral, we used the physical H-O-H bond angle coordinates and a grid
defined by 
$r_{OH}$=$\{$1.41, 1.61, 1.81, 2.01, 2.21, 2.41, 2.61, 3.01, 3.41 $a_0\}$,
and $\theta$=$\{$60, 75, 90, 105, 120, 135, 150, 165$^\circ\}$.
The CI calculation
was performed on each point on this grid.

The energy of the neutral CI calculation at the equilibrium geometry
($r_1$=$r_2$=1.81$a_0$, $\theta$=105$^\circ$) was -76.2900969 hartree,
yielding vertical excitation energies for the resonances of 
7.054, 9.431, and 13.258eV.  (No zero-point energy is included.)

\subsection{Global representation}

To construct a global representation of these five CI potential-energy surfaces\textemdash 
the $^2B_1$, the diabatic $^2A_1$ and $^2B_2$, the patching $^2A_1$ surface, and the ground
state surface\textemdash
a reference potential was first fit to the data, then subtracted from the computed points;
the remainder was then fit with three-dimensional cubic splines.  The sum of the reference
fit plus splined remainder comprises the global fit, which coincides exactly with
the calculated points.

The functional forms of the reference potential $V_{res}$ for the four
resonance curves and $V_{neut}$ for the neutral are given in the EPAPS archive\cite{epaps}.  Root-mean-square
errors of each fit were on the order of 0.1eV.

The errors of each of the analytic fits were fit to cubic splines.
Since we did not calculate the full grid of points, a multi-step
splining proceedure was required.  First, a series of one-dimensional
splines, in the $\theta_{HHO}$ direction and then along the $r_{HH}$ and $r_{OH}$
directions, was performed to obtain the splined error at the remaining grid
points.  Second, the full grid of data thus constructed was 
fit to three-dimensional cubic splines, and added to the analytic fit to obtain the
global representation.  This proceedure yields the spline surface 
 $V_S$.  
The global representation is $V_{res}+V_S$ for the resonances or
$V_{neut}+V_S$ for the neutral.

\subsection{Representation of electronic coupling term and transformation of width to diabatic basis}

The electronic coupling matrix element between the $^2A_1$ and $^2B_2$ states was represented by a fit to a polynomial times
Gaussian expansion in a rotated coordinate system.  The explicit form can be found in the EPAPS
archive\cite{epaps}.
The  RMS error of this fit was 0.05eV.

This global fit of the coupling matrix element has a small remainder.   Thus,
while the diabatic $^2A_1$ and $^2B_2$ surfaces pass exactly through the calculated points,
the coupling surface, and thus the adiabatic surfaces obtained by a diagonalization
of the electronic Hamiltonian thereby constructed, do not do so precisely.

\subsection{Patching of the surfaces}

The potential-energy surfaces constructed from the main CI calculation 
appear to reproduce the known features and all but one of the two-body
asymptotes of the physical system, without recourse to an overall
vertical adjustment in the relative position of the neutral an anion
curves, or any other ad hoc adjustment.  However, there are two regions
in which the current configuration interaction treatment fails to
reproduce the physical energetics: for the high-energy H$_2$($\sigma_g \sigma_u$) $^3\Sigma_u$ + O$^-$ 
asymptote of the 2 $^2A'$ ($^2A_1$) 
surface, and in the three-body breakup region for all three resonances.

Since the goal of the present study is to present the most physically accurate
theoretical treatment of dissociative electron attachment within the
local complex potential model, we correct these flaws in the surfaces
by employing a patching procedure.  In both cases a second surface with
the desired characteristics is constructed and patched to the errant area.
This patching is performed on the global fits $V_{res}+V_S$, not upon
the original data points.  The patching is performed by taking either the
maximum of the original and the patching surface (for the diabatic $^2A_1$
surface), or the minimum (for the three-body asymptotes), and smoothing
the resultant cusps with a simple mathematical formula.  This
formula perserves the surfaces identically in the unpatched regions, and
is described below.

\subsubsection{$^2A_1$ patching surface}

All three roots of the main CI calculation correlate to H$_2$+O$^-$ in that
arrangement channel.  However, the correct asymptote of the 2 $^2A'$ state in
that arrangement is H$_2$(1$\sigma_g$1$\sigma_u$)+O$^-$~\cite{haxton3}.  Therefore, we
performed an additional CI calculation that correlates to this state in that
arrangment, and patched this surface to the diabatic $^2A_1$ surface produced from
the main CI, thereby correcting it.  The diabatic $^2B_2$ surface is left unchanged.

The Gaussian basis and orbitals for this CI were exactly the same as those for
the main CI, including the rotation of the 4$a_1$ orbital; the only difference
was the choice of configurations.  We included all single and double
excitations from the configuration [H$_2$O]4$a_1^1$, keeping the 1$a_1$ orbital
doubly occupied, and the 4$a_1$ orbital never doubly occupied.  We took the
lowest
root of this CI.  In the O+H$_2$
arrangement, the (diabatic) 4$a_1$ orbital correlates to the H$_2$ 1$\sigma_g$ orbital, and
therefore the lowest root is O$^-$+H$_2$ (1$\sigma_g$1$\sigma_u$).  Elsewhere, 
the lowest root corresponds to a discretized continuum state of H$_2$O+$e^-$, and 
lies below the diabatic $^2A_1$ surface.
We performed this calculation on the same grid as the main CI calculation.
We constructed a global representation of the patching surface employing 
our function $V_{res}$ and a splined
residual $V_S$, just as we did for the resonance surfaces.

This surface and the diabatic $^2A_1$ surface 
intersect, and the upper surface contains
the proper asymptotes for the physical $^2A_1$ surface.  It also contains cusps
where the surfaces intersect.  These cusps were smoothed by the following
prescription.  Given the separation $\Delta E$ and the average $\overline{E}$
of these two surfaces, 
\begin{equation}
\begin{split}
\Delta E & = \left\vert E_{A_1} - E_{patching} \right\vert \\
\overline{E} & = (E_{A_1} + E_{patching})/2, 
\end{split}
\end{equation}
and a geometry-dependent minimum separation A (in eV) defined as 
\begin{equation}
A = 0.5+12\exp(-2r_{HH}/3 + 1) ,
\end{equation}
the patched $^2A_1$ surface was obtained via

\begin{widetext}
\begin{equation}
\begin{split}
 E_{A_1} \longrightarrow   \begin{cases}
  \overline{E} +  \frac{A}{2-\sqrt{2}} - \sqrt{\frac{A^2}{(2\sqrt{2}-2)^2} -
  (\frac{\Delta E}{2})^2}  & \ \ \Delta E \le \frac{A}{2-\sqrt{2}} \\
 \max(E_{A_1},E_{patching}) & \ \ \Delta E \ge \frac{A}{2-\sqrt{2}}.
\end{cases} 
\end{split}
\label{patcheqn}
\end{equation}
\end{widetext}

\subsubsection{Three-body asymptote patching}

\begin{figure*}
\begin{center}
\begin{tabular}{ccc}
%
%
\resizebox{0.65\columnwidth}{!}{\includegraphics{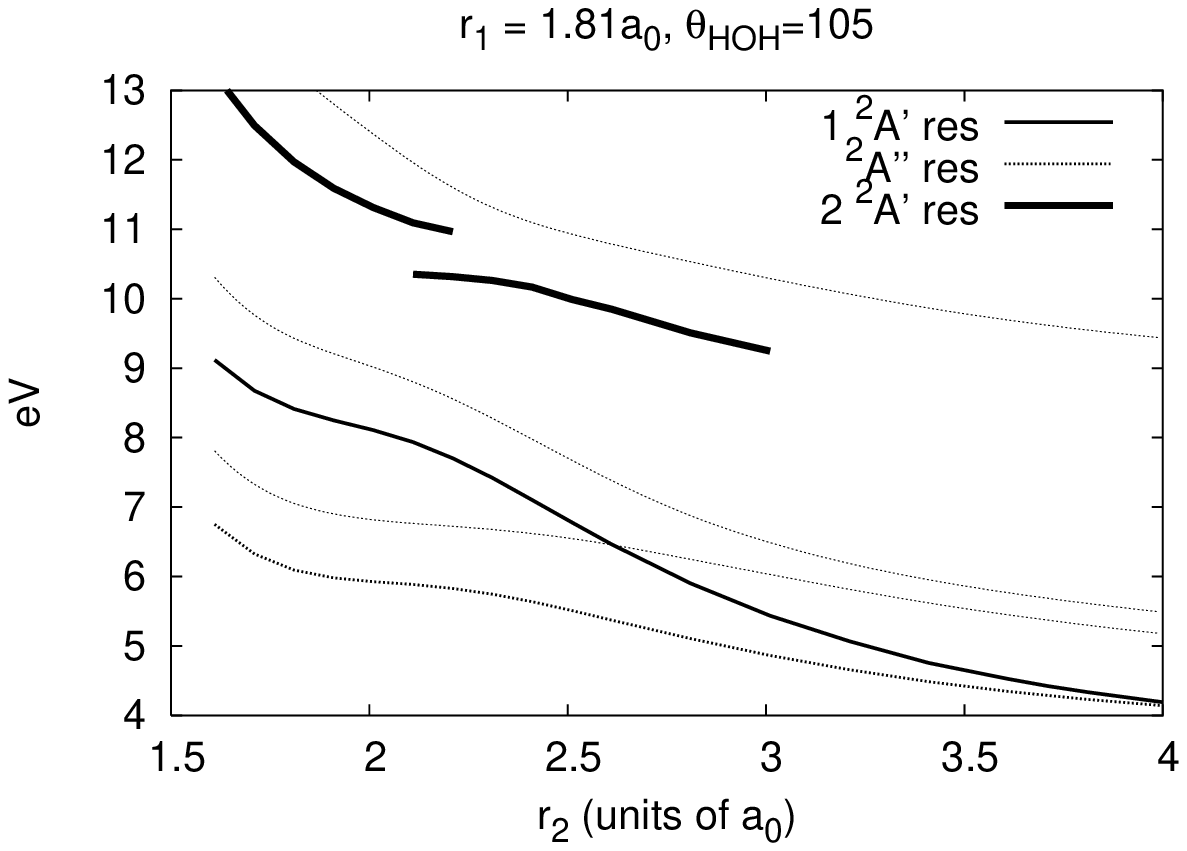}} &
\resizebox{0.65\columnwidth}{!}{\includegraphics{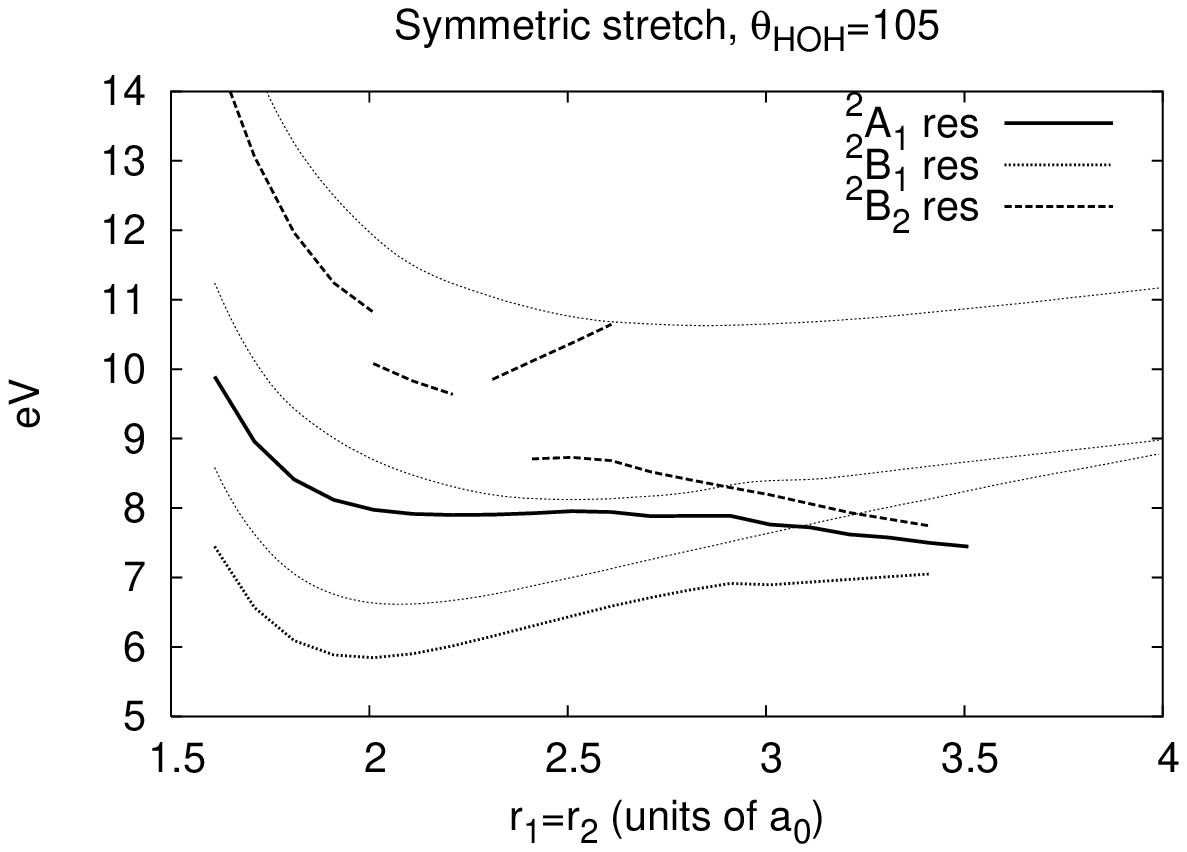}} &
\resizebox{0.65\columnwidth}{!}{\includegraphics{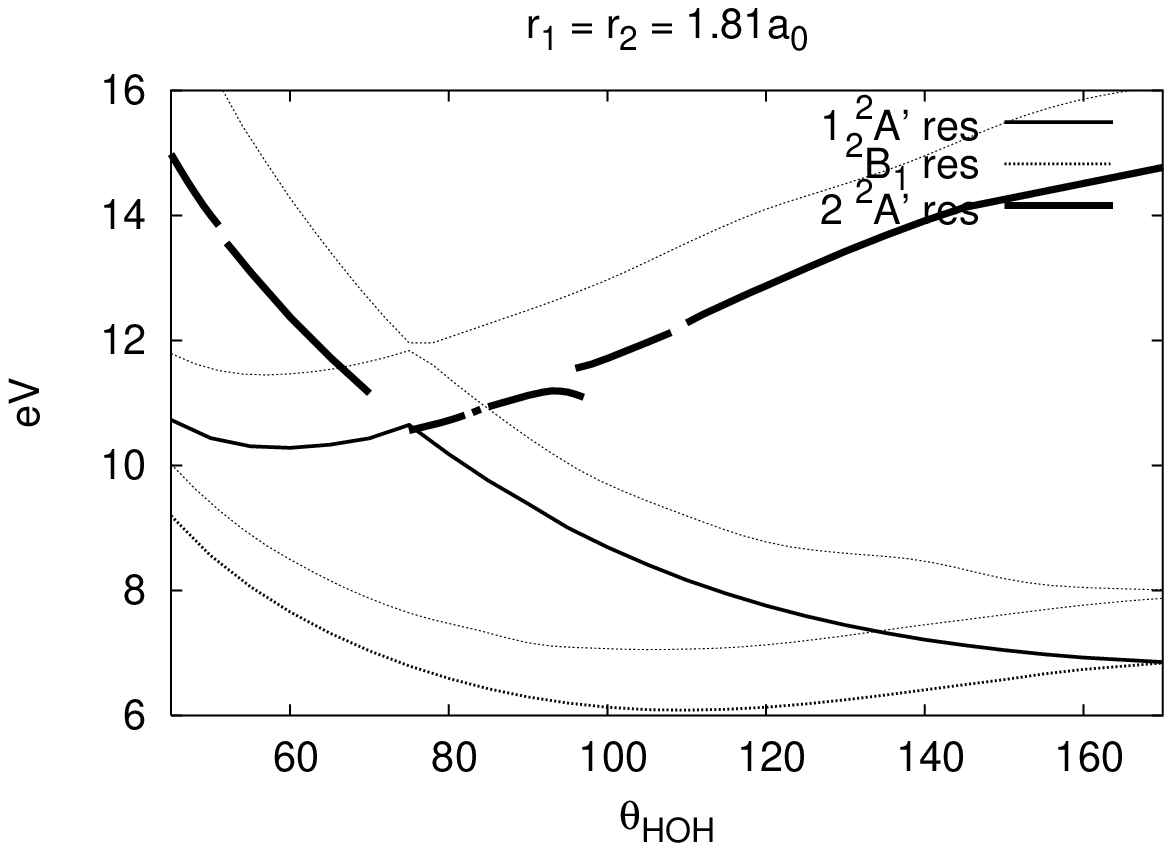}} \\
\resizebox{0.65\columnwidth}{!}{\includegraphics{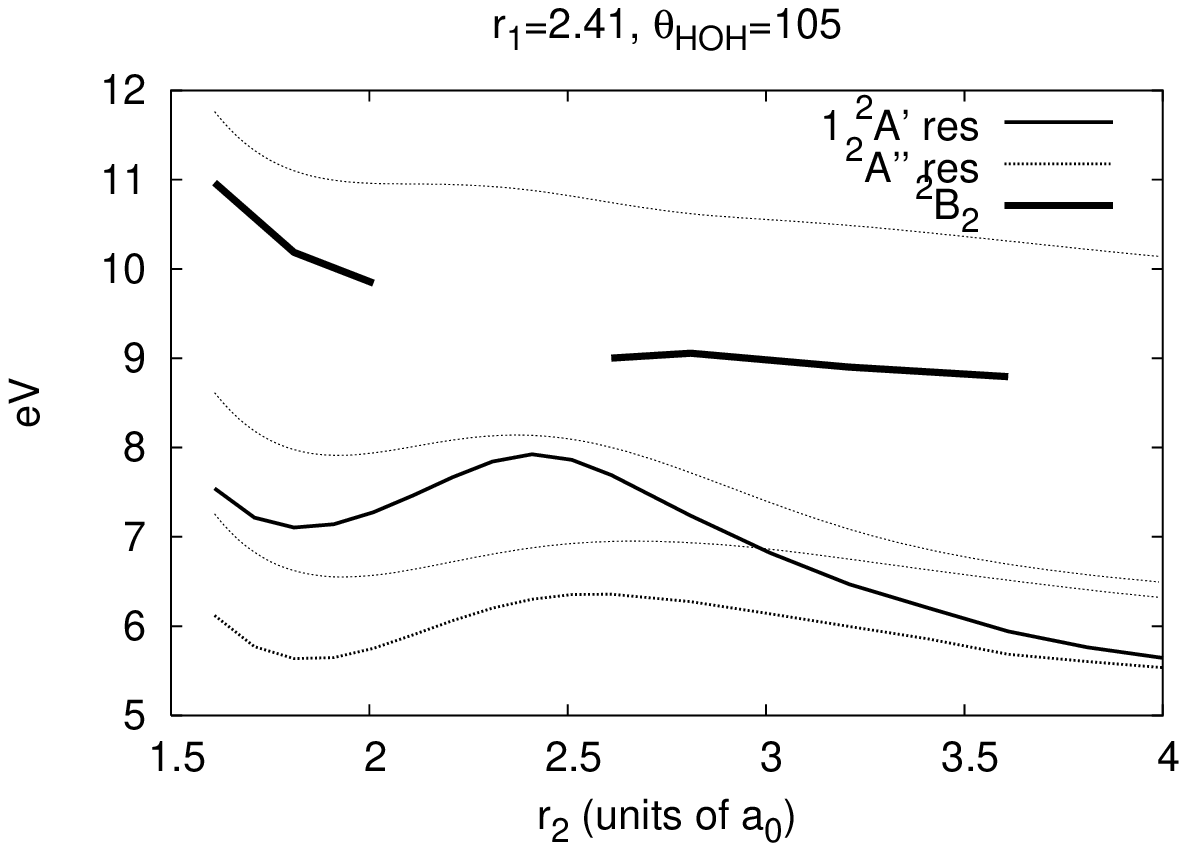}} &
\resizebox{0.65\columnwidth}{!}{\includegraphics{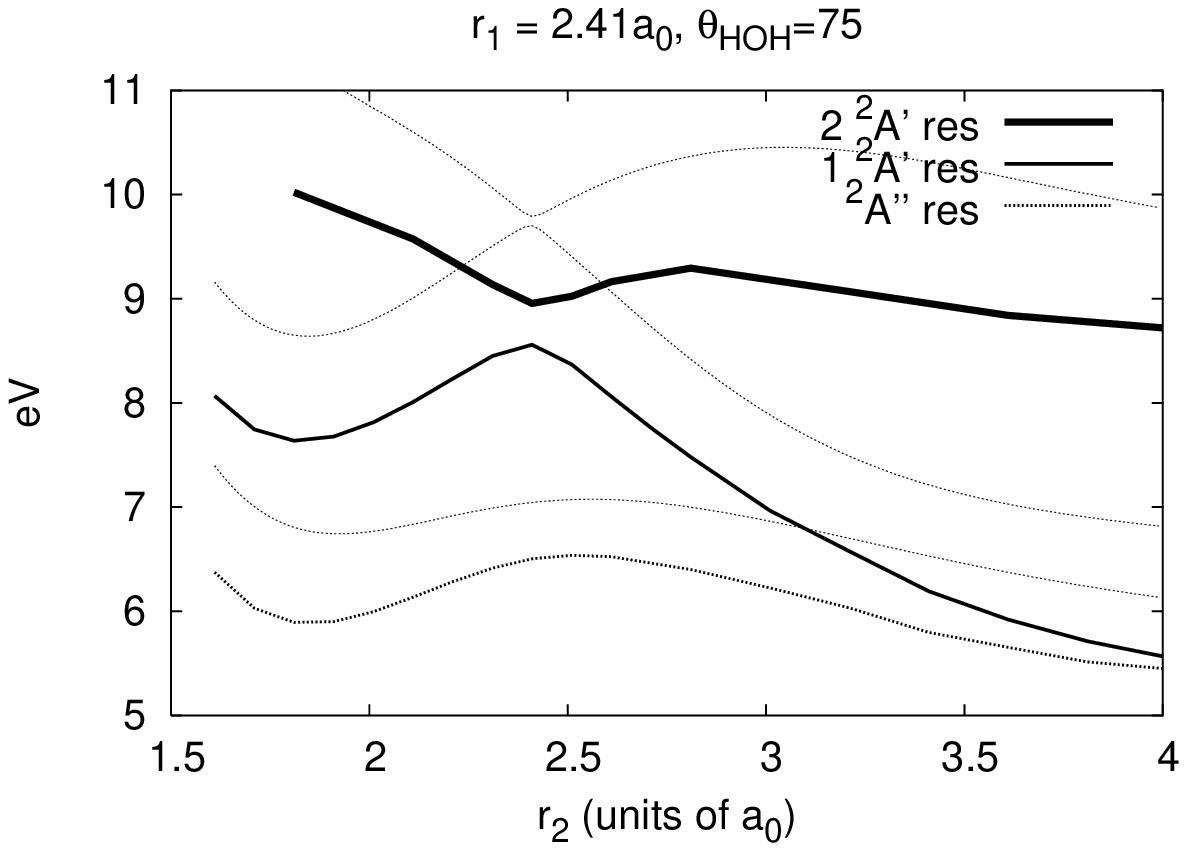}} &
\resizebox{0.65\columnwidth}{!}{\includegraphics{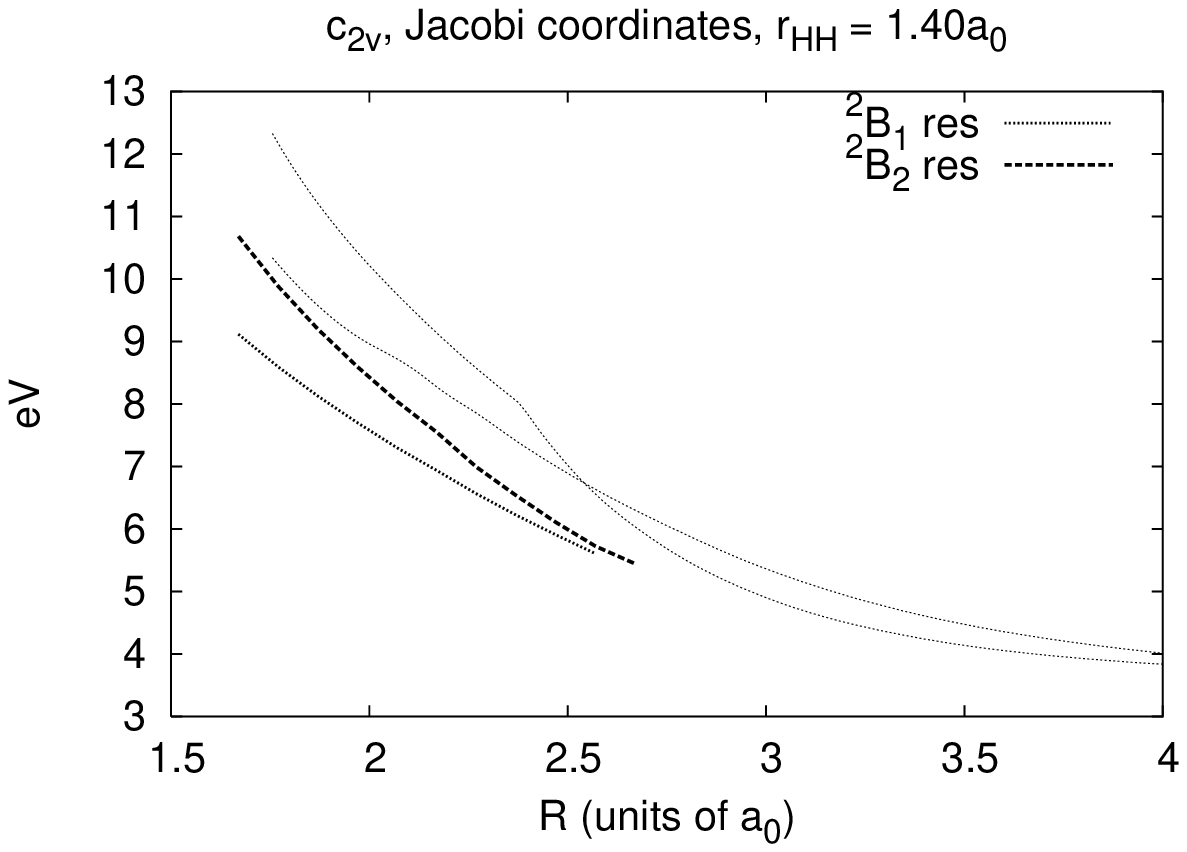}} 
\end{tabular}
\end{center}
\caption{\label{comparison}
Comparison of scattering calculations and CI results along various cuts.  Bold lines: results of scattering calculations; thin lines, CI results.}
\end{figure*}

In order to patch the three-body asymptotes of the constructed global
representations, we defined the following two functions
of the OH bond lengths $r_1$ and $r_2$:
\begin{equation}
\begin{split}
p_{low}\left(x\right) & = 8.4287 + 5.8739  \exp\left[-1.4838\left(x-1.81\right)\right] \\
p_{high}\left(x\right) & = 10.7140 + 3.5886  \exp\left[-2.4287\left(x-1.81\right)\right] \\
x & = \mathrm{min}(r_1,r_2) ,
\end{split}
\end{equation}
in eV, which functions have the same gradient as the $^2B_2$ surface in the
symmetric stretch coordinate but are 1.5eV above it at the equilibrium
geometry of the neutral.  In the three-body breakup region, 
these functions approach the values of their constant terms, which are chosen
to represent the physical asymptotes of the system.

The function $p_{low}$ is used to patch the asymptotes of the $^2B_1$ and 
the 1~$^2A'$ surfaces.  The value of its constant term, 8.43eV, is in-between
the asymptotes H$^-$+H+O$^-$ at 8.75eV and H+H+O$^-$ at 8.04eV.
While, as discussed in Ref.~\cite{haxton3}, the proper asymptote
of these adiabatic electronic states is the lower of these, the higher
may be reached by an excursion into the OH+H$^-$ two-body breakup
channel that does not rigorously follow the adiabatic state in becoming
H$^-$+H+O.  Therefore, we choose a compromise between these values
to represent the physical system.  A more accurate treatment would
use two surfaces, but we expect the present treatment to be sufficient
for determining the two-body DEA cross sections.

The function $p_{high}$ is used to patch the diabatic $^2B_2$ surface.
Its asymptote is chosen as the energy of
H$^-$+H+O($^1D$).  
Unlike the $^2B_1$ and 1~$^2A'$ surfaces, the $^2B_2$
(2 $^2A'$) adiabatic surface is inherently double-valued within the
inner regions of nuclear configuration space, and has two proper
three-body asymptotes; 
H+H+O$^-$ is the other asymptote.
We have not attempted 
to characterize the full double-valued $^2B_2$ surface.  The patching
surface $p_{high}$ corrects the $^2B_2$ three-body asymptote to the higher
of its two physical values, in order to reproduce the OH($^2\Sigma$)+H$^-$
two-body asymptote accurately.  Therefore, the repulsive wall in the H$_2$+O$^-$
potential well extends above its physical value.
It is hoped that this treatment does
not alter the dynamics leading to the two-body dissociation channels
to a significant degree.

We combine the functions $p_{high}$ and $p_{low}$ with the resonance
surfaces using the same equation, Eq.(\ref{patcheqn}), as we used
for the $^2A_1$ patching, with a constant value of $A$=1.0eV.

\subsection{Transformation of width to diabatic basis}

In order to transform the width surfaces produced from the scattering calculations, 
which are constructed in the adiabatic
1~$^2A'$ and 2 $^2A'$ basis, to the diabatic basis, the adiabatic-to-diabatic transformation
matrix was constructed using the \textit{patched} diabatic $^2A_1$ surface and the
coupling calculated from the original diabatization of the unpatched surface.  This
adiabatic-to-diabatic transformation matrix is therefore different from the transpose of the
one that
diabatized the adiabatic states from the main CI calculation.  The diabatic width
surfaces are defined as
\begin{equation}
\left( \begin{array}{cc}
\Gamma_{^2\mathrm{A}_1} & \Gamma_{\mathrm{C}} \\
\Gamma_{\mathrm{C}}  & \Gamma_{^2\mathrm{B}_2} \\
\end{array}\right) = 
U^T\left( \begin{array}{cc}
\Gamma_{1~^2\mathrm{A}'} &  0 \\
  0          & \Gamma_{2~^2\mathrm{A}'} \\
\end{array}\right) U ,
\end{equation}
with the adiabatic-to-diabatic transformation matrix U expressed in terms of the
angle $\theta'$,
\begin{equation}
U = \left( \begin{array}{cc}
\cos(\frac{\theta'}{2}) & \sin(\frac{\theta'}{2}) \\
-\sin(\frac{\theta'}{2}) & \cos(\frac{\theta'}{2}) \\
\end{array}\right) , 
\end{equation}
which is defined in terms of the patched surface ${V_{^2\mathrm{A}_1'}}$ as
\begin{equation}
\cot(\theta') = \frac{{V_{^2\mathrm{A}_1'}} - V_{^2\mathrm{B}_2}}{2\mathrm{C}}.
\end{equation}
This angle $\theta'$ is different than the original angle $\theta$ that diagonalized
the reflection operator,
\begin{equation}
\cot(\theta) = \frac{V_{^2\mathrm{A}_1} - V_{^2\mathrm{B}_2}}{2\mathrm{C}},
\end{equation}
where $V_{^2\mathrm{A}_1}$ is the original unpatched diabatic $^2A_1$ surface.
In particular, in the O$^-$+H$_2$ asymptote,
the patching vastly increases the difference between the $^2A_1$ and $^2B_2$ surfaces, and 
thus the new adiabatic-to-diabatic transformation matrix is nearly unity there.

\subsection{Comparison with complex Kohn results}

The results of the CI calculations are compared with the resonance locations obtained
from the complex Kohn calculation, along various cuts, in Fig. \ref{comparison}.
In most cases this comparison is quite favorable, although some differences are apparent.
These results are presented in terms of the original, unpatched CI surfaces.

The top three cuts, which each contain the equilibrium geometry of the neutral,
are in excellent agreement, except for the 2 $^2A'$ surface for the symmetric stretch cut.
Along this cut, the $^2B_2$ resonance has branched into the two components of the double-valued
$^2B_2$ shape-Feshbach state, as discussed in Ref. \cite{haxton3}.

We include the cuts at $r_1$=2.41$a_0$ in Fig. \ref{comparison} because these geometries
are relevant to the wavepacket dynamics on the 2 $^2A'$ surface.  The gradient of the real component of that surface,
as well as the behavior of the imaginary component, takes the propagated wavepacket through these cuts.
The cut at $r_1$=2.41$a_0$, $\theta_{HOH}$=75$^\circ$ is near the conical intersection.
Along this cut, the behavior of the CI surfaces mirrors the behavior of the Kohn resonances,
though the agreement is not as good at $\theta_{HOH}$=105$^\circ$.

\section{Description of the complex potential surfaces} \label{surfsect}

Several views of the diabatic surfaces can be found in the 
EPAPS archive\cite{epaps}.  Here we show one in
the vicinity of the conical intersection, and then focus on the
adiabatic surfaces only.

\subsection{Views of the conical intersection}

\begin{figure}[t]
\begin{center}
\begin{tabular}{c}
\resizebox{0.65\columnwidth}{!}{\includegraphics*[1.05in,1.0in][5.2in,3.85in]{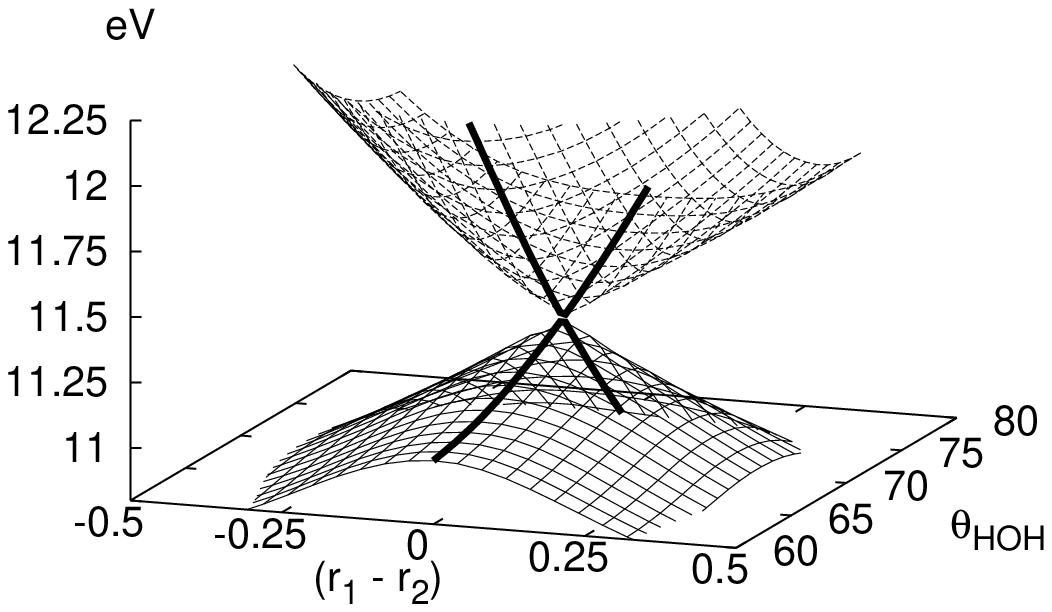}} \\
\resizebox{0.65\columnwidth}{!}{\includegraphics*[1.05in,1.0in][5.2in,3.75in]{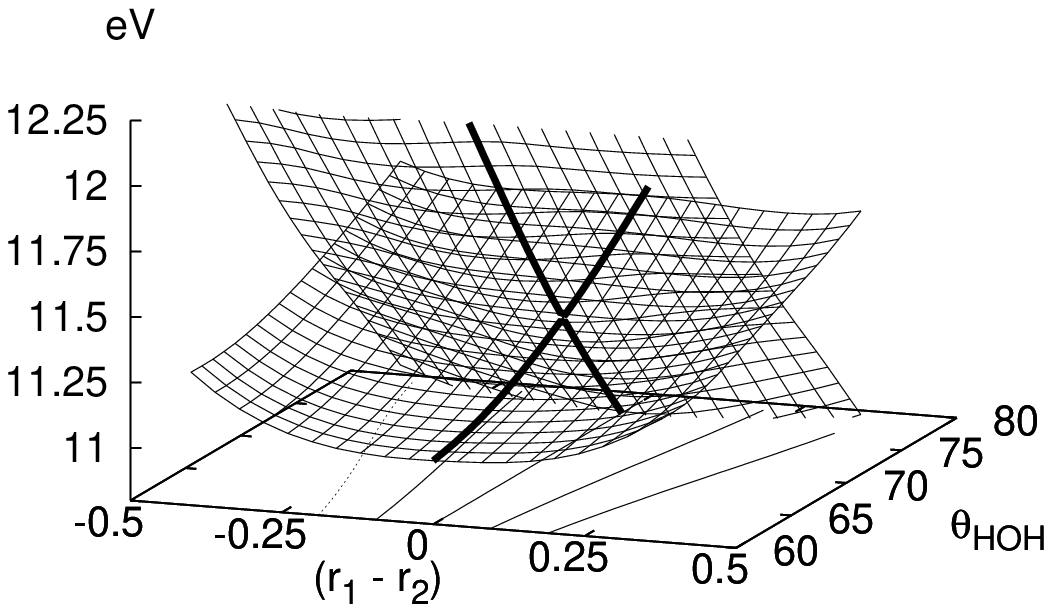}} \\
\end{tabular}
\end{center}
\caption[Conical intersection]{
Cut of conical intersection at $r_1+r_2$=3.62$a_0$: adiabatic representation, top; diabatized surfaces and coupling, bottom.  Top, values of 1 and 2 $^2A'$ potential surfaces as constructed from 
global representations $V_{res}+V_S$ of diabatic $^2A_1$ and $^2B_2$ surfaces and coupling.  
The potentials
are plotted with respect to asymmetric stretch, units of bohr, 
and bending angle in degrees.  The $^2A_1$ and
$^2B_2$ surfaces along $C_{2v}$ geometry ($r_1$=$r_2$) are marked with bold lines.  Bottom, fitted diabatized
surfaces.  
The coupling is plotted as contours at bottom, contours every 0.25eV.}
\label{conicalpic} 
\end{figure}

A plot of the conical intersection is shown in Fig. \ref{conicalpic}.
In this figure, the symmetric stretch coordinate is held constant
at $r_1+r_2$=3.62$a_0$; we see one cut of the potential-energy
surface that intersects the conical intersection seam at a point,
at a bond angle of approximately $\theta_{HOH}$=72$^\circ$.

\subsection{Energetics of the adiabatic surfaces}

\begin{figure}[t]
\begin{center}
\begin{tabular}{cc}
\includegraphics*[width=0.95\columnwidth]{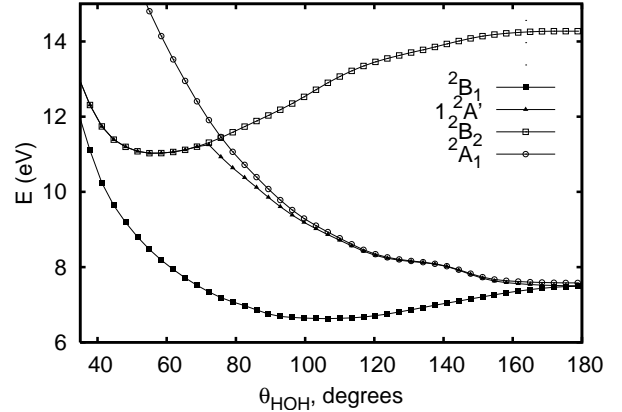} &
\end{tabular}
\end{center}
\caption[Bending potentials for the resonance curves]{
Potential-energy curves of calculated Feshbach resonances at $r_1$=$r_2$=1.81$a_0$, as constructed by global representation. }
\label{Bendfig}
\end{figure}

The real parts of the resonance surfaces $V_{res}+V_S$ with patching
are plotted in Fig. \ref{Bendfig}, fixing the bond lengths at their
equilibrium values and varying the bond angle.  The initial state starts 
at $\theta_{HOH}$=104.5$^\circ$, at which geometry the resonances are
6.63, 9.01, and 12.83eV above the ground vibrational state of the neutral.
The conical intersection is apparent at approximately $\theta_{HOH}$=76$^\circ$,
where the $^2A_1$ and $^2B_2$ surfaces intersect.  The 1~$^2A'$ surface is
plotted separately and is lower than the diabatic surfaces near the conical
intersection.  This is a consequence of the three-body patching proceedure
and the fact that different functions $p_{high}$ and $p_{low}$ are used
to patch the 1~$^2A'$ and $^2B_2$ surfaces.  The patching of the $^2B_2$
surface is apparent as the slight downward kink of the surface near 115$^\circ$,
at the edge of the Franck-Condon region of the neutral.  The bump in the
1~$^2A'$ and $^2A_1$ surfaces near 140$^\circ$ is a localized artifact
of the spline which thwarted removal.

The vertical transition energy for the $^2B_1$ state is very near the experimental
peak maximum for DEA via this resonance, while the vertical transition energies for the $^2A_1$ 
and $^2B_2$ states  exceed the
experimental DEA peak positions by $\sim$0.4eV and $\sim$1eV, respectively.  We must point out, however, that the location of the experimental peaks in the DEA
cross section do not necessarily coincide with the vertical transition
energies, especially for the upper states with their shorter lifetimes.
As we will see in paper II, the vertical transition energies of the upper resonance states 
are probably closer to their appropriate physical values
than than these comparisons would suggest.

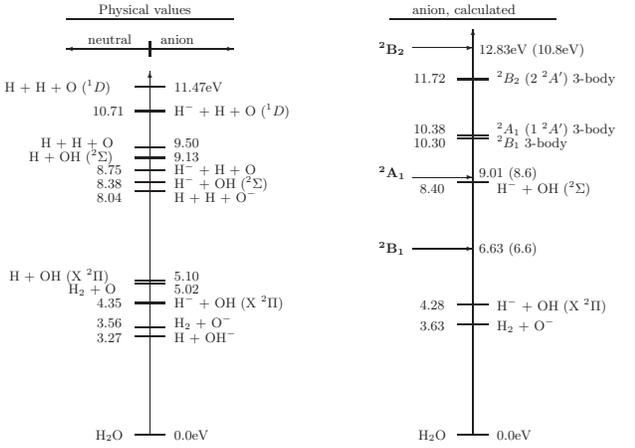
\begin{figure}[t]
  \begin{center}
\begin{tabular}{cc}

\resizebox{0.48\columnwidth}{!}{

    \begin{picture}(200,300)(0,0)

\put(45,285){\thicklines{\line(1,0){110}}}
      \put(66,288){Physical values}

      \put(100,10){\vector(0,1){240}}

      \put(90, 10){\thicklines{\line(1, 0){20}}}
      \put(90, 75){\thicklines{\line(1, 0){20}}}
      \put(90, 81){\thicklines{\line(1, 0){20}}}
      \put(90, 97){\thicklines{\line(1, 0){20}}}
      \put(90, 110){\thicklines{\line(1, 0){20}}}
      \put(90, 112){\thicklines{\line(1, 0){20}}}
      \put(90, 171){\thicklines{\line(1, 0){20}}}
      \put(90, 177){\thicklines{\line(1, 0){20}}}
      \put(90, 185){\thicklines{\line(1, 0){20}}}
      \put(90, 193){\thicklines{\line(1, 0){20}}}
      \put(90, 200){\thicklines{\line(1, 0){20}}}
      \put(90, 224){\thicklines{\line(1, 0){20}}}
      \put(90, 240){\thicklines{\line(1, 0){20}}}

      \put(100,265){\vector(-1,0){55}}
      \put(100,265){\vector(1,0){55}}
      \put(100,260){\thicklines{\line(0,1){10}}}

      \put(59,268){neutral}
      \put(107,268){anion}

      \put(64, 7){H$_2$O}

      \put(116, 71){H + OH$^-$}
      \put(116, 81){H$_2$ + O$^-$}
      \put(116, 94){H$^-$ + OH (X $^2\Pi$)}

      \put(46, 103){H$_2$ + O}
      \put(7, 112){H + OH  (X $^2\Pi$)}

      \put(116, 164){H + H + O$^-$}
      \put(116, 173){H$^-$ + OH ($^2\Sigma$)}
      \put(116, 182){H$^-$ + H + O}

      \put(20, 191){H + OH ($^2\Sigma$)}
      \put(28, 200){H + H + O}

      \put(116, 221){H$^-$ + H + O ($^1D$)}

      \put(4, 237){H + H + O ($^1D$)}

      \put(116, 7){0.0eV}

      \put(65, 71){3.27}
      \put(65, 81){3.56}
      \put(65, 94){4.35}

      \put(116, 103){5.02}
      \put(116, 112){5.10}

      \put(65, 164){8.04}
      \put(65, 173){8.38}
      \put(65, 182){8.75}

      \put(116, 191){9.13}
      \put(116, 200){9.50}

      \put(62, 221){10.71}

      \put(116, 237){11.47eV}

    \end{picture}
}
&
\resizebox{0.48\columnwidth}{!}{
    \begin{picture}(200,290)(0,0)
      \put(100,10){\vector(0,1){268}}

      \put(90, 10){\thicklines{\line(1, 0){20}}}
      \put(90, 83){\thicklines{\line(1, 0){20}}}
      \put(90, 96){\thicklines{\line(1, 0){20}}}
      \put(90, 177){\thicklines{\line(1, 0){20}}}

      \put(90, 206){\thicklines{\line(1, 0){20}}}
      \put(90, 208){\thicklines{\line(1, 0){20}}}
      \put(90, 245){\thicklines{\line(1, 0){20}}}

\put(45,285){\thicklines{\line(1,0){110}}}
      \put(60,288){anion, calculated}

      \put(64, 7){H$_2$O}

      \put(116, 79){H$_2$ + O$^-$}
      \put(116, 92){H$^-$ + OH (X $^2\Pi$)}
      \put(116, 170){H$^-$ + OH ($^2\Sigma$)}

      \put(116, 200){$^2B_1$ 3-body}
      \put(116, 209){$^2A_1$ (1 $^2A'$) 3-body}
      \put(116, 243){$^2B_2$ (2 $^2A'$) 3-body}

      \put(116, 7){0.0eV}

      \put(65, 79){3.63}
      \put(65, 93){4.28}
      \put(65, 170){8.40}

      \put(61, 200){10.30}
      \put(61, 209){10.38}
      \put(61, 243){11.72}

      \put(104, 130){6.63 (6.6)}
      \put(104, 180){9.01 (8.6)}
      \put(104, 262){12.83eV (10.8eV)}

      \put(60, 133){\vector(1,0){40}}
      \put(60, 180){\vector(1,0){40}}
      \put(60, 266){\vector(1,0){40}}

      \put(38, 130){$\mathbf{^2B_1}$}
      \put(38, 180){$\mathbf{^2A_1}$}
      \put(38, 262){$\mathbf{^2B_2}$}

    \end{picture}
}

\end{tabular}
  \end{center}
\caption{\textbf{Left:} physical
thresholds\cite{energy1, energy2, energy3, energy4, energy5, energy6, energy7} 
of one- and two-body breakup channels, relative to ground state neutral H$_2$O, 
relevant to dissociative electron attachment to H$_2$O.  Vibrational ground 
states where applicable\textemdash zero point energies are included.
\textbf{Right:} results of the present configuration
interaction calculations on the anions, relative to the calculated ground vibrational state
of the neutral.  The vertical transition energies from the ground vibrational state
of the neutral to each CI surface are marked with arrows, with experimental peak maxima
in parenthesis for comparison.
The three-body asymptotes labeled in the right panel correspond to values
of global fits of potential-energy surfaces at ($r_1$, $r_2$, $\theta$)=(10.0$a_0$, 10.0$a_0$, 60$^\circ$). }
\label{energetics}
\end{figure}

Globally, these potential-energy surfaces appear to reproduce the essential
energetics of the underlying physical states.  Only in the case of the three-body
breakup region is this agreement the result of an \textit{ad hoc} procedure;
elsewhere, the potential-energy surfaces represent the results of \textit{ab initio}
calculations.  In the case of the two-body asymptotes, we have been fortunate to
obtain very good agreement with the proper energetics.  The energetics of the system
of three coupled Feshbach resonances as calculated is summarized in Fig. \ref{energetics}.
On the left of this figure are the accepted values for the differences in internal energy among
each of these species, obtained from Refs.\cite{energy1, energy2,  energy3,  energy4,  energy5,  energy6,  energy7}.
On the right are the results of the present calculations.  The three-body asymptotes listed there
correspond to the unpatched values of the configuration-interaction surfaces at the geometry
$(r_1,r_2,\theta)=(10.0a_0, 10.0a_0, 60^\circ)$.

\subsection{Asymptotes of the adiabatic surfaces}

\begin{figure}
\begin{center}
\begin{tabular}{c}
\includegraphics*[width=0.95\columnwidth]{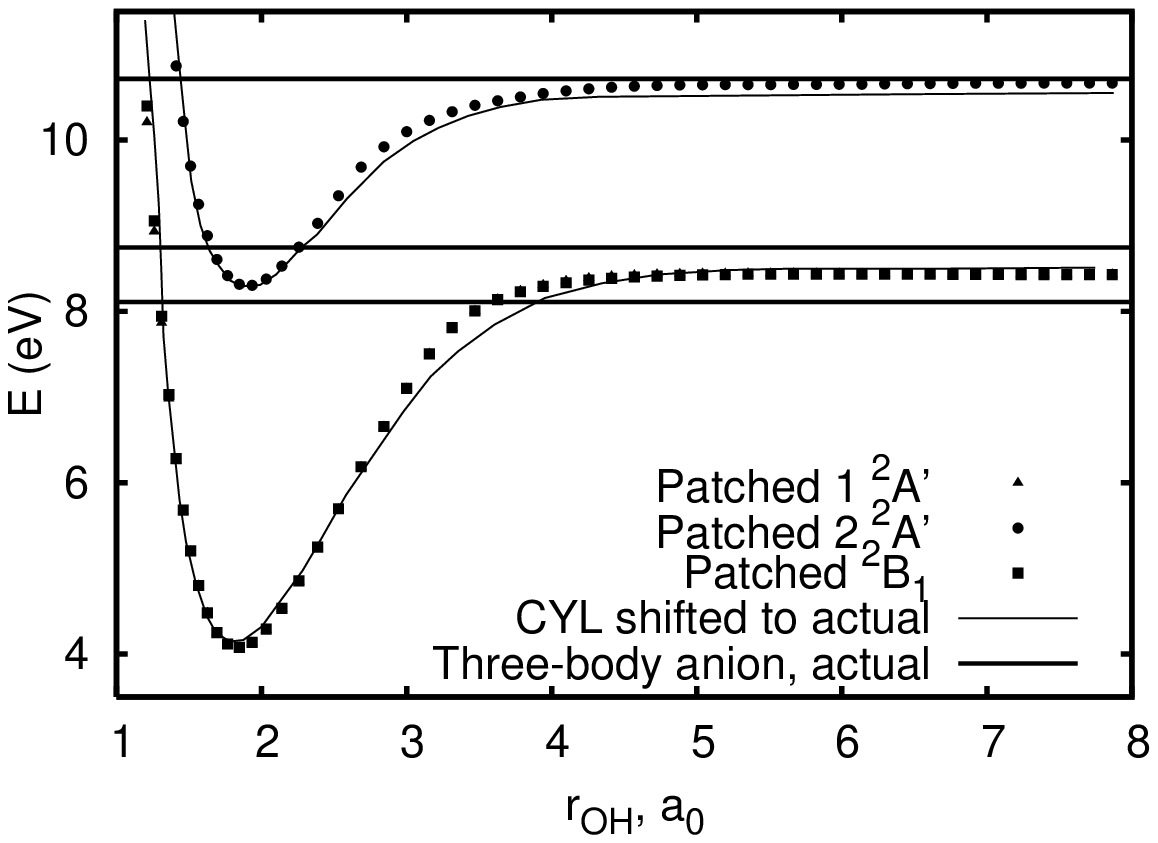} \\
\includegraphics*[width=0.95\columnwidth]{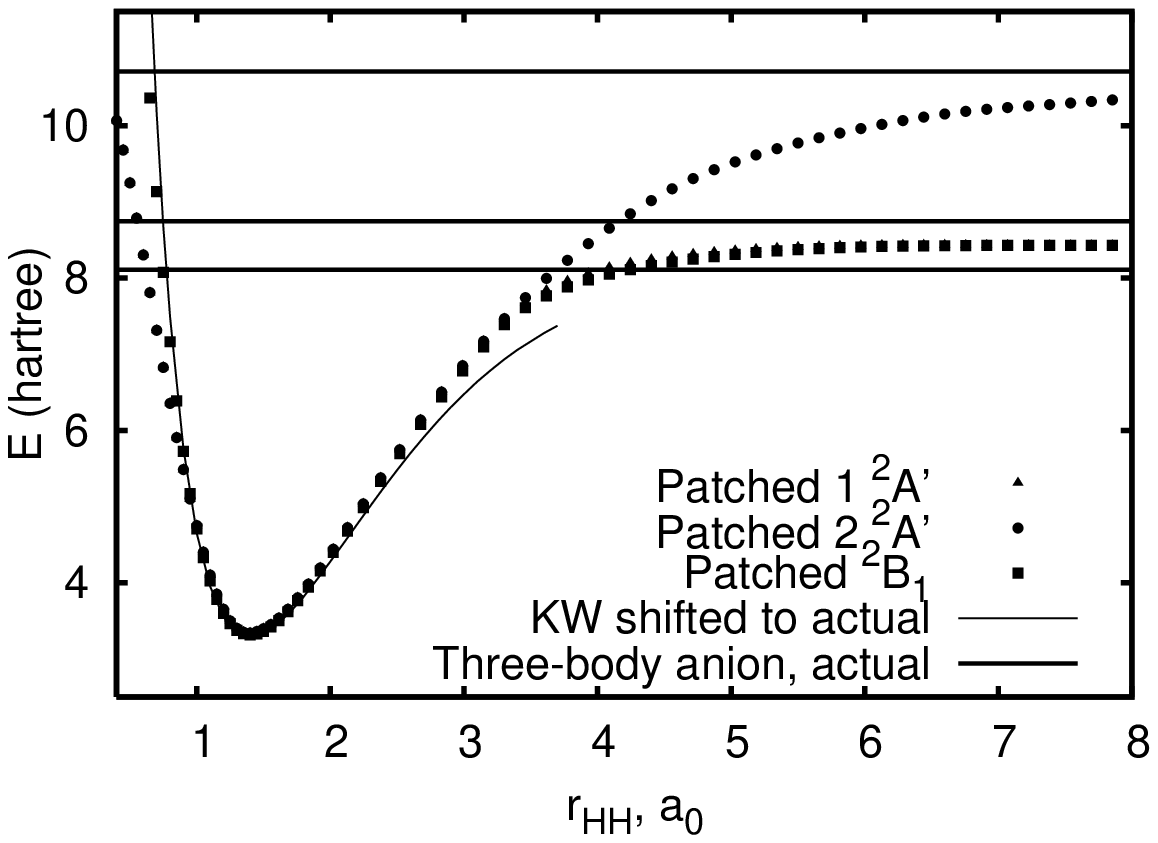} \\
\end{tabular}
\end{center}
\caption[Asymptotes of the calculated curves]{Diatomic potential curve in OH - H asymptote (top) or H$_2$ - O asymptote (bottom) of calculated Feshbach resonance curves at R=10.0$a_0$ (dots); benchmark theoretical calculations on OH ($^2\Pi$ / $^2\Sigma$)~\cite{CYL} and H$_2$~\cite{KW} (thin lines) and proper three-body asymptotes (thick lines), with zero point energy oriented relative to our calculated ground state H$_2$O zero point energy by accepted values. }
\label{asymptotesfig}
\end{figure}

The two-body 
asymptotes of the potential-energy surfaces are plotted in Fig. \ref{asymptotesfig}
and compared with benchmark theoretical calculations on the diatomic fragments.
The dots in this figure represent the values of the global representations of the
potential-energy surfaces evaluated along
the cut in Jacobi coordinates at $R$=10.0$a_0$, $\gamma$=90$^\circ$.  Also plotted in
this figure are the values of benchmark calculations for the diatomic H$_2$~\cite{KW} and
OH (X $^2\Pi$)~\cite{CYL} fragments, which are shifted so that their zero-point energies
lie at the accepted energy above our calculated ground vibrational state energy of neutral H$_2$O;
the solid horizontal lines represent the accepted energy of the three-body channels, again
shifted to correspond to our ground state H$_2$O energy.  The theoretical calculations of
Chu, Yoshimine and Liu~\cite{CYL} slightly underestimate the true dissociation energy of OH (X $^2\Pi$)
and ($^2\Sigma$).
As is clear from the comparison with these benchmark calculations, the energetics of the two-body asymptotes of
these anion surfaces are reproduced extremely well by our calculations.
The ground vibrational state of each agrees with the accepted value to within 0.08eV.
The three-body asymptotes have been adjusted by the patching procedure to correspond
with the appropriate values.

\subsection{Complete views of the complex-valued adiabatic surfaces}

\begin{figure*}
\begin{center}
\begin{tabular}{ccc}
\resizebox{0.65\columnwidth}{!}{\includegraphics{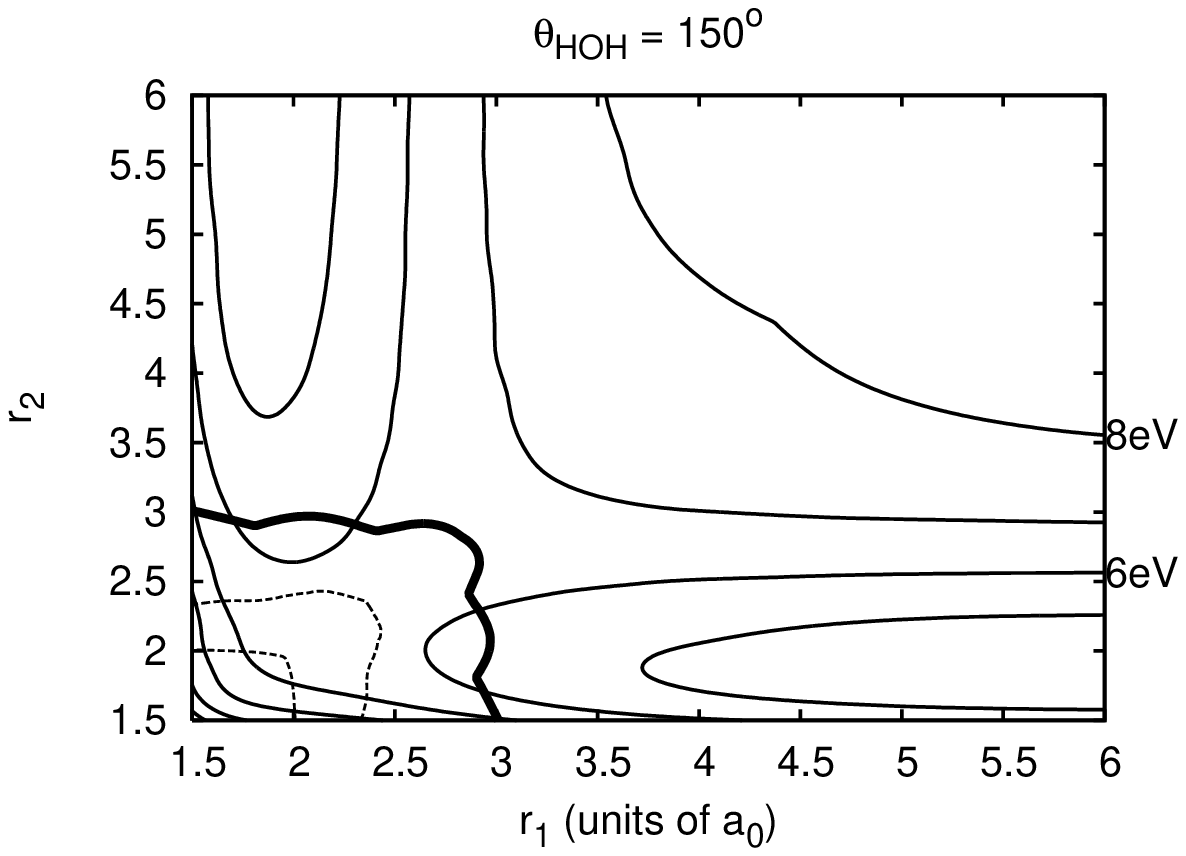}} &
\resizebox{0.65\columnwidth}{!}{\includegraphics{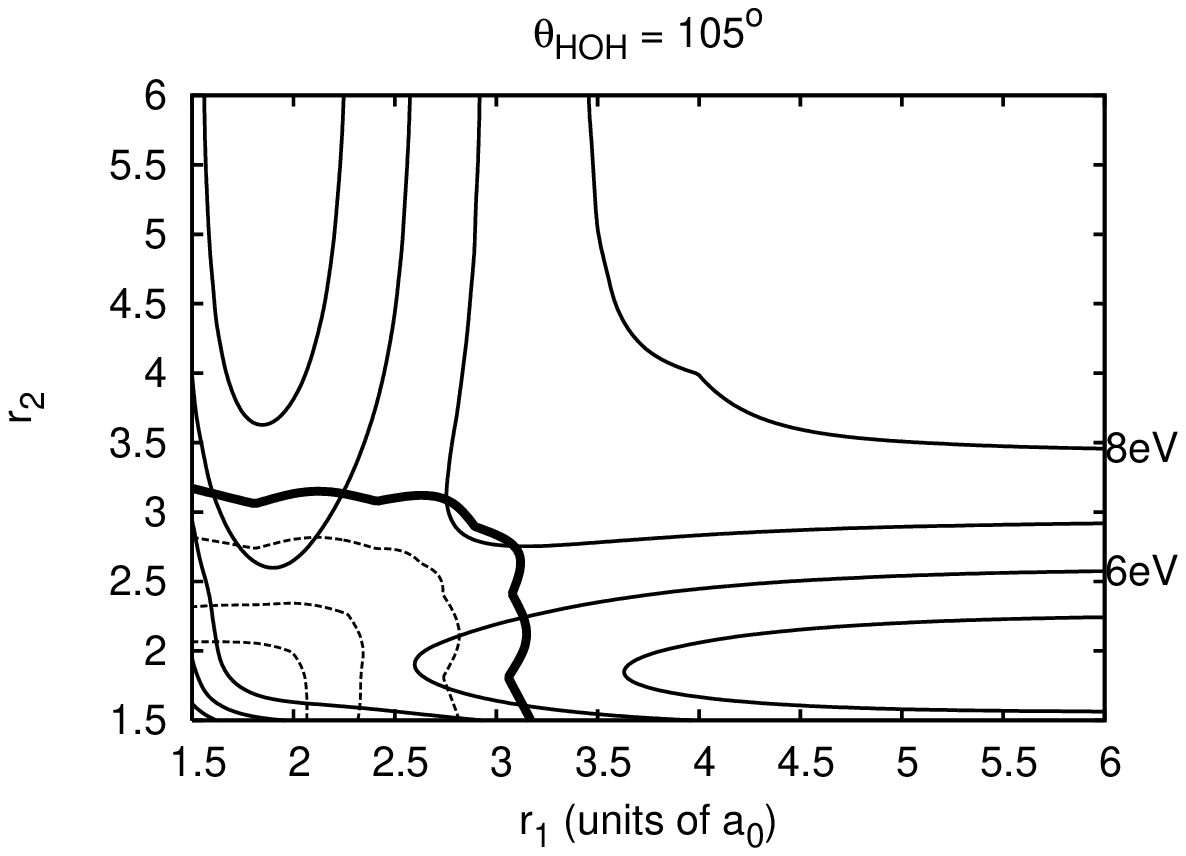}} &
\resizebox{0.65\columnwidth}{!}{\includegraphics{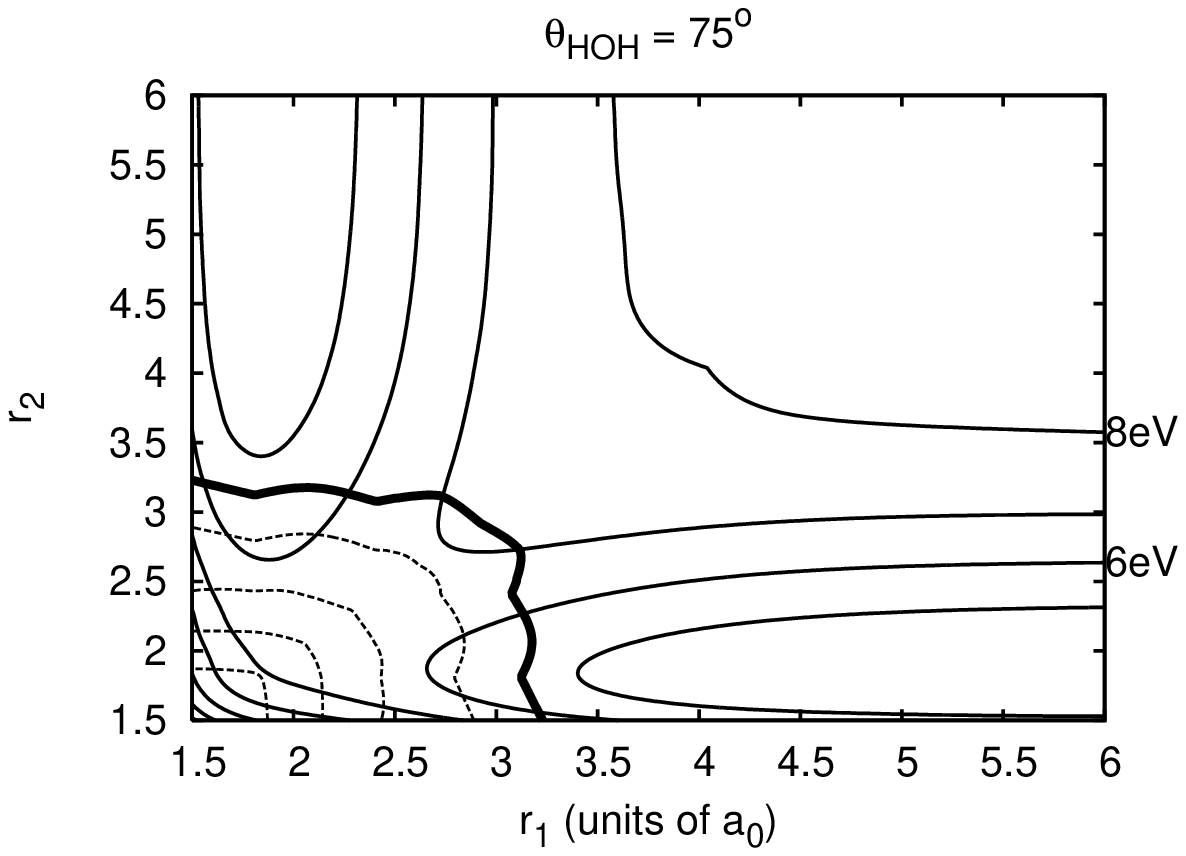}} \\
\resizebox{0.65\columnwidth}{!}{\includegraphics{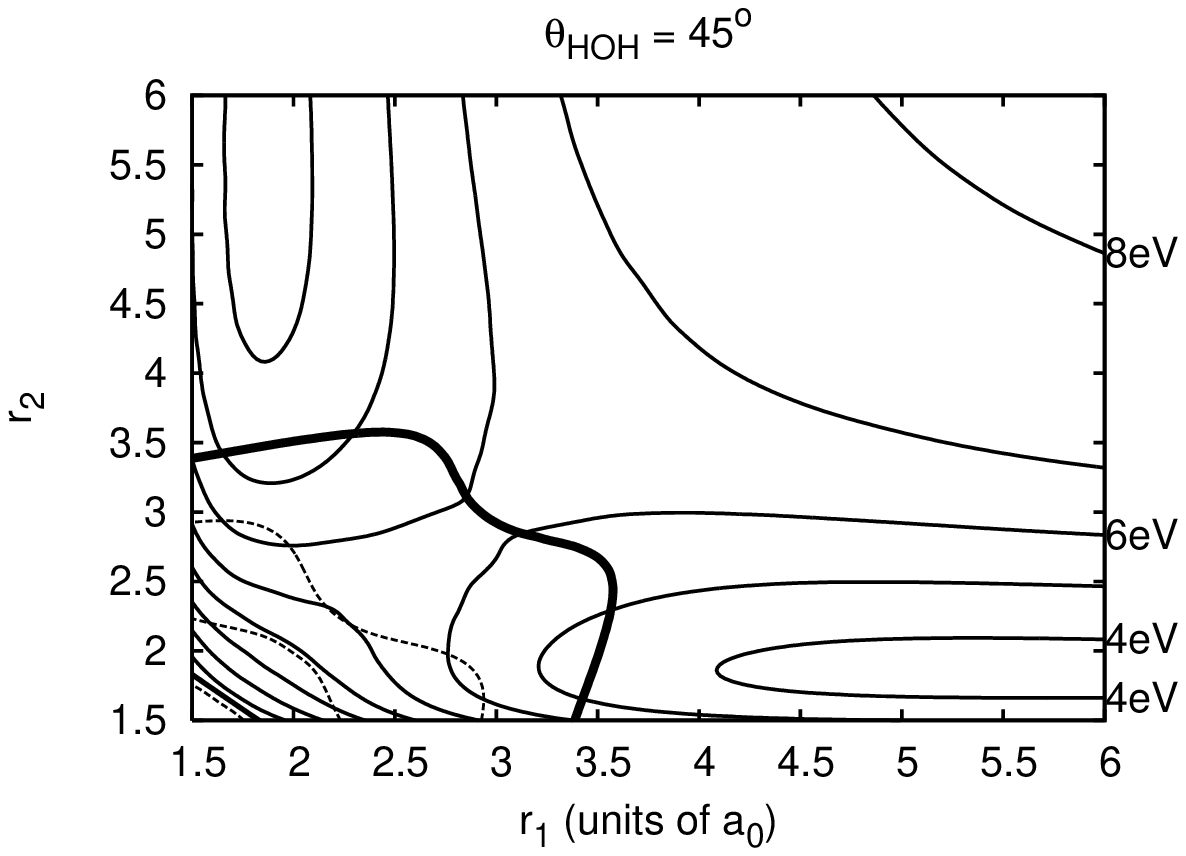}} &
\resizebox{0.65\columnwidth}{!}{\includegraphics{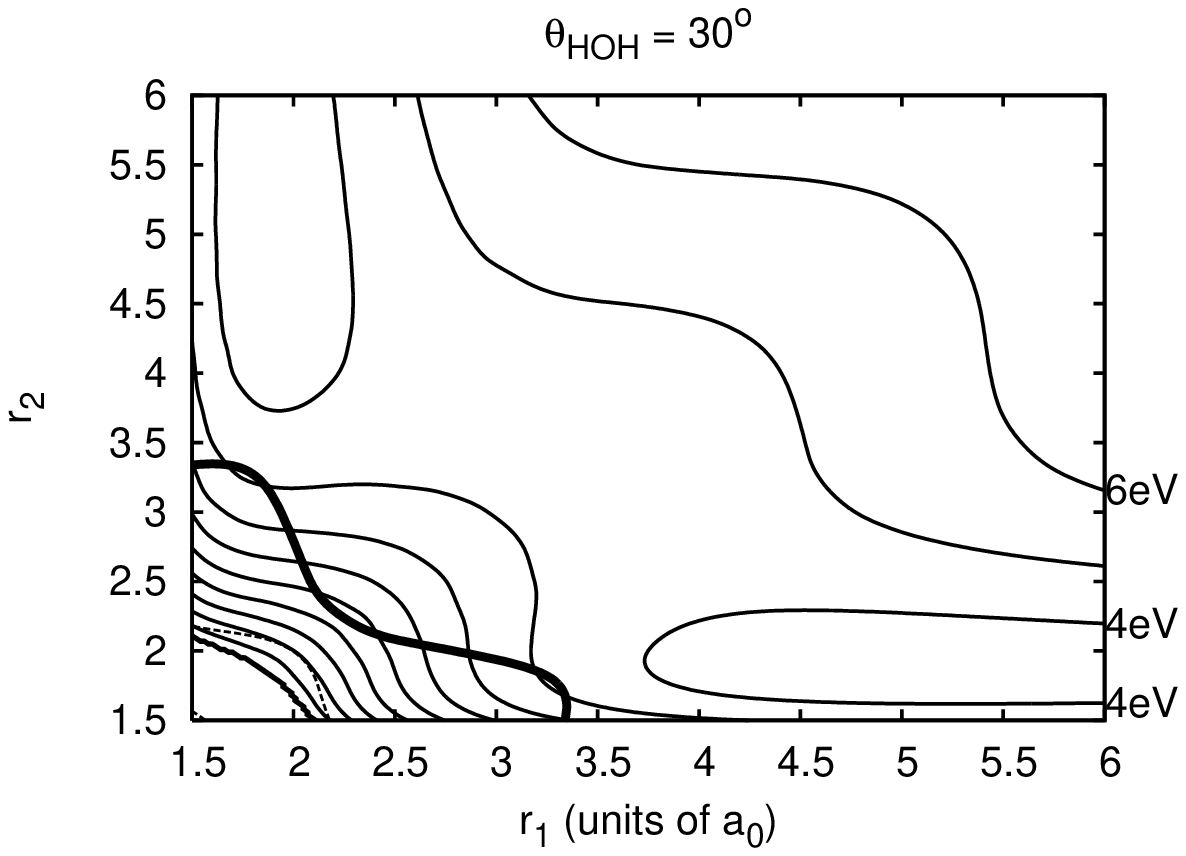}} &
\resizebox{0.65\columnwidth}{!}{\includegraphics{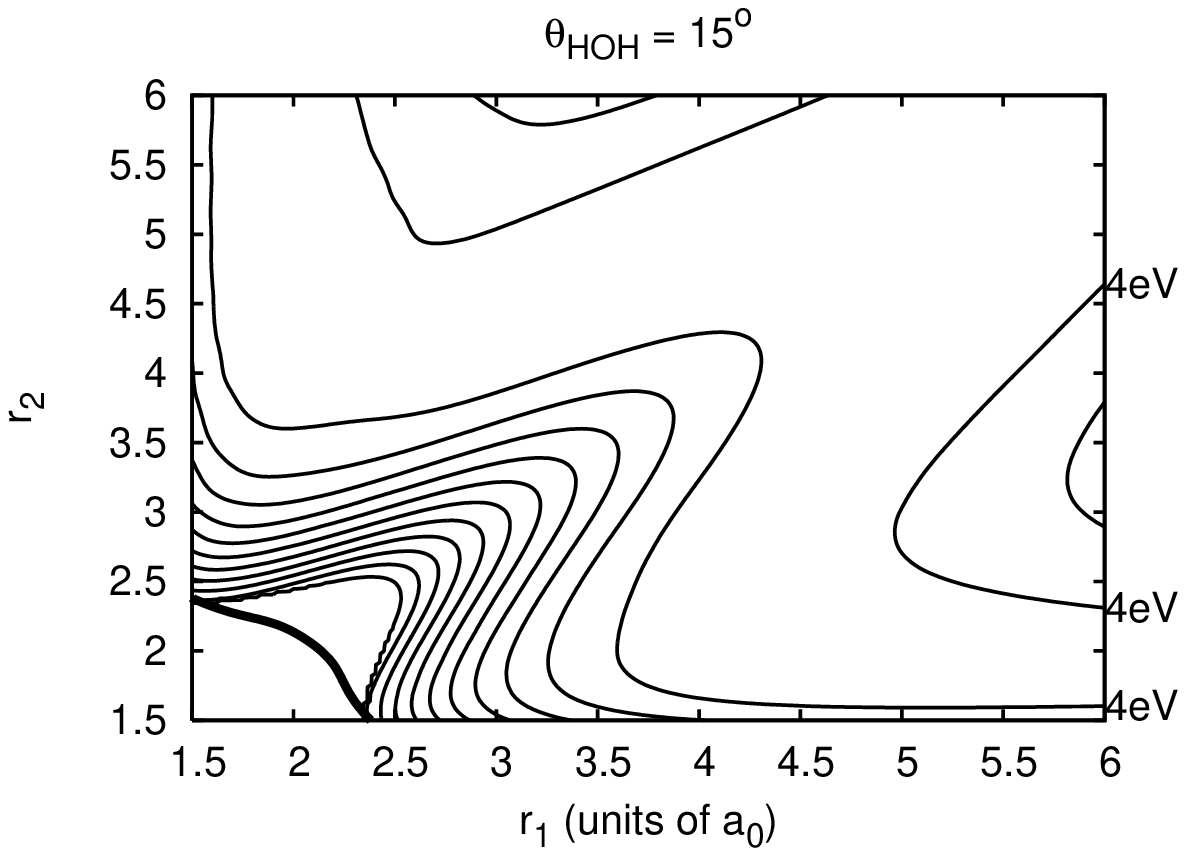}} 
\end{tabular}
\end{center}
\caption{\label{2b1surf}
Complete view of $^2B_1$ surface.  Real part: solid contours, 1.0eV spacing.
Imaginary part ($\Gamma/2$): dashed contours, 1meV spacing.  The bold contour at 1meV
is the lowest contour line for the imaginary part.}
\end{figure*}

\begin{figure*}
\begin{center}
\begin{tabular}{ccc}
\resizebox{0.65\columnwidth}{!}{\includegraphics{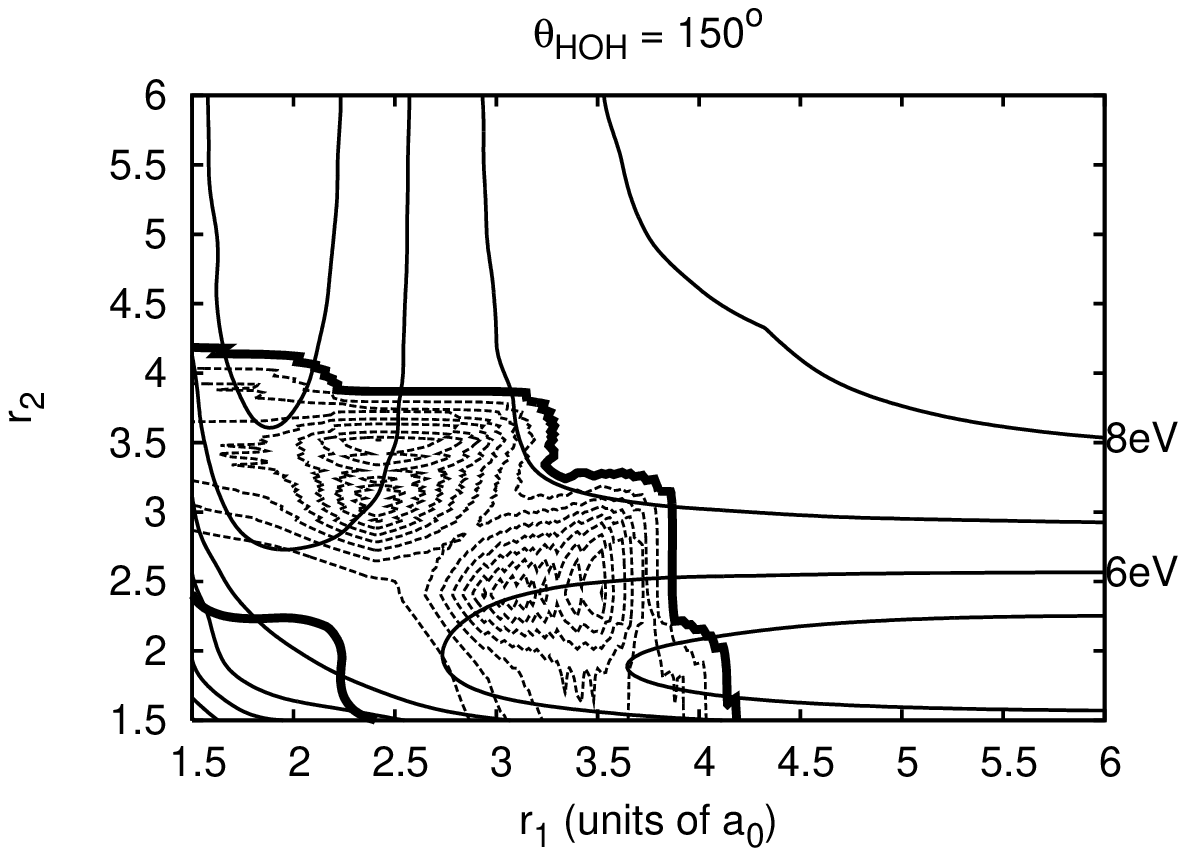}} &
\resizebox{0.65\columnwidth}{!}{\includegraphics{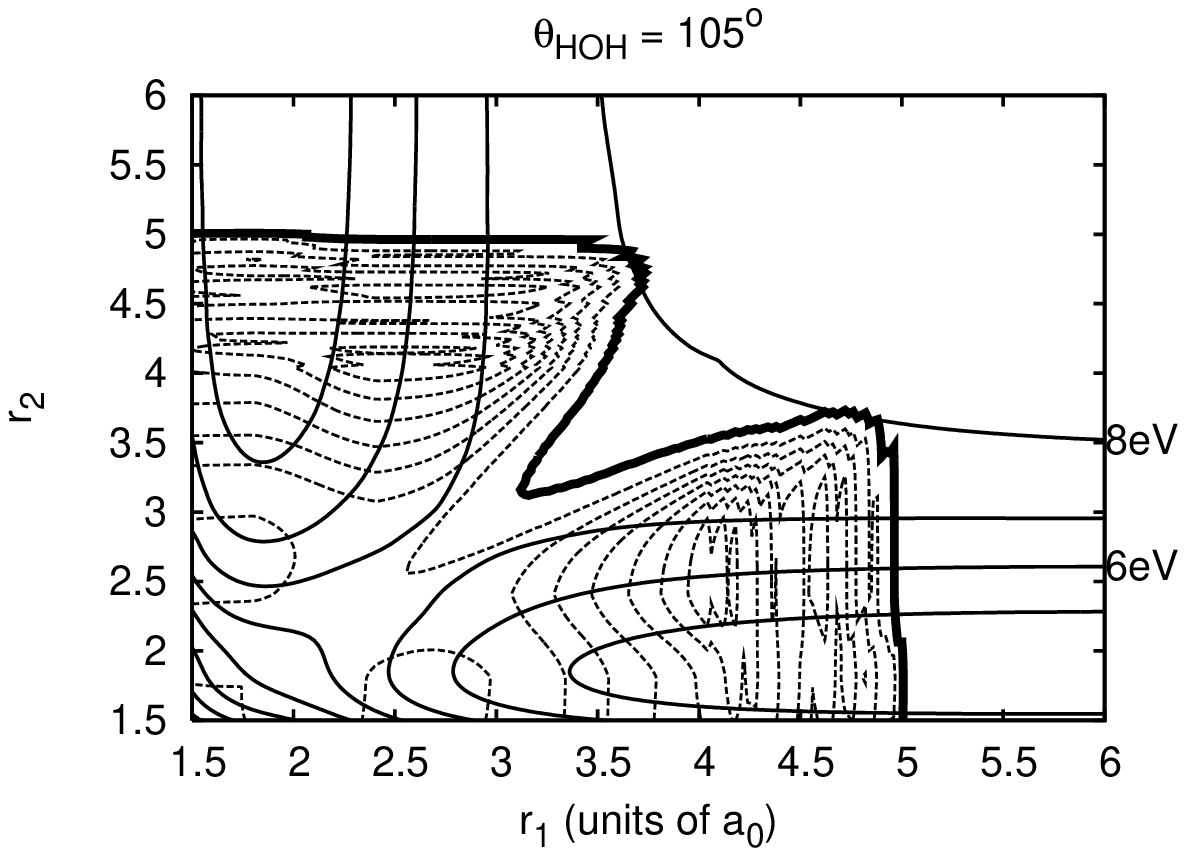}} &
\resizebox{0.65\columnwidth}{!}{\includegraphics{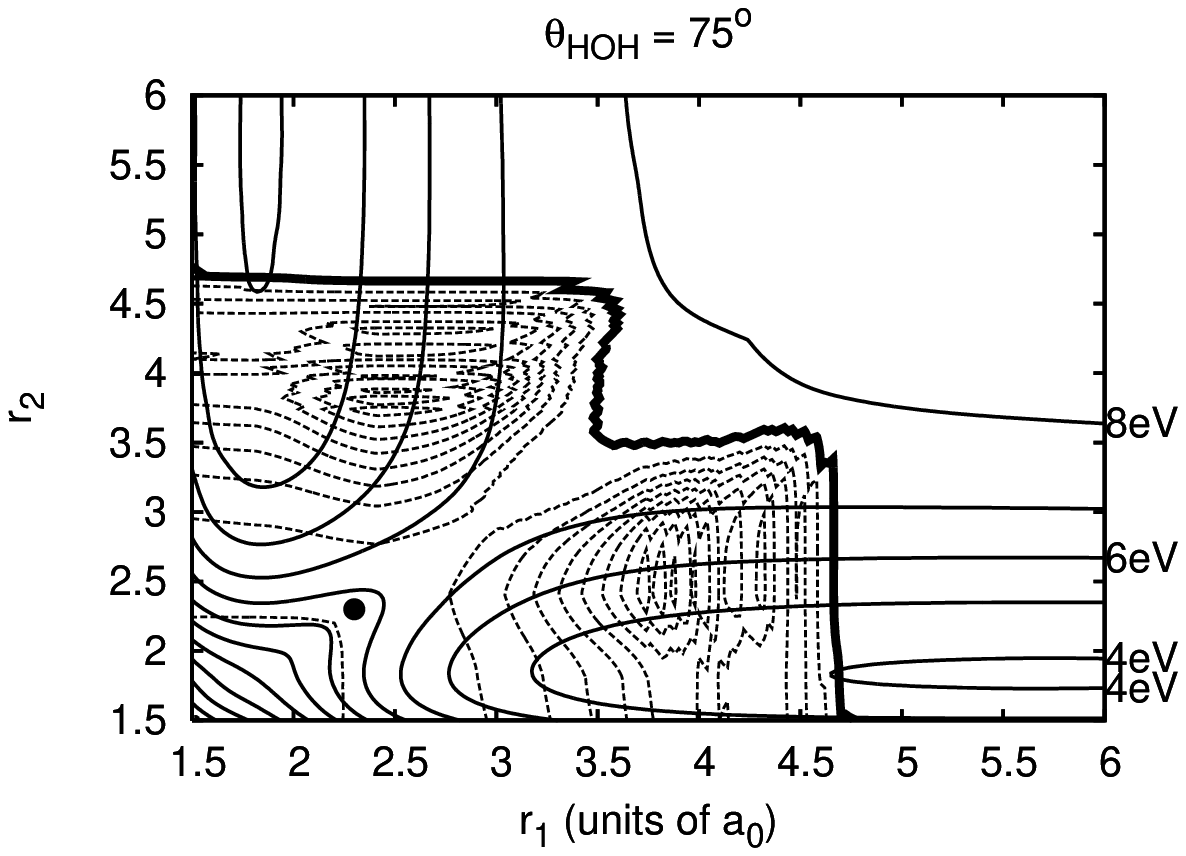}} \\
\resizebox{0.65\columnwidth}{!}{\includegraphics{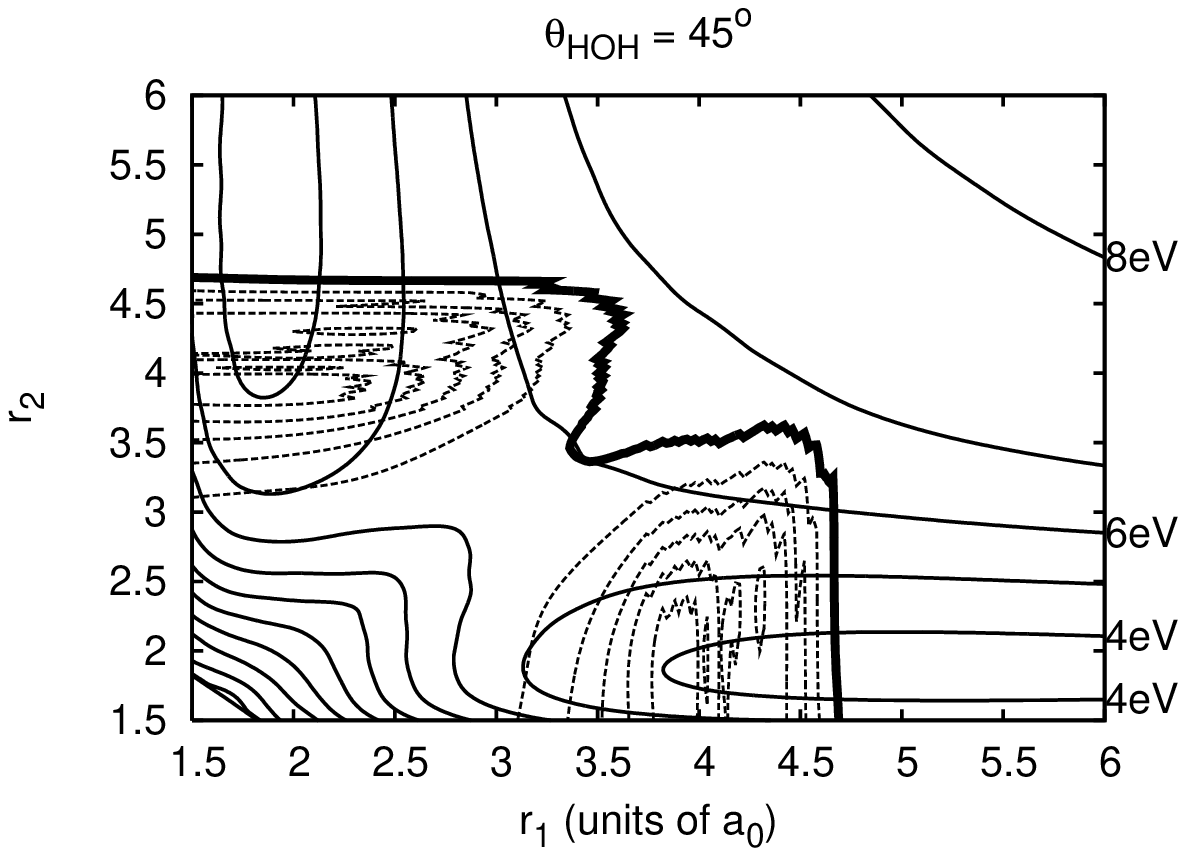}} &
\resizebox{0.65\columnwidth}{!}{\includegraphics{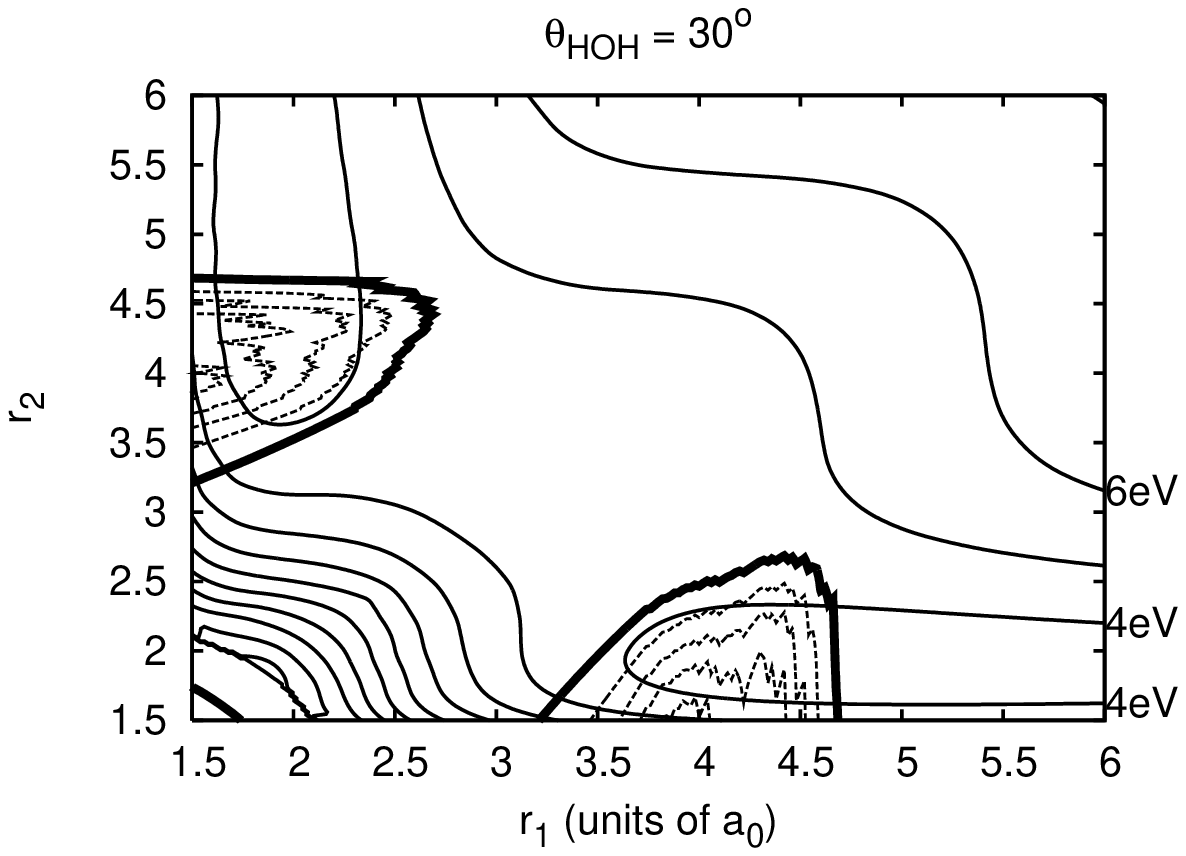}} &
\resizebox{0.65\columnwidth}{!}{\includegraphics{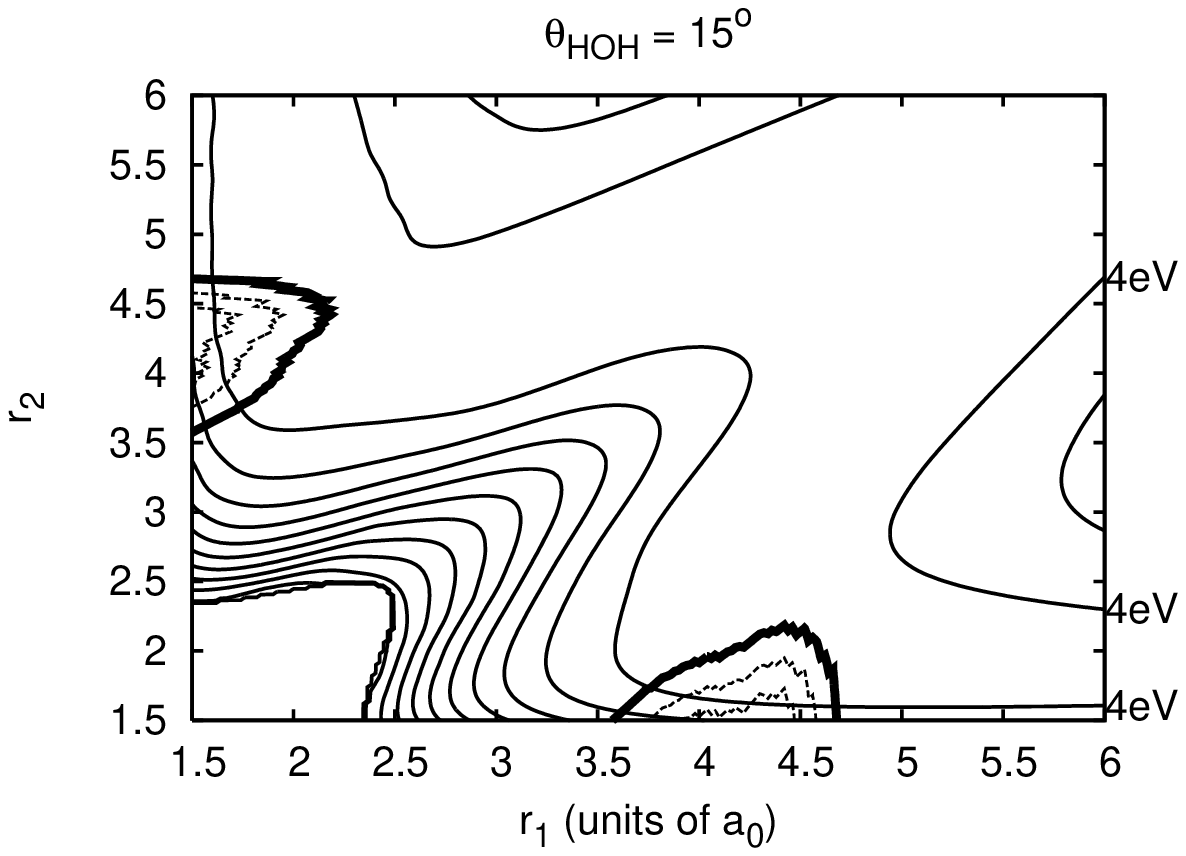}} 
\end{tabular}
\end{center}
\caption{\label{1apsurf}
Complete view of 1~$^2A'$ surface.  Real part: solid contours, 1.0eV spacing.
Imaginary part ($\Gamma/2$): dashed contours, 5meV spacing.  The bold contour at 5meV
is the lowest contour line for the imaginary part.
Dot, 75$^\circ$: intersection of conical intersection seam with this cut.}
\end{figure*}

\begin{figure*}
\begin{center}
\begin{tabular}{ccc}
\resizebox{0.65\columnwidth}{!}{\includegraphics{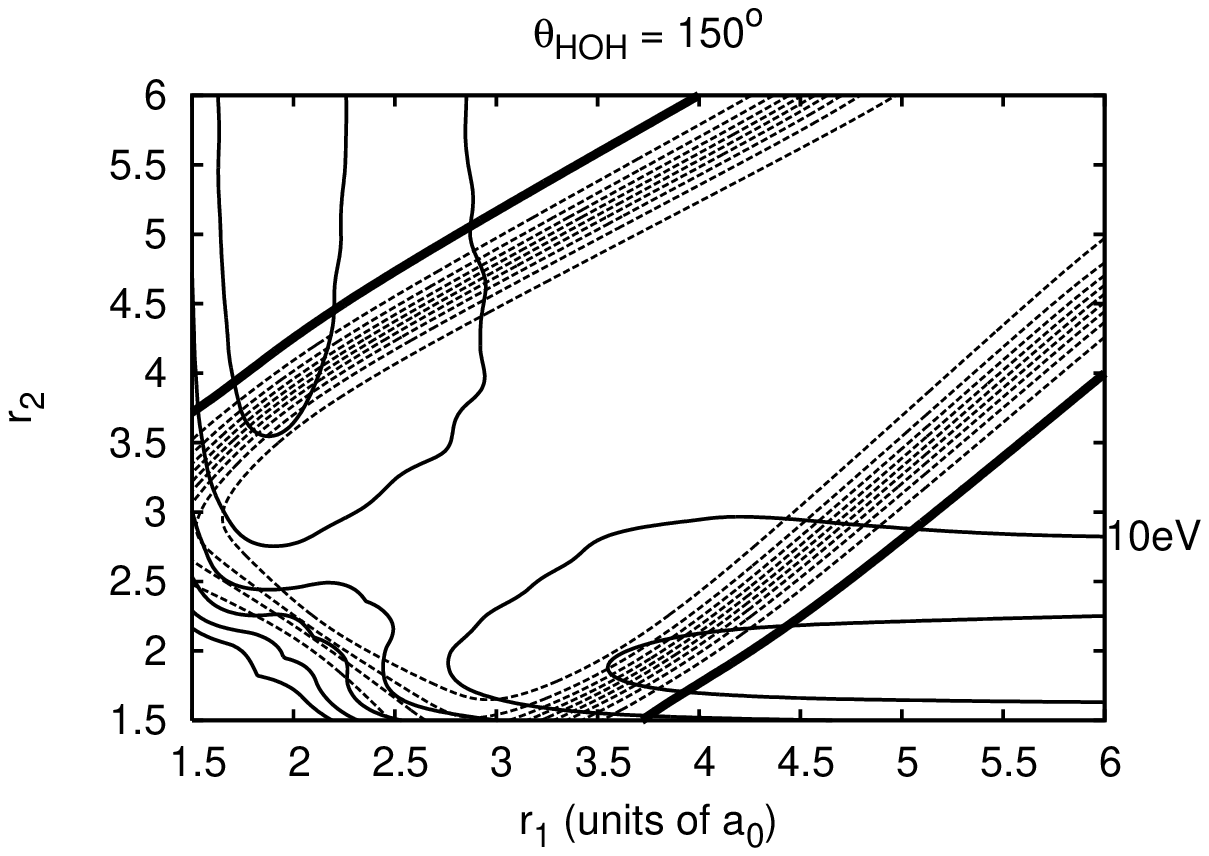}} &
\resizebox{0.65\columnwidth}{!}{\includegraphics{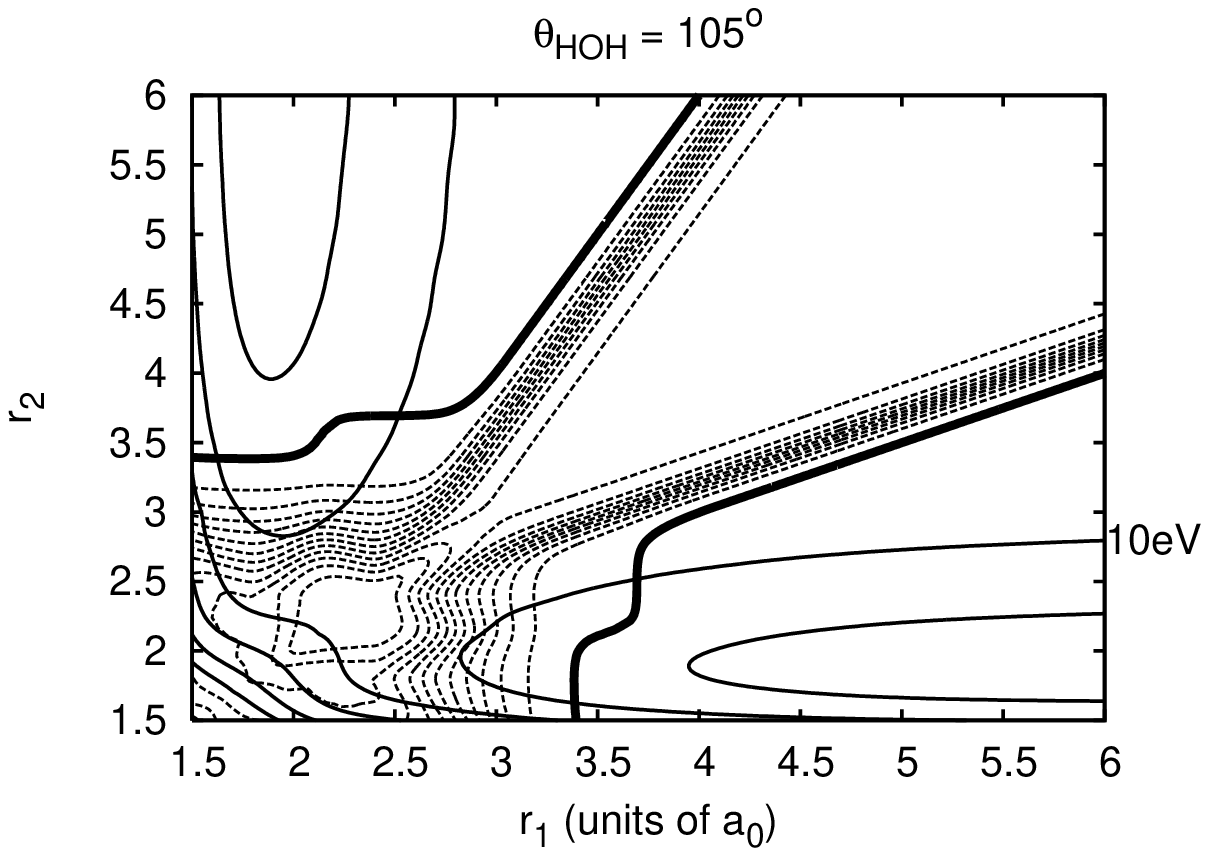}} &
\resizebox{0.65\columnwidth}{!}{\includegraphics{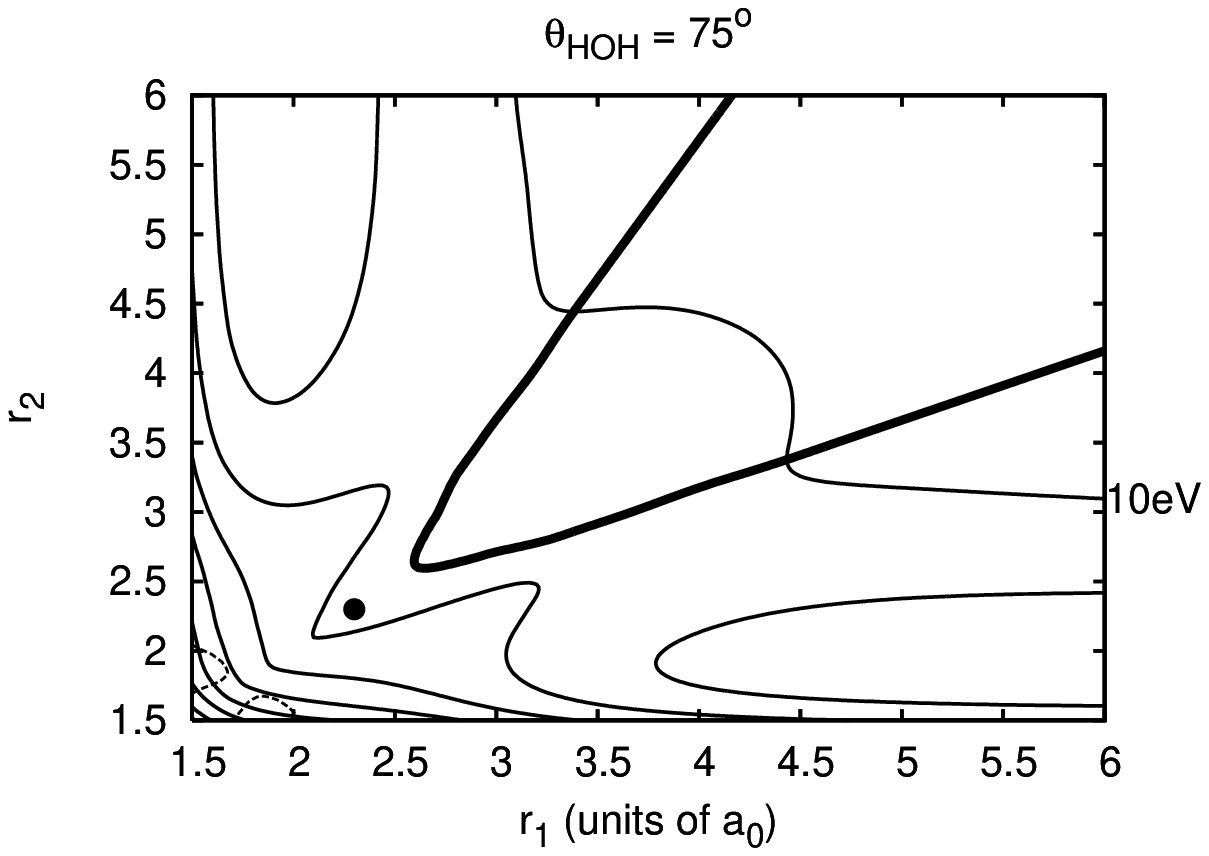}} \\
\resizebox{0.65\columnwidth}{!}{\includegraphics{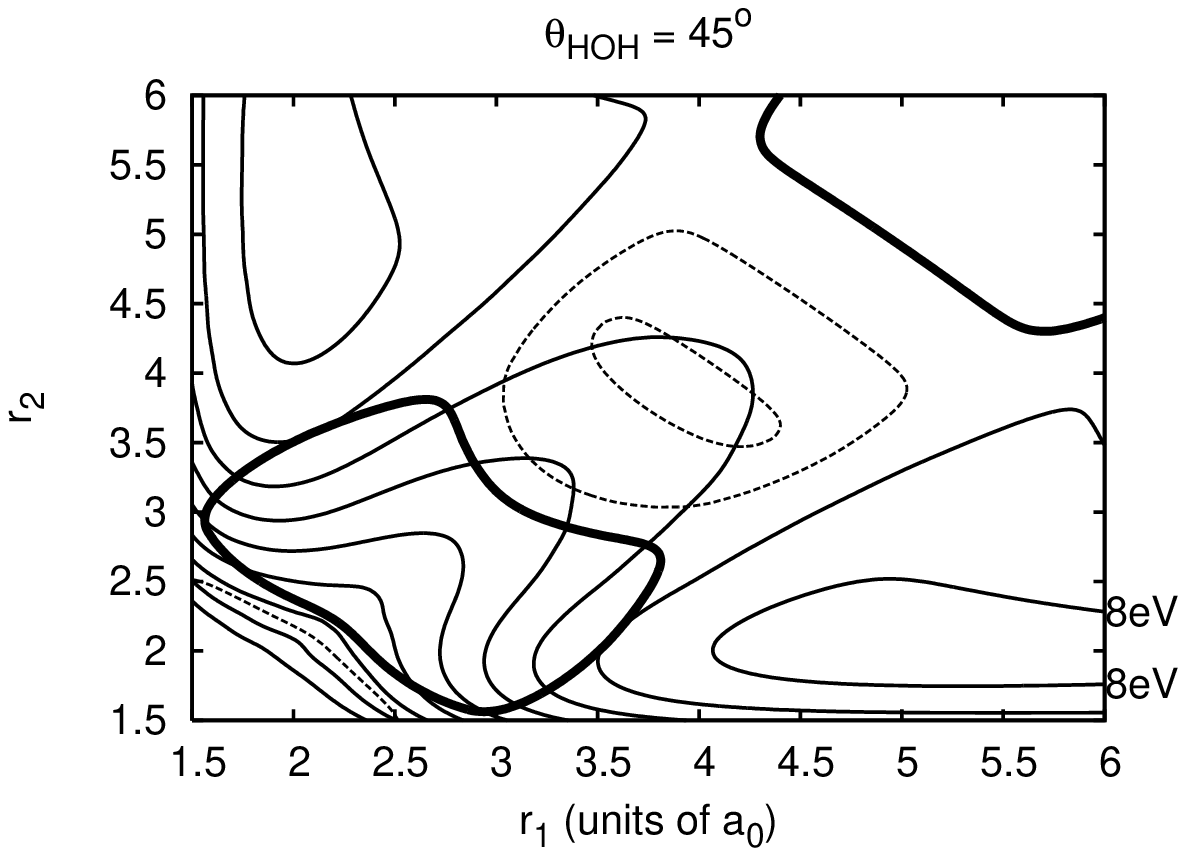}} &
\resizebox{0.65\columnwidth}{!}{\includegraphics{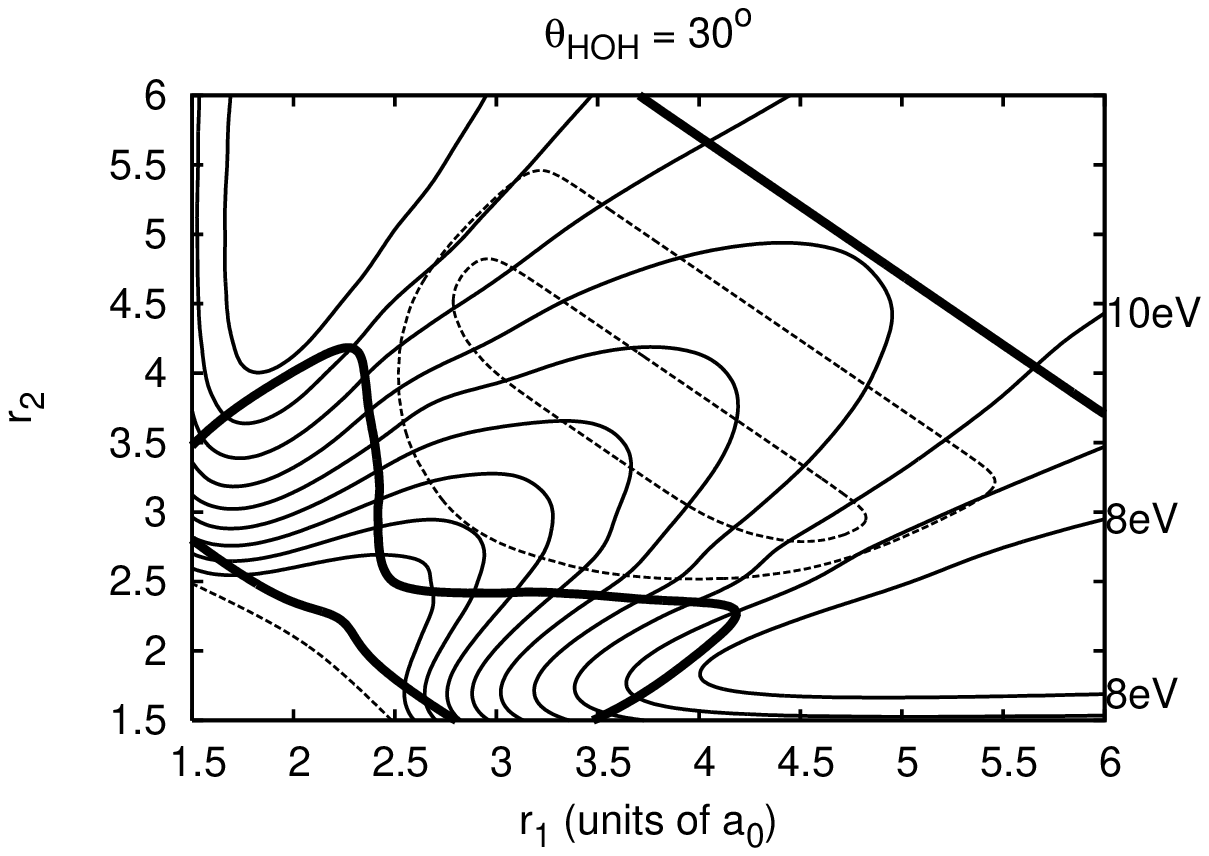}} &
\resizebox{0.65\columnwidth}{!}{\includegraphics{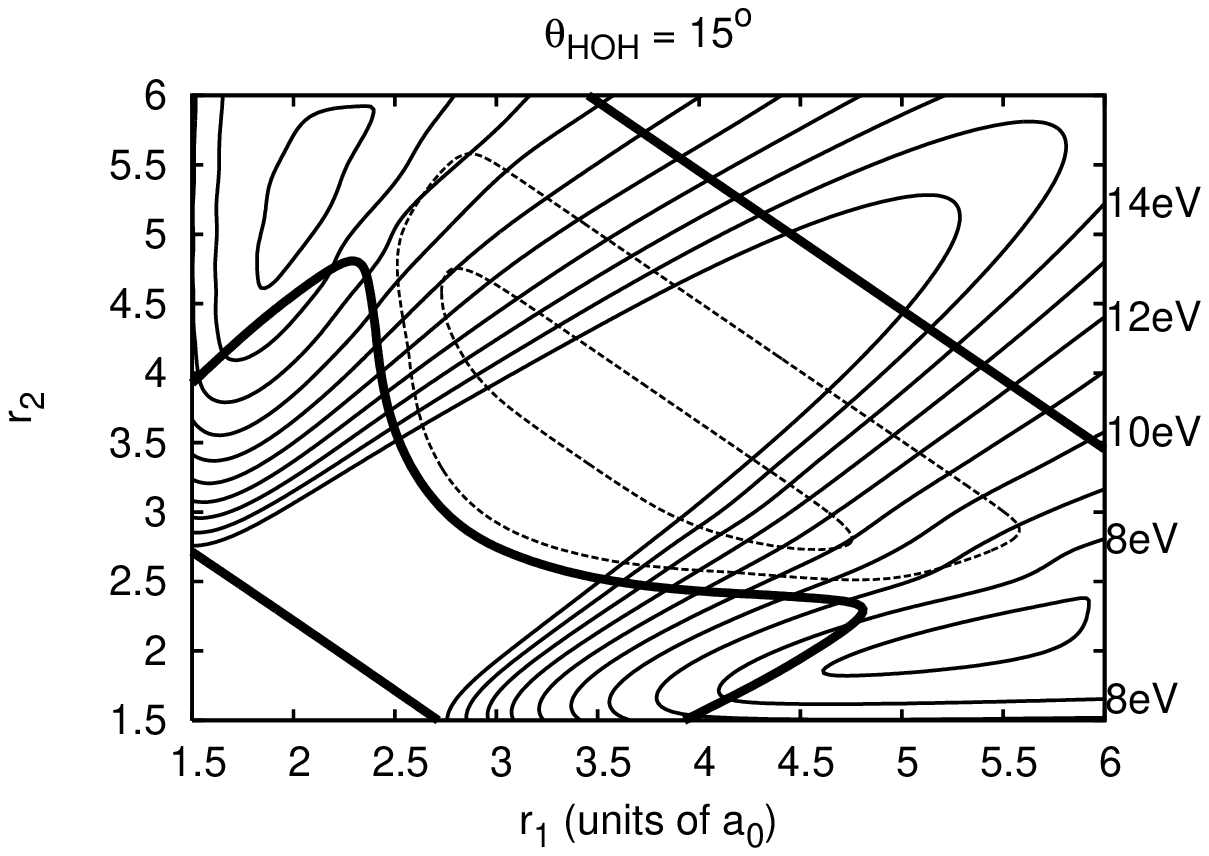}} 
\end{tabular}
\end{center}
\caption{\label{2apsurf}
Complete view of 2~$^2A'$ surface.  Real part: solid contours, 1.0eV spacing.
Imaginary part ($\Gamma/2$): dashed contours, 10meV spacing.  The bold contour at 10meV
is the lowest contour line for the imaginary part.
Dot, 75$^\circ$: intersection of conical intersection seam with this cut.}
\end{figure*}

Complete views of the global fits of the real and imaginary components of the 
adiabatic potential-energy surfaces
are shown in Figs. \ref{2b1surf}-\ref{2apsurf}.
The real and imaginary components are plotted together with contour lines,
as a function of bond length, for various bond angles.  For the imaginary
component, the contour line nearest zero is bold, and subsequent contour
lines depict the magnitude of the imaginary component increasing.
The contour lines for the real part, some of which are marked on the perimeter
of each panel, correspond to the energy above the
ground state H$_2$O energy as calculated with our neutral H$_2$O potential-energy surface, with the zero point energy included, and are therefore consistent
with Fig. \ref{energetics}.

In the view of the $^2B_1$ surface in Fig. \ref{2b1surf}, one can see
the two OH+H$^-$ channels at bottom right and upper left of each panel.
In the bottom row of panels, which show the surface as the bond angle
$\theta_{HOH}$ is decreased, one can see the potential well which
corresponds to the H$_2$+O$^-$ channel develop along the symmetric
stretch diagonal.  This channel is the lowest energy asymptote; it
reaches below 4eV, as can be seen
in the final panel at 15$^\circ$.
The bottoms of the H$^-$+OH wells in
the upper panels are above 4eV.  Although the OH+H$^-$ 
well does reach below 4eV in the panels at $\theta_{HOH}$=30$^\circ$
and 15$^\circ$, it does so only at small OH-H separations, at which 
geometry there is a local minimum due to the dipole-anion interaction.

The $^2B_1$ resonance energy is relatively flat with bending angle
near the equilibrium geometry of the neutral, and this fact is
apparent in the similar shape and value of the contour lines
at small $r_1$ and $r_2$ in Fig. \ref{2b1surf} from $\theta$=150$^\circ$
to $\theta$=75$^\circ$.  At $\theta$=45$^\circ$ and beyond, the 
bending potential becomes repulsive and the contour lines move towards
larger $r_1$ and $r_2$.  The flatness of the bending curve will cause
the dissociating wavepacket to make minimal excursions beyond the
cut at $\theta$=105$^\circ$, and in particular, only a very small
fragment of the dissociating wavepacket will reach the O$^-$+H$_2$
potential well at small $\theta$. The imaginary component of the $^2B_1$ 
surface is small and localized near the Franck-Condon
region.

The entire 1~$^2A'$ surface is shown in Fig. \ref{1apsurf}.  
The real part of this surface is similar to that of the $^2B_1$ surface,
having the same asymptotes, but is everywhere higher (except at linear
$\theta_{HOH}$=180$^\circ$ where they are degenerate), and
is highest above the $^2B_1$ surface along the conical intersection seam,
that intersects this figure at $\theta_{HOH}$=75$^\circ$ at 
($r_1$=$r_2$=2.3$a_0$), which point is marked with a dot in the figure.  
The conical intersection seam is roughly parallel to the
grain of the cut, and falls within the plane $r_1$=$r_2$.  
Thus, the dot marks the point where the 1~$^2A'$ surface meets the 2~$^2A'$
surface along this cut.  Everywhere else, it is below.

Near the Franck-Condon region of the neutral, 
the behavior of the 1~$^2A'$ surface with respect to bending angle is different
from the relatively flat $^2B_1$ surface.  In the cut at $\theta$=150$^\circ$
these two surfaces are nearly degenerate, being exactly degenerate at linear geometry.
As the bending angle is decreased, however, the 1~$^2A'$ surface rises in energy,
and the contour lines in Fig. \ref{1apsurf}
buckle outward; one such contour line reaches a point at the conical intersection
at $\theta$=75$^\circ$.  From this angle, the contour lines near the Franck-Condon
region relax slightly downward going to $\theta$=45$^\circ$, which behavior traces
the small well on the 1~$^2A'$ surface apparent in Fig. \ref{Bendfig} around 
$\theta$=55$^\circ$.  At smaller angles, the H$_2$+O$^-$ potential well
develops.

The imaginary component
of this state is much different from that of the $^2B_1$ state, being in general
higher within the Franck-Condon region, and having the large (0.1eV) peaks
in the exit wells just as the resonance becomes bound as H$^-$+OH.  These
peaks are a consequence of the A' symmetry of the resonance, as discussed in 
Ref.~\cite{haxton3}.  They will lead to a large rate of autodetachment
in this channel, and an isotope effect in the cross sections for H$_2$O versus D$_2$O
that is disproportionate to the entrance
amplitude, as discussed in Ref. \cite{haxton2}.  They also may portend the
breakdown of the LCP model near these geometries, for they indicate that the
1~$^2A'$ state may exist as a virtual state there.  We examine
this issue more fully in paper II.

The 2~$^2A'$ surface is shown in Fig. \ref{2apsurf} and is different from the
other surfaces in every region.  Its lowest points are along the well of the 
the H$^-$+OH($^2\Sigma$)
channel where the surface drops to 8eV.  This channel has a broader and shallower potential-energy well than the H$^-$+OH($^2\Pi$) asymptote of the other resonances, with
a minimum at approximately $r_{OH}$=1.95$a_0$.  This surface 
exhibits a broad plateau in the $(r_1,r_2)$ plane at $\theta_{HOH}$=75$^\circ$
and 10eV.  At this angle 
there is a dimple near the conical intersection at ($r_1$=$r_2$=2.3$a_0$), which is again
marked with a dot.
There is no H$_2$+O$^-$ well evident along the symmetric stretch direction at $\theta_{HOH}$=15$^\circ$
because in this arrangement the 2~$^2A'$ state is unbound as H$_2$ (triplet 1$\sigma_g$1$\sigma_u$)+O$^-$;
H$_2$ (triplet 1$\sigma_g$1$\sigma_u$) has a dissociative potential-energy curve.  Thus, the panel
at $\theta_{HOH}$=15$^\circ$ shows a cut along the top of this repulsive wall, going
up past the last contour line at 16eV; this repulsive wall extends infinitely in the
symmetric stretch direction at ever smaller $\theta_{HOH}$.  Near the Franck-Condon region, one can see that the 
real part of the surface is high ($\sim$15eV) at $\theta_{HOH}$=150$^\circ$, and slopes downward going
from panel to panel to $\theta_{HOH}$=75$^\circ$, as the contour lines near $r_1$=$r_2$=1.81$a_0$
go from being convex to concave.  Moving on to the panels at 45$^\circ$, 30$^\circ$ and 15$^\circ$,
the repulsive wall of H$_2$(1$\sigma_g$1$\sigma_u$)+O$^-$ then quickly develops.

This surface, constructed from the CI calculation, is single-valued, though the physical surface
is not.  The $^2B_2$ shape resonance curve which intersects that of the $^2B_2$ Feshbach resonance 
in branch-point fashion\cite{haxton3} has been
omitted from the present treatment.  The physical 2~$^2A'$ surface contains not only the 
H$_2$(1$\sigma_g$1$\sigma_u$)+O$^-$ asymptote (overall $^2A_1$ symmetry) at small H-H bond distances, but also the
metastable  H$_2^-$(1$\sigma_g^2$ 1$\sigma_u$)+O($^1D$) asymptote (overall $^2B_2$ symmetry) as well; the two
three-body asymptotes of this state are thus H+H$^-$+O($^1D$) and H+H+O$^-$ (degenerate with 1~$^2A'$).
We patch the three-body breakup region of the 
$^2B_2$ surface to its upper physical asymptote at 10.71eV.

The imaginary component of the 2~$^2A'$ state takes a large value ($\sim$0.12eV) along the symmetric stretch direction for
$\theta_{HOH}$=105$^\circ$ and 150$^\circ$ but otherwise is smaller; its representation 
is dominated by the interpolation between the large-valued regions and the small-valued regions.  In particular,
in the H$^-$+OH ($^2\Sigma$) exit well the imaginary component drops to approximately 0.02eV by 
($r_1$=1.8$a_0$, $r_2$=3.5$a_0$).  
The interpolation is physical when $r_1$ or $r_2$ is small.
However, in the three-body channel, we interpolate between physically distinct sheets of the 
$^2A'$ manifold.  Thus, the imaginary component drops from 100meV near the diagonal for 
$r_1 \approx r_2 > 3.0a_0$ going from $\theta_{HOH}$=105$^\circ$ to $\theta_{HOH}$=75$^\circ$,
but this behavior is unphysical, and corresponds to the interpolation between the two sheets of $^2B_2$
symmetry.  The imaginary component remains small along the diagonal going from $\theta_{HOH}$=75$^\circ$ to 45$^\circ$ as the conical intersection is passed and the symmetry of the 2~$^2A'$ state changes from $^2B_2$ to $^2A_1$.

\section{Conclusion}

We have calculated the potential-energy surfaces necessary for a description
of dissociative eletron attachment to $\rm{H_2O}$.  In paper II, these surfaces are used in a study of the nuclear dynamics in the local complex potential model, and the cross sections for dissociative eletron attachment are calculated.

\begin{acknowledgments}

This work was performed under the auspices of the US Department of Energy
by the University of California Lawrence Berkeley National Laboratory
under Contract DE-AC02-05CH11231 and
was supported by the U.S. DOE Office of Basic Energy
Sciences, Division of Chemical Sciences. 

\end{acknowledgments}


\begin{thebibliography}{61}
\expandafter\ifx\csname natexlab\endcsname\relax\def\natexlab#1{#1}\fi
\expandafter\ifx\csname bibnamefont\endcsname\relax
  \def\bibnamefont#1{#1}\fi
\expandafter\ifx\csname bibfnamefont\endcsname\relax
  \def\bibfnamefont#1{#1}\fi
\expandafter\ifx\csname citenamefont\endcsname\relax
  \def\citenamefont#1{#1}\fi
\expandafter\ifx\csname url\endcsname\relax
  \def\url#1{\texttt{#1}}\fi
\expandafter\ifx\csname urlprefix\endcsname\relax\def\urlprefix{URL }\fi
\providecommand{\bibinfo}[2]{#2}
\providecommand{\eprint}[2][]{\url{#2}}

\bibitem[{\citenamefont{Birtwistle and
  Herzenberg}(1971)}]{BirtwistleHerzenberg}
\bibinfo{author}{\bibfnamefont{D.~T.} \bibnamefont{Birtwistle}}
  \bibnamefont{and}
  \bibinfo{author}{\bibfnamefont{A.}~\bibnamefont{Herzenberg}},
  \bibinfo{journal}{J. Phys. B} \textbf{\bibinfo{volume}{4}},
  \bibinfo{pages}{53} (\bibinfo{year}{1971}).

\bibitem[{\citenamefont{Dube and Herzenberg}(1979)}]{DubeHerzenberg}
\bibinfo{author}{\bibfnamefont{L.}~\bibnamefont{Dube}} \bibnamefont{and}
  \bibinfo{author}{\bibfnamefont{A.}~\bibnamefont{Herzenberg}},
  \bibinfo{journal}{Phys. Rev. A} \textbf{\bibinfo{volume}{20}},
  \bibinfo{pages}{194} (\bibinfo{year}{1979}).

\bibitem[{\citenamefont{Bardsley and Wadehra}(1983)}]{BardsleyWadehra}
\bibinfo{author}{\bibfnamefont{J.~N.} \bibnamefont{Bardsley}} \bibnamefont{and}
  \bibinfo{author}{\bibfnamefont{J.~M.} \bibnamefont{Wadehra}},
  \bibinfo{journal}{J. Chem. Phys.} \textbf{\bibinfo{volume}{78}},
  \bibinfo{pages}{7227} (\bibinfo{year}{1983}).

\bibitem[{\citenamefont{O'Malley and Taylor}(1968)}]{OmalleyTaylor}
\bibinfo{author}{\bibfnamefont{T.~F.} \bibnamefont{O'Malley}} \bibnamefont{and}
  \bibinfo{author}{\bibfnamefont{H.~S.} \bibnamefont{Taylor}},
  \bibinfo{journal}{Phys. Rev.} \textbf{\bibinfo{volume}{176}},
  \bibinfo{pages}{207} (\bibinfo{year}{1968}).

\bibitem[{\citenamefont{O'Malley}(1966)}]{Omalley}
\bibinfo{author}{\bibfnamefont{T.~F.} \bibnamefont{O'Malley}},
  \bibinfo{journal}{Phys. Rev.} \textbf{\bibinfo{volume}{150}},
  \bibinfo{pages}{14} (\bibinfo{year}{1966}).

\bibitem[{\citenamefont{Haxton et~al.}(2007)\citenamefont{Haxton, Rescigno, and
  McCurdy}}]{paper2}
\bibinfo{author}{\bibfnamefont{D.~J.} \bibnamefont{Haxton}},
  \bibinfo{author}{\bibfnamefont{T.~N.} \bibnamefont{Rescigno}},
  \bibnamefont{and} \bibinfo{author}{\bibfnamefont{C.~W.}
  \bibnamefont{McCurdy}}, \bibinfo{journal}{Phys. Rev. A}
  \textbf{\bibinfo{volume}{75}}, \bibinfo{pages}{012711}
  (\bibinfo{year}{2007}).

\bibitem[{\citenamefont{Lozier}(1930)}]{Lozier}
\bibinfo{author}{\bibfnamefont{W.~N.} \bibnamefont{Lozier}},
  \bibinfo{journal}{Phys. Rev.} \textbf{\bibinfo{volume}{36}},
  \bibinfo{pages}{1417} (\bibinfo{year}{1930}).

\bibitem[{\citenamefont{Fedor and \textit{et al.}}(2006)}]{fedor}
\bibinfo{author}{\bibfnamefont{J.}~\bibnamefont{Fedor}} \bibnamefont{and}
  \bibinfo{author}{\bibnamefont{\textit{et al.}}}, \bibinfo{journal}{J. Phys B}
  \textbf{\bibinfo{volume}{39}}, \bibinfo{pages}{3935} (\bibinfo{year}{2006}).

\bibitem[{\citenamefont{Buchel'nikova}(1959)}]{Buchel}
\bibinfo{author}{\bibfnamefont{I.~S.} \bibnamefont{Buchel'nikova}},
  \bibinfo{journal}{Zh. Eksp. Teor. Fiz.} \textbf{\bibinfo{volume}{35}},
  \bibinfo{pages}{1119} (\bibinfo{year}{1959}).

\bibitem[{\citenamefont{Schultz}(1966)}]{Schultz}
\bibinfo{author}{\bibfnamefont{G.~J.} \bibnamefont{Schultz}},
  \bibinfo{journal}{J. Chem. Phys.} \textbf{\bibinfo{volume}{44}},
  \bibinfo{pages}{3856} (\bibinfo{year}{1966}).

\bibitem[{\citenamefont{Compton and Christophorou}(1967)}]{compton}
\bibinfo{author}{\bibfnamefont{R.~N.} \bibnamefont{Compton}} \bibnamefont{and}
  \bibinfo{author}{\bibfnamefont{L.~G.} \bibnamefont{Christophorou}},
  \bibinfo{journal}{Phys. Rev.} \textbf{\bibinfo{volume}{154}},
  \bibinfo{pages}{110} (\bibinfo{year}{1967}).

\bibitem[{\citenamefont{Melton}(1972)}]{Melton}
\bibinfo{author}{\bibfnamefont{C.~E.} \bibnamefont{Melton}},
  \bibinfo{journal}{J. Chem. Phys.} \textbf{\bibinfo{volume}{57}},
  \bibinfo{pages}{4218} (\bibinfo{year}{1972}).

\bibitem[{\citenamefont{Sanche and Schultz}(1972)}]{sancheschultz}
\bibinfo{author}{\bibfnamefont{L.}~\bibnamefont{Sanche}} \bibnamefont{and}
  \bibinfo{author}{\bibfnamefont{G.~J.} \bibnamefont{Schultz}},
  \bibinfo{journal}{J. Chem. Phys.} \textbf{\bibinfo{volume}{58}},
  \bibinfo{pages}{479} (\bibinfo{year}{1972}).

\bibitem[{\citenamefont{Trajmar and Hall}(1974)}]{tjhall}
\bibinfo{author}{\bibfnamefont{S.}~\bibnamefont{Trajmar}} \bibnamefont{and}
  \bibinfo{author}{\bibfnamefont{R.~I.} \bibnamefont{Hall}},
  \bibinfo{journal}{J. Phys. B.} \textbf{\bibinfo{volume}{7}},
  \bibinfo{pages}{L458} (\bibinfo{year}{1974}).

\bibitem[{\citenamefont{Beli\`c et~al.}(1981)\citenamefont{Beli\`c, Landau, and
  Hall}}]{belic}
\bibinfo{author}{\bibfnamefont{D.~S.} \bibnamefont{Beli\`c}},
  \bibinfo{author}{\bibfnamefont{M.}~\bibnamefont{Landau}}, \bibnamefont{and}
  \bibinfo{author}{\bibfnamefont{R.~I.} \bibnamefont{Hall}},
  \bibinfo{journal}{J. Phys. B.} \textbf{\bibinfo{volume}{14}},
  \bibinfo{pages}{175} (\bibinfo{year}{1981}).

\bibitem[{\citenamefont{Curtis and Walker}(1992)}]{curtiswalker}
\bibinfo{author}{\bibfnamefont{M.~G.} \bibnamefont{Curtis}} \bibnamefont{and}
  \bibinfo{author}{\bibfnamefont{I.~C.} \bibnamefont{Walker}},
  \bibinfo{journal}{J. Chem. Soc. Faraday Trans.}
  \textbf{\bibinfo{volume}{88}}, \bibinfo{pages}{2805} (\bibinfo{year}{1992}).

\bibitem[{\citenamefont{Claydon et~al.}(1971)\citenamefont{Claydon, Segal, and
  Taylor}}]{Claydon}
\bibinfo{author}{\bibfnamefont{C.~R.} \bibnamefont{Claydon}},
  \bibinfo{author}{\bibfnamefont{G.~A.} \bibnamefont{Segal}}, \bibnamefont{and}
  \bibinfo{author}{\bibfnamefont{H.~S.} \bibnamefont{Taylor}},
  \bibinfo{journal}{J. Chem. Phys} \textbf{\bibinfo{volume}{54}},
  \bibinfo{pages}{3799} (\bibinfo{year}{1971}).

\bibitem[{\citenamefont{Jungen et~al.}(1979)\citenamefont{Jungen, Vogt, and
  Staemmler}}]{Jungen}
\bibinfo{author}{\bibfnamefont{M.}~\bibnamefont{Jungen}},
  \bibinfo{author}{\bibfnamefont{J.}~\bibnamefont{Vogt}}, \bibnamefont{and}
  \bibinfo{author}{\bibfnamefont{V.}~\bibnamefont{Staemmler}},
  \bibinfo{journal}{Chem. Phys.} \textbf{\bibinfo{volume}{37}},
  \bibinfo{pages}{49} (\bibinfo{year}{1979}).

\bibitem[{\citenamefont{Gil et~al.}(1994)\citenamefont{Gil, Rescigno, McCurdy,
  and Lengsfeld~III}}]{Gil}
\bibinfo{author}{\bibfnamefont{T.~J.} \bibnamefont{Gil}},
  \bibinfo{author}{\bibfnamefont{T.~N.} \bibnamefont{Rescigno}},
  \bibinfo{author}{\bibfnamefont{C.~W.} \bibnamefont{McCurdy}},
  \bibnamefont{and} \bibinfo{author}{\bibfnamefont{B.~H.}
  \bibnamefont{Lengsfeld III}}, \bibinfo{journal}{Phys. Rev. A}
  \textbf{\bibinfo{volume}{49}}, \bibinfo{pages}{2642} (\bibinfo{year}{1994}).

\bibitem[{\citenamefont{Morgan}(1998)}]{Morgan}
\bibinfo{author}{\bibfnamefont{L.~A.} \bibnamefont{Morgan}},
  \bibinfo{journal}{J. Phys. B} \textbf{\bibinfo{volume}{31}},
  \bibinfo{pages}{5003} (\bibinfo{year}{1998}).

\bibitem[{\citenamefont{Gorfinkel et~al.}(2002)\citenamefont{Gorfinkel, Morgan,
  and Tennyson}}]{Gorfinkiel}
\bibinfo{author}{\bibfnamefont{J.~D.} \bibnamefont{Gorfinkel}},
  \bibinfo{author}{\bibfnamefont{L.~A.} \bibnamefont{Morgan}},
  \bibnamefont{and} \bibinfo{author}{\bibfnamefont{J.}~\bibnamefont{Tennyson}},
  \bibinfo{journal}{J. Phys. B} \textbf{\bibinfo{volume}{35}},
  \bibinfo{pages}{543} (\bibinfo{year}{2002}).

\bibitem[{\citenamefont{Haxton et~al.}(2003{\natexlab{a}})\citenamefont{Haxton,
  Zhang, McCurdy, and Rescigno}}]{haxton1}
\bibinfo{author}{\bibfnamefont{D.~J.} \bibnamefont{Haxton}},
  \bibinfo{author}{\bibfnamefont{Z.}~\bibnamefont{Zhang}},
  \bibinfo{author}{\bibfnamefont{C.~W.} \bibnamefont{McCurdy}},
  \bibnamefont{and} \bibinfo{author}{\bibfnamefont{T.~N.}
  \bibnamefont{Rescigno}}, \bibinfo{journal}{Phys. Rev. A}
  \textbf{\bibinfo{volume}{69}}, \bibinfo{pages}{062713}
  (\bibinfo{year}{2003}{\natexlab{a}}).

\bibitem[{\citenamefont{Haxton et~al.}(2003{\natexlab{b}})\citenamefont{Haxton,
  Zhang, Meyer, Rescigno, and McCurdy}}]{haxton2}
\bibinfo{author}{\bibfnamefont{D.~J.} \bibnamefont{Haxton}},
  \bibinfo{author}{\bibfnamefont{Z.}~\bibnamefont{Zhang}},
  \bibinfo{author}{\bibfnamefont{H.-D.} \bibnamefont{Meyer}},
  \bibinfo{author}{\bibfnamefont{T.~N.} \bibnamefont{Rescigno}},
  \bibnamefont{and} \bibinfo{author}{\bibfnamefont{C.~W.}
  \bibnamefont{McCurdy}}, \bibinfo{journal}{Phys. Rev. A}
  \textbf{\bibinfo{volume}{69}}, \bibinfo{pages}{062714}
  (\bibinfo{year}{2003}{\natexlab{b}}).

\bibitem[{\citenamefont{Haxton et~al.}(2006)\citenamefont{Haxton, Rescigno, and
  McCurdy}}]{haxton4}
\bibinfo{author}{\bibfnamefont{D.~J.} \bibnamefont{Haxton}},
  \bibinfo{author}{\bibfnamefont{T.~N.} \bibnamefont{Rescigno}},
  \bibnamefont{and} \bibinfo{author}{\bibfnamefont{C.~W.}
  \bibnamefont{McCurdy}}, \bibinfo{journal}{Phys. Rev. A}
  \textbf{\bibinfo{volume}{73}}, \bibinfo{pages}{062724}
  (\bibinfo{year}{2006}).

\bibitem[{\citenamefont{Haxton et~al.}(2005)\citenamefont{Haxton, Rescigno, and
  McCurdy}}]{haxton3}
\bibinfo{author}{\bibfnamefont{D.~J.} \bibnamefont{Haxton}},
  \bibinfo{author}{\bibfnamefont{T.~N.} \bibnamefont{Rescigno}},
  \bibnamefont{and} \bibinfo{author}{\bibfnamefont{C.~W.}
  \bibnamefont{McCurdy}}, \bibinfo{journal}{Phys. Rev. A}
  \textbf{\bibinfo{volume}{72}}, \bibinfo{pages}{022705}
  (\bibinfo{year}{2005}).

\bibitem[{\citenamefont{Kohn}(1948)}]{kohn1}
\bibinfo{author}{\bibfnamefont{W.}~\bibnamefont{Kohn}}, \bibinfo{journal}{Phys.
  Rev.} \textbf{\bibinfo{volume}{74}}, \bibinfo{pages}{1763}
  (\bibinfo{year}{1948}).

\bibitem[{\citenamefont{Nesbet}(1968)}]{kohn2}
\bibinfo{author}{\bibfnamefont{R.~K.} \bibnamefont{Nesbet}},
  \bibinfo{journal}{Phys. Rev.} \textbf{\bibinfo{volume}{175}},
  \bibinfo{pages}{134} (\bibinfo{year}{1968}).

\bibitem[{\citenamefont{Nesbet}(1969)}]{kohn3}
\bibinfo{author}{\bibfnamefont{R.~K.} \bibnamefont{Nesbet}},
  \bibinfo{journal}{Phys. Rev.} \textbf{\bibinfo{volume}{179}},
  \bibinfo{pages}{60} (\bibinfo{year}{1969}).

\bibitem[{\citenamefont{Hazi et~al.}(1981)\citenamefont{Hazi, Rescigno, and
  Kurilla}}]{Kurilla}
\bibinfo{author}{\bibfnamefont{A.~U.} \bibnamefont{Hazi}},
  \bibinfo{author}{\bibfnamefont{T.}~\bibnamefont{Rescigno}}, \bibnamefont{and}
  \bibinfo{author}{\bibfnamefont{M.}~\bibnamefont{Kurilla}},
  \bibinfo{journal}{Phys. Rev. A} \textbf{\bibinfo{volume}{23}},
  \bibinfo{pages}{1089} (\bibinfo{year}{1981}).

\bibitem[{\citenamefont{Miller and op~de Haar}(1987)}]{kohn5}
\bibinfo{author}{\bibfnamefont{W.~H.} \bibnamefont{Miller}} \bibnamefont{and}
  \bibinfo{author}{\bibfnamefont{B.~M. D. D.~J.} \bibnamefont{op~de Haar}},
  \bibinfo{journal}{J. Chem. Phys.} \textbf{\bibinfo{volume}{86}},
  \bibinfo{pages}{6213} (\bibinfo{year}{1987}).

\bibitem[{\citenamefont{Schneider and Rescigno}(1988)}]{kohn6}
\bibinfo{author}{\bibfnamefont{B.~I.} \bibnamefont{Schneider}}
  \bibnamefont{and} \bibinfo{author}{\bibfnamefont{T.~N.}
  \bibnamefont{Rescigno}}, \bibinfo{journal}{Phys. Rev. A}
  \textbf{\bibinfo{volume}{47}}, \bibinfo{pages}{3749} (\bibinfo{year}{1988}).

\bibitem[{\citenamefont{Zhang et~al.}(1988)\citenamefont{Zhang, Chu, and
  Miller}}]{kohn7}
\bibinfo{author}{\bibfnamefont{J.~Z.~H.} \bibnamefont{Zhang}},
  \bibinfo{author}{\bibfnamefont{S.-I.} \bibnamefont{Chu}}, \bibnamefont{and}
  \bibinfo{author}{\bibfnamefont{W.~H.} \bibnamefont{Miller}},
  \bibinfo{journal}{J. Chem. Phys.} \textbf{\bibinfo{volume}{88}},
  \bibinfo{pages}{6233} (\bibinfo{year}{1988}).

\bibitem[{\citenamefont{Lengsfeld III and Rescigno}(1991)}]{kohn8}
\bibinfo{author}{\bibfnamefont{B.~H.} \bibnamefont{Lengsfeld III}} \bibnamefont{and}
  \bibinfo{author}{\bibfnamefont{T.~N.} \bibnamefont{Rescigno}},
  \bibinfo{journal}{Phys. Rev. A} \textbf{\bibinfo{volume}{44}},
  \bibinfo{pages}{2913} (\bibinfo{year}{1991}).

\bibitem[{\citenamefont{Rescigno
  et~al.}(1995{\natexlab{a}})\citenamefont{Rescigno, McCurdy, Orel, and
  Lengsfeld III}}]{rmo95}
\bibinfo{author}{\bibfnamefont{T.~N.} \bibnamefont{Rescigno}},
  \bibinfo{author}{\bibfnamefont{C.~W.} \bibnamefont{McCurdy}},
  \bibinfo{author}{\bibfnamefont{A.~E.} \bibnamefont{Orel}}, \bibnamefont{and}
  \bibinfo{author}{\bibfnamefont{B.~H.} \bibnamefont{Lengsfeld III}}, in
  \emph{\bibinfo{booktitle}{Computational Methods for Electron-Molecule
  Collisions}}, edited by \bibinfo{editor}{\bibfnamefont{W.~M.}
  \bibnamefont{Huo}} \bibnamefont{and} \bibinfo{editor}{\bibfnamefont{F.~A.}
  \bibnamefont{Gianturco}} (\bibinfo{publisher}{Plenum}, \bibinfo{address}{New
  York}, \bibinfo{year}{1995}{\natexlab{a}}).

\bibitem[{\citenamefont{Rescigno
  et~al.}(1995{\natexlab{b}})\citenamefont{Rescigno, Lengsfeld III, and McCurdy}}]{rlm95}
\bibinfo{author}{\bibfnamefont{T.~N.} \bibnamefont{Rescigno}},
  \bibinfo{author}{\bibfnamefont{B.~H.} \bibnamefont{Lengsfeld III}},
  \bibnamefont{and} \bibinfo{author}{\bibfnamefont{C.~W.}
  \bibnamefont{McCurdy}}, in \emph{\bibinfo{booktitle}{Modern Electronic
  Structure Theory}}, edited by \bibinfo{editor}{\bibfnamefont{D.~R.}
  \bibnamefont{Yarkony}} (\bibinfo{publisher}{World Scientific},
  \bibinfo{address}{Singapore}, \bibinfo{year}{1995}{\natexlab{b}}),
  vol.~\bibinfo{volume}{1}, pp. \bibinfo{pages}{501--588}.

\bibitem[{\citenamefont{Feshbach}(1962)}]{Feshbach}
\bibinfo{author}{\bibfnamefont{H.}~\bibnamefont{Feshbach}},
  \bibinfo{journal}{Ann. Phys.} \textbf{\bibinfo{volume}{19}},
  \bibinfo{pages}{287} (\bibinfo{year}{1962}).

\bibitem[{\citenamefont{Newton}(1982)}]{Newton}
\bibinfo{author}{\bibfnamefont{R.~G.} \bibnamefont{Newton}},
  \emph{\bibinfo{title}{Scattering Theory of Particles and Waves}}
  (\bibinfo{publisher}{Springer-Verlag}, \bibinfo{address}{New York},
  \bibinfo{year}{1982}), \bibinfo{edition}{2nd} ed.

\bibitem[{\citenamefont{van Harrevelt and van Hemert}(2000)}]{robsurf}
\bibinfo{author}{\bibfnamefont{R.}~\bibnamefont{van Harrevelt}}
  \bibnamefont{and} \bibinfo{author}{\bibfnamefont{M.~C.} \bibnamefont{van
  Hemert}}, \bibinfo{journal}{J. Chem. Phys.} \textbf{\bibinfo{volume}{112}},
  \bibinfo{pages}{5777} (\bibinfo{year}{2000}).

\bibitem[{\citenamefont{Kendall et~al.}(1992)\citenamefont{Kendall,
  T.~H.~Dunning, and Harrison}}]{Dunning}
\bibinfo{author}{\bibfnamefont{R.~A.} \bibnamefont{Kendall}},
  \bibinfo{author}{\bibfnamefont{J.}~\bibnamefont{T.~H.~Dunning}},
  \bibnamefont{and} \bibinfo{author}{\bibfnamefont{R.~J.}
  \bibnamefont{Harrison}}, \bibinfo{journal}{J. Chem. Phys.}
  \textbf{\bibinfo{volume}{96}}, \bibinfo{pages}{6796} (\bibinfo{year}{1992}).

\bibitem[{\citenamefont{Chipman}(1989)}]{Chipman}
\bibinfo{author}{\bibfnamefont{D.}~\bibnamefont{Chipman}},
  \bibinfo{journal}{Theor. Chim. Acta} \textbf{\bibinfo{volume}{76}},
  \bibinfo{pages}{73} (\bibinfo{year}{1989}).

\bibitem[{epa()}]{epaps}
\bibinfo{note}{EPAPS Document No. E-PLRAAN-75-005702 for subroutines that
  generate the real and imaginary components of the constructed potential
  energy surfaces. For more information on EPAPS, see
  http://www.aip.org/pubservs/epaps.html}.

\bibitem[{\citenamefont{Nestmann and Peyerimhoff}(1990)}]{nestmann}
\bibinfo{author}{\bibfnamefont{B.~M.} \bibnamefont{Nestmann}} \bibnamefont{and}
  \bibinfo{author}{\bibfnamefont{S.~D.} \bibnamefont{Peyerimhoff}},
  \bibinfo{journal}{J. Phys. B} \textbf{\bibinfo{volume}{23}},
  \bibinfo{pages}{L773} (\bibinfo{year}{1990}).

\bibitem[{\citenamefont{Chu}(1974)}]{OH1974}
\bibinfo{author}{\bibfnamefont{S.-I.} \bibnamefont{Chu}}, \bibinfo{journal}{J.
  Chem. Phys.} \textbf{\bibinfo{volume}{61}}, \bibinfo{pages}{5389}
  (\bibinfo{year}{1974}).

\bibitem[{\citenamefont{Domcke and Stock}(1997)}]{diareview}
\bibinfo{author}{\bibfnamefont{W.}~\bibnamefont{Domcke}} \bibnamefont{and}
  \bibinfo{author}{\bibfnamefont{G.}~\bibnamefont{Stock}},
  \bibinfo{journal}{Adv. Chem. Phys.} \textbf{\bibinfo{volume}{100}},
  \bibinfo{pages}{1} (\bibinfo{year}{1997}).

\bibitem[{\citenamefont{Yarkony}(1996)}]{yarkonyconsequence}
\bibinfo{author}{\bibfnamefont{D.~R.} \bibnamefont{Yarkony}},
  \bibinfo{journal}{J. Chem. Phys.} \textbf{\bibinfo{volume}{105}},
  \bibinfo{pages}{10456} (\bibinfo{year}{1996}).

\bibitem[{\citenamefont{Yarkony}(2001)}]{yarkonyCW}
\bibinfo{author}{\bibfnamefont{D.~R.} \bibnamefont{Yarkony}},
  \bibinfo{journal}{J. Phys. Chem.} \textbf{\bibinfo{volume}{105}},
  \bibinfo{pages}{6277} (\bibinfo{year}{2001}).

\bibitem[{\citenamefont{Sadygov and Yarkony}(1998)}]{atdtpoisson}
\bibinfo{author}{\bibfnamefont{R.~G.} \bibnamefont{Sadygov}} \bibnamefont{and}
  \bibinfo{author}{\bibfnamefont{D.~R.} \bibnamefont{Yarkony}},
  \bibinfo{journal}{J. Chem. Phys.} \textbf{\bibinfo{volume}{109}},
  \bibinfo{pages}{20} (\bibinfo{year}{1998}).

\bibitem[{\citenamefont{Mead}(1992)}]{meadrev}
\bibinfo{author}{\bibfnamefont{C.~A.} \bibnamefont{Mead}},
  \bibinfo{journal}{Rev. Mod. Phys.} \textbf{\bibinfo{volume}{64}},
  \bibinfo{pages}{51} (\bibinfo{year}{1992}).

\bibitem[{\citenamefont{Baer}(2002)}]{baerrev}
\bibinfo{author}{\bibfnamefont{M.}~\bibnamefont{Baer}},
  \bibinfo{journal}{Physics Reports} \textbf{\bibinfo{volume}{358}},
  \bibinfo{pages}{75} (\bibinfo{year}{2002}).

\bibitem[{\citenamefont{Macias and Riera}(1978)}]{propertydia}
\bibinfo{author}{\bibfnamefont{A.}~\bibnamefont{Macias}} \bibnamefont{and}
  \bibinfo{author}{\bibfnamefont{A.}~\bibnamefont{Riera}}, \bibinfo{journal}{J.
  Phys. B} \textbf{\bibinfo{volume}{11}}, \bibinfo{pages}{L489}
  (\bibinfo{year}{1978}).

\bibitem[{\citenamefont{Werner and Meyer}(1981)}]{dipoledia}
\bibinfo{author}{\bibfnamefont{H.-J.} \bibnamefont{Werner}} \bibnamefont{and}
  \bibinfo{author}{\bibfnamefont{W.}~\bibnamefont{Meyer}}, \bibinfo{journal}{J.
  Chem. Phys.} \textbf{\bibinfo{volume}{74}}, \bibinfo{pages}{5802}
  (\bibinfo{year}{1981}).

\bibitem[{\citenamefont{van Harrevelt and van Hemert}(2001)}]{aband_photo}
\bibinfo{author}{\bibfnamefont{R.}~\bibnamefont{van Harrevelt}}
  \bibnamefont{and} \bibinfo{author}{\bibfnamefont{M.~C.} \bibnamefont{van
  Hemert}}, \bibinfo{journal}{J. Chem. Phys.} \textbf{\bibinfo{volume}{114}},
  \bibinfo{pages}{9453} (\bibinfo{year}{2001}).

\bibitem[{\citenamefont{Wilkinson}(1963)}]{energy1}
\bibinfo{author}{\bibfnamefont{P.~G.} \bibnamefont{Wilkinson}},
  \bibinfo{journal}{Astrophys. J.} \textbf{\bibinfo{volume}{138}},
  \bibinfo{pages}{778} (\bibinfo{year}{1963}).

\bibitem[{\citenamefont{Carlone and Dalby}(1969)}]{energy2}
\bibinfo{author}{\bibfnamefont{C.}~\bibnamefont{Carlone}} \bibnamefont{and}
  \bibinfo{author}{\bibfnamefont{F.~W.} \bibnamefont{Dalby}},
  \bibinfo{journal}{Can. J. Phys.} \textbf{\bibinfo{volume}{47}},
  \bibinfo{pages}{1945} (\bibinfo{year}{1969}).

\bibitem[{\citenamefont{Russic and \textit{et al.}}(2002)}]{energy3}
\bibinfo{author}{\bibfnamefont{B.}~\bibnamefont{Russic}} \bibnamefont{and}
  \bibinfo{author}{\bibnamefont{\textit{et al.}}}, \bibinfo{journal}{J. Phys.
  Chem. A} \textbf{\bibinfo{volume}{106}}, \bibinfo{pages}{2727}
  (\bibinfo{year}{2002}).

\bibitem[{\citenamefont{Herzberg}(1969)}]{energy4}
\bibinfo{author}{\bibfnamefont{G.}~\bibnamefont{Herzberg}},
  \bibinfo{journal}{Phys. Rev. Lett.} \textbf{\bibinfo{volume}{23}},
  \bibinfo{pages}{1081} (\bibinfo{year}{1969}).

\bibitem[{\citenamefont{Smith et~al.}(1997)\citenamefont{Smith, Kim, and
  Lineberger}}]{energy5}
\bibinfo{author}{\bibfnamefont{J.~R.} \bibnamefont{Smith}},
  \bibinfo{author}{\bibfnamefont{J.~B.} \bibnamefont{Kim}}, \bibnamefont{and}
  \bibinfo{author}{\bibfnamefont{W.~C.} \bibnamefont{Lineberger}},
  \bibinfo{journal}{Phys. Rev. A} \textbf{\bibinfo{volume}{55}},
  \bibinfo{pages}{2036} (\bibinfo{year}{1997}).

\bibitem[{\citenamefont{Lykke et~al.}(1991)\citenamefont{Lykke, Murray, and
  Lineberger}}]{energy6}
\bibinfo{author}{\bibfnamefont{K.~R.} \bibnamefont{Lykke}},
  \bibinfo{author}{\bibfnamefont{K.~K.} \bibnamefont{Murray}},
  \bibnamefont{and} \bibinfo{author}{\bibfnamefont{W.~C.}
  \bibnamefont{Lineberger}}, \bibinfo{journal}{Phys. Rev. A}
  \textbf{\bibinfo{volume}{43}}, \bibinfo{pages}{6104} (\bibinfo{year}{1991}).

\bibitem[{\citenamefont{Valli et~al.}(1999)\citenamefont{Valli, Blondel, and
  Delsart}}]{energy7}
\bibinfo{author}{\bibfnamefont{C.}~\bibnamefont{Valli}},
  \bibinfo{author}{\bibfnamefont{C.}~\bibnamefont{Blondel}}, \bibnamefont{and}
  \bibinfo{author}{\bibfnamefont{C.}~\bibnamefont{Delsart}},
  \bibinfo{journal}{Phys. Rev. A} \textbf{\bibinfo{volume}{59}},
  \bibinfo{pages}{3809} (\bibinfo{year}{1999}).

\bibitem[{\citenamefont{Chu et~al.}(1974)\citenamefont{Chu, Yoshimine, and
  Liu}}]{CYL}
\bibinfo{author}{\bibfnamefont{S.-I.} \bibnamefont{Chu}},
  \bibinfo{author}{\bibfnamefont{M.}~\bibnamefont{Yoshimine}},
  \bibnamefont{and} \bibinfo{author}{\bibfnamefont{B.}~\bibnamefont{Liu}},
  \bibinfo{journal}{J. Chem. Phys.} \textbf{\bibinfo{volume}{61}},
  \bibinfo{pages}{5389} (\bibinfo{year}{1974}).

\bibitem[{\citenamefont{Kolos and Wolniewicz}(1964)}]{KW}
\bibinfo{author}{\bibfnamefont{W.}~\bibnamefont{Kolos}} \bibnamefont{and}
  \bibinfo{author}{\bibfnamefont{L.}~\bibnamefont{Wolniewicz}},
  \bibinfo{journal}{J. Chem. Phys.} \textbf{\bibinfo{volume}{41}},
  \bibinfo{pages}{3663} (\bibinfo{year}{1964}).

\end{thebibliography}
\end{document}